\documentclass{LMCS}

\usepackage{step-indexing}
\usepackage{enumerate,hyperref}
\theoremstyle{plain}

\def\doi{5 (4:2) 2009}
\lmcsheading%
{\doi}
{1--48}
{}
{}
{Mar.~\phantom{0}3, 2009}
{Dec.~17, 2009}
{}   

\begin{document}
 
\title[A Step-indexed Semantics of Imperative Objects]{A Step-indexed Semantics of Imperative Objects\rsuper*}

\author[C.~Hri\c{t}cu]{C\u{a}t\u{a}lin Hri\c{t}cu\rsuper a}	
\address{{\lsuper a}Department of Computer Science, Saarland University, Saarbr\"{u}cken, Germany}	
\email{\texttt{hritcu@cs.uni-sb.de}}  

\author[J.~Schwinghammer]{Jan Schwinghammer\rsuper b}	
\address{{\lsuper b}Programming Systems Lab, Saarland University, Saarbr\"{u}cken, Germany}	
\email{\texttt{jan@ps.uni-sb.de}}  




\titlecomment{{\lsuper*}A preliminary version of this paper was presented at the
International Workshop on Foundations of Object-Oriented Languages (FOOL'08),
13 January 2008, San Francisco, California}


\begin{abstract}
\noindent
Step-indexed semantic interpretations of types were proposed as an alternative to
purely syntactic proofs of type safety using subject reduction. The types are
interpreted as sets of values indexed by the number of computation steps for
which these values are guaranteed to behave like proper elements of the type.
Building on work by Ahmed, Appel and others, we introduce a step-indexed
semantics for the imperative object calculus of Abadi and Cardelli. Providing a
semantic account of this calculus using more `traditional', domain-theoretic
approaches has proved challenging due to the combination of dynamically allocated
objects, higher-order store, and an expressive type system. Here we show that,
using step-indexing, one can interpret a rich type discipline with object types,
subtyping, recursive and bounded quantified types in the presence of state.
\end{abstract}

\maketitle
 
\section{Introduction}
\label{sec:introduction}

\noindent The \emph{imperative object calculus} of Abadi and Cardelli is a very
small, yet very expressive object-oriented language \cite{Abadi:Cardelli:96}.
Despite the extreme simplicity of its syntax, the calculus models many important
concepts of object-oriented programming,
as well as the often subtle interaction between them. 
In particular it raises interesting and non-trivial
questions with respect to typing.

In contrast to the more common class-based object-oriented languages, in the
imperative object 
calculus every object comes equipped with its own set of methods that can be 
updated at run-time. As a consequence, the methods need to reside in the store, 
\IE the store is \emph{higher-order}. Moreover, objects are
\emph{allocated dynamically} and aliasing is possible. 
Dynamically-allocated, higher-order store is present in different forms
in many practical programming languages (\EG pointers to functions in C and
 general references in SML), but it considerably complicates
the construction of adequate semantic models in which one can reason about the
behaviour of programs
(as pointed out for instance by Reus \cite{Reus:03}). 

Purely syntactic arguments such as subject reduction suffice for proving the 
soundness of traditional type systems. However, once such type systems are 
turned into powerful specification languages, like the logic of objects of Abadi
and Leino \cite{Abadi:Leino:04} or the hybrid type
system of Flanagan \ETAL
\cite{Flanagan:Freund:Tomb:06}, purely syntactic arguments 
seem no longer appropriate. 
The meaning of assertions is no longer obvious,
since they have to describe the code on the heap. 
We believe that specifications of program behaviour should have a meaning 
independent of the particular proof system on which syntactic 
preservation proofs rely, as also argued by Benton \cite{Benton:05} and by Reus
and Schwinghammer \cite{Reus:Schwinghammer:06}.

In the case of specifications one would ideally prove soundness with respect to
a semantic 
model that makes a clear distinction between semantic validity and derivability 
using the syntactic rules. However, building such semantic models is challenging,
and there is currently no fully satisfactory semantic account of the imperative
object calculus:

\begin{enumerate}[\hbox to8 pt{\hfill}]
\item\noindent{\hskip-12 pt\bf Denotational semantics:}\
Domain-theoretic models have been employed in proving the soundness of the
logic of Abadi and Leino \cite{Reus:Schwinghammer:06, Reus:Streicher:04}.
However, the existing techniques fall short of providing
convincing models of \emph{typed} objects: Reus and Streicher 
\cite{Reus:Streicher:04} consider an
untyped semantics, and the model presented by Reus and Schwinghammer 
\cite{Reus:Schwinghammer:06}
 handles neither second-order types, nor subtyping in depth.
Due to the dynamically-allocated higher-order store present in the
imperative object calculus,
the models rely on techniques for recursively defined domains in functor 
categories \cite{Levy:02,Pitts:96}. This makes them {complex}, and 
establishing properties even for specific programs often requires a 
substantial effort.

\item\noindent{\hskip-12 pt\bf Equational reasoning:}\
Gordon \ETAL \cite{Gordon:Hankin:Lassen:97} develop reasoning principles for establishing the contextual 
equivalence of untyped objects, and apply 
them to prove correctness of a compiler
optimization. Jeffrey and Rathke \cite{Jeffrey:Rathke:05} 
consider a concurrent variant of the calculus and characterize may-testing 
equivalence in terms of the trace sets generated by a labeled transition 
system. In both cases the semantics is limited to
equational reasoning, \IE establishing contextual equivalences between
programs. In theory, this can be used to verify a program
by showing it  equivalent to one that is trivially correct and acts
as a specification. However, this can be more cumbersome in practice than using
program logics, the established formalism for specifying and proving the
correctness of programs.

\item\noindent{\hskip-12 pt\bf Translations:}\
Abadi \ETAL \cite{Abadi:Cardelli:Viswanathan:96} give an adequate encoding of the imperative object calculus
into a lambda calculus with records, references, recursive and existential types
and subtyping.
Together with an interpretation of this target language, an adequate model for
the imperative object calculus could, in principle, be obtained.
However, we are not aware of any worked-out adequate domain-theoretic models for
general references and impredicative second-order types.
Even if such a model was given, it would still be preferable to have a
self-contained semantics for the
object calculus, without the added
complexity of the (non-trivial) translation.
\end{enumerate}

\noindent A solution to the problem of finding adequate models of
objects could be the step-indexed semantic models of types, introduced
by Appel and McAllester \cite{Appel:McAllester:01} as an alternative
to subject reduction proofs.  Such models are based directly on the
operational semantics, and are more easy to construct than the
existing domain-theoretic models. The types are simply interpreted as
sets of syntactic values indexed by a number of computation
steps. Intuitively, a term belongs to a certain type if it behaves
like an element of that type for any number of steps.  Every type is
built as a sequence of increasingly accurate semantic approximations,
which allows one to easily deal with recursion. Type safety is an
immediate consequence of this interpretation of types, and the
semantic counterparts of the usual typing rules are proved as
independent lemmas, either directly or by induction on the index.
Ahmed \ETAL \cite{Ahmed:04, Ahmed:Appel:Virga:03} successfully applied
this generic technique to a lambda calculus with general references,
impredicative polymorphism and recursive types.

In this paper we further extend the semantics of Ahmed \ETAL with object types and
subtyping, and we use the resulting interpretation to prove the soundness of an
expressive type system for the imperative object calculus. The main
contribution of our work is the novel semantics of object types.
We extend this semantics
in two orthogonal ways.
First, we adapt it to self types, \IE recursive object types that validate the
usual subtyping rules as well as strong typing rules with structural assumptions.
Second, we study a natural generalization of object types that results in
simpler and more expressive typing rules.

Even though in this paper we are concerned with the safety of a type system,
the step-indexing technique is not restricted to types,
and has already been used for equational reasoning \cite{Ahmed:Dreyer:Rossberg:09,Ahmed:06, 
Appel:McAllester:01} and for proving the
soundness of Hoare-style program logics of low-level languages \cite{Benton:05, Benton:06}.
We expect therefore that it will eventually become possible to use a
step-indexed model to prove the soundness of more expressive program logics
for the imperative object calculus.

\subsection*{Outline} 
The next section introduces the syntax, operational semantics, and type
system that we consider for the imperative object calculus.
In Section~\ref{sec:model} we present a step-indexed semantics for this
calculus.
In particular, we define the interpretations of types and establish
their semantic properties.
In Section~\ref{sec:soundness} these properties are used to prove the soundness
of the type system.
Section~\ref{sec:self-types} studies self types, while
Section~\ref{sec:gen-obj} discusses a natural generalization of object types.
Section~\ref{sec:related} gives a comparison to related work and
Section~\ref{sec:conclusion} concludes.
The Appendix presents the proofs of the most interesting typing and
subtyping lemmas for object types, while an earlier technical report contains
additional proofs \cite{Hritcu:Schwinghammer:TR2007}.

\section{The Imperative Object Calculus}
\label{sec:calculus}

\noindent We recall the syntax of the imperative object calculus with
recursive and second-order types, and introduce a small-step
operational semantics for this calculus that is equivalent to the
big-step semantics given by Abadi and Cardelli
\cite{Abadi:Cardelli:96}.

\subsection{Syntax}

\begin{figure}
\centering
\begin{align*} 
A, B, C 
&\bnfeq X \hmid \stopp \hmid \sbott \hmid 
\sarr{A}{B} && \text{(type expressions)}\\
 &\hbnfmid \sobjtv{A}{d} \hmid
 \smu{X}{A} \\
  &\hbnfmid \sforall{X}{A}{B} \hmid \sexists{X}{A}{B} \\[1mm]
\tvar &\bnfeq \tinv \hmid \tcov \hmid \tcon&& \text{(variance annotations)}\\[1mm]
a,b
& \bnfeq  x && \text{(variable)}\\
& \hbnfmid  \sobj{A}{d} && \text{(object creation)}\\
& \hbnfmid  a.\m && \text{(method invocation)}\\
& \hbnfmid  \supd{a}{\m}{\sself{x}{A}{b}} && \text{(method update)}\\
& \hbnfmid  \sclone{a} && \text{(shallow copy)}\\[1mm]
& \hbnfmid  \slam{x}{A}{b} && \text{(procedure)}\\
& \hbnfmid  \sapp{a}{b} & &\text{(application)}\\[1mm]
& \hbnfmid  \sfoldd{A}{b} & &\text{(recursive  folding)}\\
& \hbnfmid  \sunfoldd{A}{b} & &\text{(recursive  unfolding)}\\[1mm]
& \hbnfmid  \stlam{X}{A}{b} && \text{(type abstraction)}\\
& \hbnfmid  \stapp{a}{A} && \text{(type application)}\\
& \hbnfmid  \spack{X}{A}{C}{a}{B} && \text{(existential package)}\\ 
& \hbnfmid  \sopen{a}{X}{A}{x}{B}{b}{C} && \text{(package opening)}
\end{align*}
\caption{\label{fig:syntax} Syntax of types and terms}
\end{figure}

Let $\msn{Var}, \msn{TVar}$ and $\msn{Meth}$ be pairwise disjoint, countably 
infinite sets of \emph{variables}, \emph{type variables} and \emph{method
names}, respectively. Let $x,y$ range over \msn{Var}, $X,Y$ range over
\msn{TVar}, and let $\m$ range over $\msn{Meth}$.
Figure~\ref{fig:syntax} defines the syntax of the types and terms of the
imperative object calculus.

Objects are unordered collections of named methods, written as $\sobj{A}{d}$.
In a method $\m=\sself{x}{A}{b}$, $\varsigma$ is a
 binder that binds the `self' argument $x$ in the method body $b$. The self 
argument can be used inside the method body for invoking the methods of the 
containing object. Methods with arguments other than self can be obtained by 
having a procedure as the method body.
The methods of an object can be invoked 
or updated, but no new methods can be added, and the existing methods cannot be 
deleted. The type of objects with methods named $\m_d$ that return results of 
type $A_d$, for $d$ in some set $D$, is written as $\sobjtv{A}{d}$, where 
$\tvar\in\left\{\tinv,\tcov,\tcon\right\}$ is a \emph{variance annotation} that indicates if 
the method is considered \emph{invoke-only} (\tcov), \emph{update-only} (\tcon), 
or if it may be used without restriction (\tinv).

While procedural abstractions are  sometimes defined in the imperative object
calculus using an additional let construct, we include them as primitives.  We
write procedures with type $\sarr{A}{B}$ as $\slam{x}{A}{b}$ and applications as $\sapp{a}{b}$, respectively.
We use $\sfoldd{A}{}$ and $\sunfoldd{A}{}$ to denote the 
isomorphism between a recursive type $\smu{X}{B}$ and its unfolding 
$\msub{B}{\mmap{X}{\smu{X}{B}}}$. Finally, we consider bounded universal and 
existential types $\sforall{X}{A}{B}$ and $\sexists{X}{A}{B}$ along with their 
introduction and elimination forms \cite{Cardelli:97}.

The set of free variables of a term $a$ is denoted by $\mfv{a}$, and
similarly the free type variables in a type $A$ by $\mfv{A}$. We identify types
and terms up to the consistent renaming of bound variables.
We use $\mmap{t}{r}$ to denote the singleton map that maps $t$ to $r$. For a
finite map $\sigma$ from variables to terms, $\msub{a}{\sigma}$ denotes the
result of capture-avoiding substitution of all  $x\in\mfv{a} \cap
\mdom{\sigma}$ by $\sigma(x)$.
The same notation is used for the substitution of type variables.
Generally, for any function $f$, the notation
$\mext{f}{t}{r}$ denotes the function that maps $t$ to $r$, and otherwise agrees
with $f$.

\subsection{Operational Semantics}
\label{subsec:operational-semantics}
Let $\msn{Loc}$ be a countably infinite set of
\emph{heap locations} ranged over by $\loc$. We extend the set of terms 
 by run-time
representations of objects \svobj{d}, associating heap locations to a set of
method names. \emph{Values} are given by the grammar:
\begin{align*}
 v \in \msn{Val} \bnfeq  \svobj{d} \mbar \slam{x}{A}{b} \mbar \sfoldd{A}{v}  
 \mbar \stlam{X}{A}{b} \mbar \spack{X}{A}{C}{v}{B}
\end{align*}
Apart from run-time objects, values consist of procedures, values of recursive
type, type abstractions and existential packages as in the call-by-value lambda
calculus. We often only consider terms 
and values
without free variables, and denote the set of these \emph{closed
terms} and \emph{closed
values} by $\msn{CTerm}$ and $\msn{CVal}$, respectively.
A \emph{program} is a closed term that does not contain any locations,
and we denote the set of all programs by $\msn{Prog}$. 
A \emph{heap} $h$ is a finite map from \msn{Loc} to \msn{CVal}\footnote{In
fact, for the purpose of modelling the imperative object calculus it would
suffice to regard procedures as the only kind of storable value.}, and we
write \msn{Heap} for the set of all heaps.

\begin{figure}

\centering
\begin{align*}
\C[\cdot]  &\bnfeq [\cdot] \hmid \C.\m \hmid 
\supd{\C}{\m}{\sself{x}{A}{b}} \hmid \sclone{\C} \hmid \sapp{\C}{b} \hmid \sapp{v}{\C} \hmid \sfoldd{A}{\C} \hmid
\sunfoldd{A}{\C}\\
&   \hbnfmid \stapp{\C}{A} \hmid  \spack{X}{A}{C}{\C}{B}
\hmid \sopen{\C}{X}{A}{x}{B}{b}{C}
\end{align*}

\caption{\label{fig:evaluation-contexts}Evaluation contexts}
\end{figure}

\begin{figure*}
\centering
\begin{align*}
&\textsc{\small{(Red-Obj)}}
&\xcfg{\h}{\sobj{A}{d}} &\tred{}{}
\xcfg{\mextx{\h}{l_d}{\slam{x_d}{A}{b_d}}{d \in D}}{\svobj{d}}\\
&&& {\quad\ \text{where $\forall d \in D.\; l_d \notin \mdom{\h}$}}\\[1mm]
&\textsc{\small{(Red-Inv)}}
&\xcfg{\h}{\svobj{d}.\m_e} &\tred{}{}
\xcfg{\h}{\sapp{\h(l_e)}{\!\svobj{d}}}\text{, if $e {\in} D$}\\[1mm]
&\textsc{\small{(Red-Upd)}}
&\xcfg{\h}{\supd{\svobj{d}}{\m_e}{\sself{x}{A}{b}}} &\tred{}{}
\xcfg{\mextx{\h}{l_e}{\slam{x}{A}{b}}}{\svobj{d}}\text{, if $e {\in} D$}\\[1mm]
&\textsc{\small{(Red-Clone)}}
&\xcfg{\h}{\sclone{\svobj{d}}} &\tred{}{}
\xcfg{\mextx{\h}{l_d'}{\h(l_d)}{d \in D}}{\svobjx{l'_d}{d}}\\
&&& {\quad\ \text{where $\forall d \in D.\; l_d' \notin \mdom{\h}$}}\\[1mm]
&\textsc{\small{(Red-Beta)}}
&\xcfg{\h}{\sapp{(\slam{x}{A}{b})}{v}} &\tred{}{}
\xcfg{\h}{\msub{b}{\mmap{x}{v}}}\\[1mm]
&\textsc{\small{(Red-Unfold)}}\hspace{-1cm}
&\xcfg{\h}{\sunfoldd{A}{(\sfoldd{B}{v}})} &\tred{}{}\xcfg{\h}{v}\\[1mm]
&\textsc{\small{(Red-TBeta)}}
&\xcfg{\h}{\stapp{(\stlam{X}{A}{b})}{B}} &\tred{}{}
\xcfg{\h}{\msub{b}{\mmap{X}{B}}}\\[1mm]
&\textsc{\small{(Red-Open)}}
&\xcfg{\h}{\sopen{v}{X}{A}{x}{B}{b}{C}}
 &\tred{}{}
 \xcfg{\h}{\msub{b}{\mTWOmap{x}{v'}{X}{C'}}}\\
&&& {\quad\ \text{where $v\meqsyn \spack{X'}{A'}{C'}{v'}{B'}$}}
\end{align*}
\caption{\label{fig:reduction}One-step reduction relation}
\end{figure*}

Figure~\ref{fig:evaluation-contexts} defines the set of \emph{evaluation
contexts}, formalizing a left-to-right, call-by-value strategy.
We write $\C[a]$ for the term obtained by plugging $a$ into the hole
$[\cdot]$ of $\C$.  The one-step reduction relation 
$\tjred{}{}{}$ is defined as the least relation on \emph{configurations} 
$\xcfg{\h}{a}\in\msn{Heap}\times\msn{CTerm}$
generated by the rules in Figure~\ref{fig:reduction} and closed under the
following context rule:
\begin{align}
\tag*{\textsc{\small{(Red-Ctx)}}}
{\tredc{\h}{a}{\h'}{a'}}\ \Longrightarrow\ 
{\tredc{\h}{\C[a]}{\h'}{\C[a']}}\text{\hfill}
\end{align}

The methods are actually stored in the heap as procedures. Object
construction allocates new heap storage for these procedures and returns a record
of references to them \textsc{(Red-Obj)}. Upon method invocation the  
corresponding stored procedure is retrieved from the heap and applied to the 
enclosing object \textsc{(Red-Inv)}. The self parameter is thus passed just like any
other procedure argument. Identifying methods and procedures makes the
`self-application' semantics of method invocation explicit, while technically
it allows us to use the step-indexed model of Ahmed \ETAL \cite{Ahmed:04,
Ahmed:Appel:Virga:03} with only few modifications.

\begin{figure}
\centering
\textbf{Subtyping} \hfill $\boxed{\ssubtype{\Gamma}{A}{B}}$ 
\begin{align*}\mrule{SubRefl}{\sgtype{\Gamma}{A}}{\ssubtype{\Gamma}{A}{A}} \qquad 
\mrule{SubTrans}{\ssubtype{\Gamma}{A}{A'} \quad 
\ssubtype{\Gamma}{A'}{B}}{\ssubtype{\Gamma}{A}{B}}\end{align*}\\[\inferenceruleskip] 
\begin{align*}\mrule{SubTop}{\sgtype{\Gamma}{A}}{\ssubtype{\Gamma}{A}{\stopp}} \qquad 
\mrule{SubBot}{\sgtype{\Gamma}{A}}{\ssubtype{\Gamma}{\sbott}{A}} 
\qquad \mrule{SubVar}{\sgenv{\Gamma_1, X \ssub A, \Gamma_2}}{\ssubtype{\Gamma_1, 
X \ssub A, \Gamma_2}{X}{A}}\end{align*}\\[\inferenceruleskip]
\begin{align*}\mrule{SubProc}{\ssubtype{\Gamma}{A'}{A} \quad 
\ssubtype{\Gamma}{B}{B'}}{\ssubtype{\Gamma}{\sarr{A}{B}}{\sarr{A'}{B'}}}\end{align*}\\[\inferenceruleskip] 
\begin{align*}\mrule{SubObj}{E \subseteq D \quad \forall e {\in} E.\; (\tvar_e \in 
\{\tcov,\tinv\} \mimpl \ssubtype{\Gamma}{A_e}{B_e}) \\ \hspace{2.25cm} \mand 
(\tvar_e \in \{\tcon,\tinv\} \mimpl 
\ssubtype{\Gamma}{B_e}{A_e})}{\ssubtype{\Gamma}{\sobjtv{A}{d}}{\sobjtv{B}{e}}}\end{align*}\\[\inferenceruleskip]
\begin{align*}\mrule{SubObjVar}{\forall 
d \in D.\; \tvar_d = \tinv \mor \tvar_d = 
\tvar'_d}{\ssubtype{\Gamma}{\sobjtv{A}{d}}{\sobjtvv{\tvar'_d}{A}{d}}}\end{align*}\\[\inferenceruleskip] 
\begin{align*}\mrule{SubRec}{\sgtype{\Gamma}{\smu{X}{A}} \quad 
\sgtype{\Gamma}{\smu{Y}{B}} \quad \ssubtype{\Gamma, Y \ssub \stopp, X \ssub 
Y}{A}{B}}{\ssubtype{\Gamma}{\smu{X}{A}}{\smu{Y}{B}}}\end{align*}\\[\inferenceruleskip] 
\begin{align*}\mrule{SubUniv}{\ssubtype{\Gamma}{A'}{A} \quad \ssubtype{\Gamma, X \ssub 
A'}{B}{B'}}{\ssubtype{\Gamma}{\sforall{X}{A}{B}}{\sforall{X}{A'}{B'}}}\end{align*}\\[\inferenceruleskip]
\begin{align*}\mrule{SubExist}{\ssubtype{\Gamma}{A}{A'} \quad \ssubtype{\Gamma, X \ssub 
A}{B}{B'}}{\ssubtype{\Gamma}{\sexists{X}{A}{B}}{\sexists{X}{A'}{B'}}}\end{align*}
\caption{\label{fig:SyntacticSubtyping} Subtyping}
\end{figure}
 
\begin{figure}
\textbf{Subsumption and axioms}\hfill $\boxed{\stype{\Gamma}{a}{A}}$
\begin{align*}\mrule{Sub}{\stype{\Gamma}{a}{A} 
\quad \ssubtype{\Gamma}{A}{B}}{\stype{\Gamma}{a}{B}} \qquad 
\mrule{Var}{\sgenv{\Gamma_1, x{:}A, \Gamma_2}}{\stype{\Gamma_1, x{:}A, 
\Gamma_2}{x}{A}}\end{align*}\\[\inferenceruleskip]
\textbf{Procedure types}\hfill\quad 
\begin{align*}\mrule{Lam}{\stype{\Gamma, 
x{:}A}{b}{B}}{\stype{\Gamma}{\slam{x}{A}{b}}{\sarr{A}{B}}} \qquad 
\mrule{App}{\stype{\Gamma}{a}{\sarr{B}{A}} \quad 
\stype{\Gamma}{b}{B}}{\stype{\Gamma}{\sapp{a}{b}}{A}}\end{align*}\\[\inferenceruleskip]
\textbf{Object types}
\qquad (where $A \meqsyn \sobjtv{A}{d}$)\hfill\quad 
\begin{align*}\mrule{Obj}{\forall d {\in} D.\; 
\stype{\Gamma, x_d{:}A}{b_d}{A_d}}{\stype{\Gamma}{\sobj{A}{d}}{A}} \qquad
\mrule{Clone}{\stype{\Gamma}{a}{A}}{\stype{\Gamma}{\sclone{a}}{A}} \end{align*}\\[\inferenceruleskip] 
\begin{align*}\mrule{Inv}{\stype{\Gamma}{a}{A} \quad e \in D  \quad \tvar_e \in 
\{\tcov,\tinv\}}{\stype{\Gamma}{a.\m_e}{A_e}}\end{align*}\\[\inferenceruleskip]
\begin{align*}\mrule{Upd}{\stype{\Gamma}{a}{A} \quad e \in D \quad \stype{\Gamma, 
x{:}A}{b}{A_e} \quad \tvar_e \in 
\{\tcon,\tinv\}}{\stype{\Gamma}{\supd{a}{\m_e}{\sself{x}{A}{b}}}{A}}\end{align*}\\[\inferenceruleskip]
\textbf{Recursive types} \hfill\quad
\begin{align*}\mrule{Unfold}{\stype{\Gamma}{a}{\smu{X}{A}}}{\stype{\Gamma}{\sunfold{X}{A}{a}}{\stsub{A}{X}{\smu{X}{A}}}}\end{align*}\\[\inferenceruleskip]
\begin{align*}\mrule{Fold}{\stype{\Gamma}{a}{\stsub{A}{X}{\smu{X}{A}}}}{\stype{\Gamma}{\sfold{X}{A}{a}}{\smu{X}{A}}}\end{align*}\\[\inferenceruleskip]
\textbf{Bounded quantified types}\hfill\quad 
\begin{align*}\mrule{TAbs}{\stype{\Gamma, X \ssub 
A}{b}{B}}{\stype{\Gamma}{\stlam{X}{A}{b}}{\sforall{X}{A}{B}}} \qquad
\mrule{TApp}{\stype{\Gamma}{a}{\sforall{X}{A}{B}} \quad 
\ssubtype{\Gamma}{A'}{A}}{\stype{\Gamma}{\stapp{a}{A'}}{\stsub{B}{X}{A'}}}\end{align*}\\[\inferenceruleskip] 
\begin{align*}\mrule{Pack}{\ssubtype{\Gamma}{C}{A} \quad 
\stype{\Gamma}{\stsub{a}{X}{C}}{\stsub{B}{X}{C}}}{\stype{\Gamma}{(\spack{X}{A}{C}{a}{B})}{\sexists{X}{A}{B}}}\end{align*}\\[\inferenceruleskip] 
\begin{align*}\mrule{Open}{\stype{\Gamma}{a}{\sexists{X}{A}{B}} \quad \sgtype{\Gamma}{C} 
\quad \stype{\Gamma, X \ssub A, 
x{:}B}{b}{C}}{\stype{\Gamma}{(\sopen{a}{X}{A}{x}{B}{b}{C})}{C}}\end{align*}
\caption{\label{fig:SyntacticTermTyping}Typing of terms}
\end{figure}

While variables are  immutable identifiers, methods can be updated
destructively. Such updates only modify the heap and leave the run-time object
unchanged \textsc{(Red-Upd)}. 
Object cloning generates a shallow copy of an object in the heap
\textsc{(Red-Clone)}. The last four rules in Figure~\ref{fig:reduction} are as
in the lambda calculus.

For $k\in\mnat$, $\tjred{}{k}{}$ denotes the $k$-step reduction relation.
We write $\mirred{\xcfg{\h}{a}}$ \,if the configuration $\xcfg{\h}{a}$ is
irreducible (\IE there exists no configuration $\xcfg{\h'}{a'}$ such that
$\tredc{\h}{a}{\h'}{a'}$).

Note that reduction is not deterministic, due to the arbitrarily chosen fresh 
locations in \textsc{{(Red-Obj)}} and \textsc{{(Red-Clone)}}. However, we still 
have that there is always at most one, uniquely determined redex. This has the 
important consequence that the reduction order is fixed. 
For example, if there is a 
reduction sequence beginning with a method invocation and ending in an
irreducible
configuration: $\mirred{\tjredc {\h_1}{a.\m}{k}{\h_2}{b}}$, then this sequence
can be split into
\begin{align*}
{\tjredc {\h_1}{a.\m}{i}{\h_1'}{a'.\m}}
\tjred{}{k-i}{\xcfg{\h_2}{b}}
\end{align*} 
where $\mirred{\tjredc{\h_1}{a}{i}{\h_1'}{a'}}$ 
for some $i\geq 0$. 
Similar decompositions into subsequences hold for reductions starting from the other term forms.

It is easy to see that the operational semantics is independent of the type 
annotations inside terms. Also the semantic types that we define in 
Section~\ref{sec:model} will not depend on the syntactic type expressions 
in the terms. In order to reduce the notational overhead and to prevent 
confusion between the syntax and semantics of types we will omit type 
annotations when presenting the step-indexed semantics. For example, 
instead of the type application $\stapp{a}{A}$ we will merely write $\ttapp{a}$.

\subsection{Type System}
\label{subsec:type-system}

The type system we consider features procedure, object, iso-recursive and
(impredicative, bounded) quantified types, as well as subtyping,
and corresponds to \FObSubRec from \cite{Abadi:Cardelli:96}.  
It is fairly standard and consists of four inductively defined typing judgments:
\begin{enumerate}[$\bullet$]
  \item $\sgenv{\Gamma}$, describing \emph{well-formed typing contexts},
  \item $\sgtype{\Gamma}{A}$, defining \emph{well-formed types},
  \item $\ssubtype{\Gamma}{A}{B}$, for \emph{subtyping} between well-formed
  types, and 
  \item  $\stype{\Gamma}{a}{A}$, for \emph{typing terms}.
\end{enumerate}
The typing context $\Gamma$ is a list containing type bindings for the
(term) variables $x{:}A$ and upper bounds for the type variables $X\ssub A$. 
A typing context is well-formed if it does not contain duplicate
bindings for (term or type) variables and all types appearing in
it are well-formed. A type is well-formed with respect
to a well-formed context $\Gamma$ if all its type variables appear in
$\Gamma$.

Figure \ref{fig:SyntacticSubtyping} defines the subtyping relation.
For the object types it allows subtyping in width:
an object type with more methods is a subtype of an object type with fewer  
methods, as long as the types of the common methods agree.
For the invoke-only 
(\tcov) and update-only methods (\tcon) in object types, covariant
respectively contravariant subtyping in depth is allowed \textsc{(SubObj)}.
Furthermore, the unrestricted methods (\tinv)  can be regarded, by subtyping, as
either invoke-only or update-only ({\sc SubObjVar}).
Since the annotations can be conveniently chosen at creation time ({\sc Obj}) 
this brings much flexibility. As explained by Abadi and Cardelli 
\cite{Abadi:Cardelli:96}, this 
allows us to distinguish in the type system  between the invocations and updates
done through the self argument, and the ones done from the outside.
The main idea is to type an object creation with an object type where all
methods are considered invariant, so that all invocations and updates through
the self argument (internal) are allowed, but have to be type preserving.
Then rules ({\sc Sub}) and ({\sc SubObjVar}) are applied and some of the methods
can become invoke-only, some others update-only.
This enables the subsequent weakening of the types of these methods using ({\sc
SubObj}).
In effect, this allows for safe and flexible subtyping of methods,
at the price of restricting update and invocation of the
methods from the outside. Nevertheless, the internal updates and invocations 
remain unrestricted.

Figure \ref{fig:SyntacticTermTyping} defines the typing relation.
The applicability of the rules for method invocation \textsc{(Inv)},
and for method update \textsc{(Upd)}, depends on the variance annotation. Also
notice that only type-preserving updates are allowed in \textsc{(Upd)}.
Finally, it is important to note that we do not give
types to heap locations, since the type system is only used to check
programs, and programs do not contain locations. In contrast, a proof of type
safety using the preservation and progress properties would require the
syntactic judgement to also depend on a heap typing since partially evaluated
terms would also need to be typed.




\section{A Step-indexed Semantics of Objects}
\label{sec:model}

Modelling higher-order store is necessarily more involved than the 
treatment of first-order storage since the semantic domains become mutually
recursive. Recall that heaps store values that may be procedures. These
in turn can be modeled as functions that take a value and the initial
heap 
as input, and return a value and the possibly modified heap upon termination. 
This suggests the following semantic domains for values and heaps,~respectively:\begin{align}
\label{eqn:semantic-domain-of-heaps}
\begin{aligned}
D_\msn{Val} &= (D_\msn{Heaps}\times D_\msn{Val}\rightharpoonup
D_\msn{Heaps}\times D_\msn{Val}) + \ldots\\
D_\msn{Heaps} &=\msn{Loc} \rightharpoonup_{\textit{fin}} D_\msn{Val}
\end{aligned}
\end{align}
A simple cardinality argument shows that there are no set-theoretic solutions 
(\IE where $D \rightharpoonup E$ denotes the set of all partial
functions from $D$ to $E$) satisfying the 
equations in~\eqref{eqn:semantic-domain-of-heaps}.
A possible solution is to use a domain-theoretic approach, as done for
the imperative object calculus by Reus and Streicher \cite{Reus:Streicher:04},
building on earlier work by Kamin and Reddy \cite{Kamin:Reddy:94}.

In a model of a typed calculus one also wants to interpret the types. 
But naively
taking a collection \msn{Type} of subsets  $\tau\subseteq D_\msn{Val}$ as 
interpretations of syntactic types does not work, since values generally 
{depend} on the heap and a typed model should guarantee that all heap 
access operations are type-correct. We are  led to the following approach:
first, in order to ensure that updates are type-preserving, we also consider
\emph{heap typings}. Heap typings are partial maps $\Psi\in\msn{HeapTyping} =
\msn{Loc}\rightharpoonup_{\textit{fin}} 
\msn{Type}$ that track the set of values that may be stored in each
heap location. Second,  the collection of types is refined to take heap
typings
into account: a type  will now consist of values paired with heap typings
that describe the necessary requirements on heaps. These ideas  suggest
that we  take
\begin{align}
\label{eqn:circular-definition-of-types}
\begin{aligned}
\msn{Type} &= \mpow{(\msn{HeapTyping}\times D_\msn{Val})}\\
\msn{HeapTyping} &=
\msn{Loc}\rightharpoonup_{\textit{fin}} \msn{Type}
\end{aligned}
\end{align}
Again, a cardinality argument shows the impossibility of defining these sets.

A final obstacle to modelling the object calculus, albeit independent of the 
higher-order nature of heaps,
is due to dynamic allocation in the heap. This results in heap
typings that may \emph{vary} in the course of a computation, reflecting the changing
`shape' of the heap. However, as is the case for many high-level languages,
the object calculus is well-behaved in this respect:
\begin{enumerate}[$\bullet$]
\item inside the language, there is no possibility of deallocating heap locations; and
\item  only weak (\IE type-preserving) updates are allowed.
\end{enumerate}
As a consequence,
\emph{extensions} are the only changes that need to be considered for heap
typings. Intuitively, values that rely on
heaps with typing $\Psi$ will also be type-correct for extended heaps, with an
extended heap typing $\Psi'\sqsupseteq\Psi$. For this reason, semantic models of dynamic allocation typically lend
themselves to a Kripke-style presentation, where all
semantic entities are indexed by \emph{possible worlds} drawn from the set of heap typings, partially
(pre-) ordered by heap typing extension \cite{Levy:02,Moggi:90,Oles:85,Pitts:Stark:98,Reddy:Yang:04}.

Rather than trying to extend the already complex domain-theoretic models to 
heap typings and dynamic allocation, we will use the step-indexing technique. 
Since this technique is based directly on the operational semantics, it
provides an alternative
 that has less mathematical overhead. In particular, there is no need to find 
semantic domains satisfying \eqref{eqn:semantic-domain-of-heaps}; we can simply 
have $D_\msn{Val} $ be the set of closed values and use syntactic 
procedures in place of set-theoretic functions. 
Moreover, it is relatively easy to also model impredicative second-order types
in the step-indexed model of Ahmed \ETAL \cite{Ahmed:04, Ahmed:Appel:Virga:03},
which is crucial for the interpretation of object types we develop below.
Although recently there has been progress in finding domain-theoretic models of
languages that combine references and polymorphic
types \cite{Birkedal:Stovring:Thamsborg:09, Bohr:07,Bohr:Birkedal:06},
the constructions are more involved.

The circularity in  \eqref{eqn:circular-definition-of-types} is resolved by 
considering a stratification  based on a notion of `$k$-step execution safety'.
The central idea is that a term has type $\tau$ with approximation $k$ if this
assumption cannot be proved wrong (in the sense of reaching a stuck state) in
any context by executing fewer than $k$ steps.
The key insight for constructing the sets satisfying 
\eqref{eqn:circular-definition-of-types} is that all operations on the heap consume one step.
Thus, in order to determine whether a  pair $\mtuple{\Psi,
v}$, where $\Psi$ is a heap typing and $v$ a value,  belongs to a type $\tau$
with approximation
$k$ it is sufficient to know the types of the stored values on which $v$ relies
(as recorded by $\Psi$) only up to level $k-1$.
The true meaning of types and heap typings is then obtained by taking the
limit over all such approximations.

For instance, if a heap typing $\Psi$ asserts that
a \textit{Bool}-returning procedure is stored at location $l$, \IE
$\Psi(l) = \left[\m{:}\textit{Bool}\right]\to\textit{Bool}$,
then it is certainly not safe to assume that the pair 
$\mtuple{\Psi,\tlam{y}{\{\m{=}l\}.\m}}$ belongs to the type of
\textit{Int}-returning procedures. However, it is not possible to contradict
this assumption by taking only two reduction steps: the first step is
consumed by the beta reduction, the second one by the method selection
$\{\m{=}l\}.\m$ in the procedure body, which involves a heap access.
In this case, there are no steps left to observe that the result of the
computation is a boolean rather than an integer. Consequently,  the value
$\tlam{y}{\{\m{=}l\}.\m}$ is in the type of \textit{Int}-returning procedures
for two computation steps, even though it does not actually
return an integer. One can of course distinguish such `false positives'
by taking more reduction steps.

The preceding considerations are now formalized, building on the model 
originally developed by Ahmed \ETAL for an ML-like language with general
references and impredicative second-order types \cite{Ahmed:04, 
Ahmed:Appel:Virga:03}. Apart from some notational differences, the definitions in 
Section~\ref{subsec:semantic-model} are the same as in \cite{Ahmed:04}.
Section~\ref{subsec:subtyping} adds subtyping, while
Section~\ref{subsec:procedure-types} deals with procedure types, and
Section~\ref{subsec:reference-types} revisits reference types.
The semantics of
object types is presented in Section~\ref{subsec:object-types}
and constitutes the main contribution of this paper.
We further deviate from \cite{Ahmed:04} by 
adding bounds to the second-order 
types in Section~\ref{subsec:bounded-quantified-types},
and by using iso-recursive instead of equi-recursive types in 
Section~\ref{subsec:recursive-types}.

\subsection{The Semantic Model}
\label{subsec:semantic-model}

To make the (circular) definition of types and heap typings from
\eqref{eqn:circular-definition-of-types} 
work, the step-indexed semantics considers triples with
an additional natural number component, representing the step index, rather than just
pairs. First, we  inductively define two families
$(\msn{PreType}_k)_{k\in\mnat}$ of \emph{pre-types}, and $(\msn{HeapPreTyping}_k)_{k\in\mnat}$ of \emph{heap pre-typings}, by
\begin{align*}
\tau\in \msn{PreType}_0\ & \mequiv\ \tau = \emptyset \\
\tau\in \msn{PreType}_{k+1}\ & \mequiv\ \tau\in\mpow{(\mnat\times(\textstyle{\bigcup_{j\leq
k}\msn{HeapPreTyping}_j})\times \msn{CVal})}\\
 &\quad \   \mand \forall
\mtuple{j, \Psi, v} \in \tau.\  j \leq k \mand \Psi \in \msn{HeapPreTyping}_j
\end{align*}
where 
$\msn{HeapPreTyping}_k  = \msn{Loc} \rightharpoonup_{\textit{fin}}
\msn{PreType}_k$. 
That is, each $\tau\in\msn{PreType}_k$ is a set of triples $\mtuple{j,\Psi,v}$
where the set $\msn{HeapPreTyping}_j$ from which the heap pre-typing $\Psi$ is
drawn depends on the index $j<k$. 
 Clearly $\msn{PreType}_{k}\subseteq\msn{PreType}_{k+1}$ and
thus 
$\msn{HeapPreTyping}_{k}\subseteq\msn{HeapPreTyping}_{k+1}$ for all $k$. 
Now it is possible to set
\begin{align*}
\tau\in \msn{PreType}\ & \mequiv\ \tau\in\mpow{(\mnat\times(\textstyle{\bigcup_{j}\msn{HeapPreTyping}_j})\times \msn{CVal})}\\
 &\quad \ \mand \forall
\mtuple{j, \Psi, v} \in \tau.\   \Psi \in \msn{HeapPreTyping}_j
\end{align*}
We call the elements of 
this set \emph{pre}-types, rather
than types, since there will be a further condition that proper types must
satisfy (this is done in Definition~\ref{def:SemanticTypes} below).
From now on, when writing $\mtuple{k,\Psi,v}$, we always
implicitly assume that $\Psi\in\msn{HeapPreTyping}_k$. By \msn{HeapPreTyping} we
denote the set $\msn{Loc} \rightharpoonup_{\textit{fin}}
\msn{PreType}$ of finite maps into pre-types. 

Each pre-type $\tau$ is a union of sets $\tau_k\in\msn{PreType}_k$
where the index appearing in  elements of $\tau_k$ is bounded by $k$.
This is made explicit by the following notion of semantic approximation and the
stratification invariant below. 

\begin{defi}[Semantic approximation]
\label{def:SemanticApproximation}
For any pre-type $\tau$ we
call $\tapprox{\tau}{k}$ the \emph{$k$-th approximation of $\tau$} and define
it as the subset containing all elements of $\tau$ that have an index 
strictly less than $k$:
$\tapprox{\tau}{k} = \mset{\mtuple{j, \Psi, v} \in \tau}{j < k}$.
This definition is lifted pointwise to the (partial) functions in
\msn{HeapPreTyping}:
$\tapprox{\Psi}{k} = \mlam{l \in \mdom{\Psi}}{\tapprox{\Psi(l)}{k}}$.
\end{defi}

\begin{prop}[Stratification]
\label{prop:stratification-invariant}
For all  $\tau\in\msn{PreType}$ and  $k\in\mnat$,
$\tapprox{\tau}{k}\in\msn{PreType}_k$.  
Moreover,  
$\tau = \bigcup_k\tapprox\tau k$. \qed
\end{prop} 

So in particular, if $\mtuple{k,\Psi,v}\in\tau$ and $l\in\mdom\Psi$ then 
$\Psi(l)\in\msn{PreType}_j$ for some $j\leq k$. This is captured by the 
following `stratification invariant', which will be satisfied by all the
constructions on (pre-) types, and which ensures the well-foundedness of the
whole construction:
\begin{quotation}
\label{quot:stratification-invariant}
\textbf{Stratification invariant.}\ 
 For all pre-types $\tau$, 
$\tapprox{\tau}{k+1}$ cannot depend on any pre-type beyond approximation $k$.
\end{quotation}

\noindent As indicated above, in order to take dynamic allocation into account
we consider
a possible worlds model. Intuitively we think of a pair $(k,\Psi)$ as describing
the \emph{state} of a heap $\h$, where $\Psi$ lists locations in $\h$ that are
guaranteed to be allocated, and contains the types of the stored values up to
approximation $k$. 
In the course of a computation, there are three different situations where the
heap state changes:
\begin{enumerate}[$\bullet$]
\item New objects are allocated on the heap, which is reflected by a
heap pre-typing $\Psi'$ with additional locations compared to $\Psi$. This operation
does not affect any of the previously stored objects, so $\Psi'$ will be an
extension of  $\Psi$.
\item The program executes for $k-j$ steps, for some $j\leq k$, without accessing the
heap. This is reflected by a heap state $(j,\tapprox{\Psi}{j})$ that `forgets'
that we have a more precise approximation, and guarantees that the heap is
safe only for $j$ execution steps.
\item The heap is updated, but in such a way that all typing guarantees of
$\Psi$ are preserved. Thus updates will be reflected by an information forgetting
extension, as in the previous case.
However, because of the step taken by the update itself, in this case we 
necessarily have that $j<k$.
\end{enumerate}
The following definition of state extension captures these possible evolutions
of a state.

\begin{defi}[State extension]
\label{def:StateExtension} 
\emph{State extension}  $\sqsubseteq$ is the relation on 
$\mnat\times\msn{HeapPreTyping}$  defined by
\begin{align*}
\textend{k}{\Psi}{j}{\Psi'}\  &\mequiv\  
j \leq k 
\mand \mdom{\Psi}\subseteq\mdom{\Psi'}\\
&\quad \  
\mand \forall l \in 
\mdom{\Psi}.\; \tapprox{\Psi'}{j}(l) = \tapprox{\Psi}{j}(l)
\end{align*}
\end{defi}

The step-indexing technique relies on the approximation of the `true' set of 
values that constitute a type, by all those values that behave accordingly
unless a certain number of computation steps are taken. Limiting the number of
available steps, we will only be able to make fewer distinctions. 
 Moreover, if for instance a
procedure relies on locations
in the heap as described by a state $(k,\Psi)$, we can safely apply the procedure after 
further allocations. In fact, if we are only interested in safely executing 
the procedure for $j<k$ steps, a heap described by state
$(j,\tapprox{\Psi}{j})$ will suffice.
These conditions are captured precisely by state extension, 
so we require our semantic types to be closed under state extension:
\begin{defi}[Semantic types and heap typings]
\label{def:SemanticTypes} The set \msn{Type} of \emph{semantic types} is the
subset of \msn{PreType} defined by 
\begin{align*}
\tau\in\msn{Type}\ & \mequiv\    
\forall k,j \geq 0.\;  \forall \Psi, \Psi'.\;  \forall v \in \msn{CVal}.\\
 &\qquad \quad \textend{k}{\Psi}{j}{\Psi'} \mand \mtuple{k,\Psi,v} \in \tau \
 \mimpl\ \mtuple{j,\Psi',v} \in \tau 
\end{align*}
We also define the set $\msn{HeapTyping}= \msn{Loc}
\rightharpoonup_{\textit{fin}}\msn{Type}$ of \emph{heap typings},  ranged over by $\Psi$
in the following, as the subset of heap pre-typings that map to semantic types.
\end{defi}
As explained by Ahmed  \cite{Ahmed:04}, this structure may be viewed as an
instance of Kripke models of
intuitionistic logic where states are the possible worlds, state extension is the 
 reachability relation between worlds, and where  closure
 under state extension corresponds to Kripke monotonicity. 

Next we define when a particular heap $h$ conforms to the requirements
expressed by a heap typing $\Psi$. This is done with respect to an approximation
index.   
\begin{defi}[Well-typed heap]
\label{def:WellTypedHeap}
A heap $\h$ is \emph{well-typed} with respect to $\Psi$ with approximation
$k$, written as $\h:_k\Psi$, if $\mdom\Psi \subseteq \mdom\h$ and 
\begin{gather*}
    \forall j < k.\;
\forall l \in \mdom\Psi.\ \mtuple{j, \tapprox{\Psi}{j}, \h(l)} \in 
\Psi(l)
\end{gather*}
\end{defi}

Semantic types only contain values, but we also need to associate types with
terms that are not values. We do this in two steps, first for closed terms, then
for arbitrary ones.
A closed term has a certain type to approximation $k$ with respect to
some heap typing $\Psi$, if in all heaps that are well-typed with respect to 
$\Psi$ the term behaves like an element of the type for $k$ computation steps.
In general, before
reducing to a value the term will execute for $j$ steps, and possibly allocate
some new heap locations in doing so.
The state describing the final heap will therefore be an extension of the state
describing
the initial heap, and it only needs to be safe for the
remaining $k-j$
steps. Similarly, the final value needs to be in the original type only for
another $k-j$ steps. The next definition makes this precise.

\begin{defi}[Closed term has semantic type]
\label{def:ClosedTermKPsiType}
We say that a closed term $a$ \emph{has type} $\tau$ with respect to the state
$(k,\Psi)$, denoted as $a :_{k,\Psi} \tau$, if and only if  
\begin{align*}
\forall j < k, \h, \h', b.\; & (\h:_k\Psi \mand \tjredc{\h}{a}{j}{\h'}{b} \mand
\mirred{\xcfg{\h'}{b}}) \\
& \mimpl \exists \Psi'.\; \textend{k}{\Psi}{k-j}{\Psi'} \mand
\h':_{k-j}\Psi' \mand \mtuple{k-j, \Psi', b} \in \tau
\end{align*}
\end{defi}

%

Even though the terms we evaluate are closed, when type-checking their
subterms we also have to reason about open terms. Typing open terms is done with respect
to a semantic type environment $\Sigma$ that maps variables to semantic types.
We reduce typing open terms to typing their closed instances obtained by
substituting all free
variables with appropriately typed, closed values. This is done by a value
environment $\sigma$ 
(a finite map from variables to closed values)
that agrees with the type environment.

\begin{defi}[Value environment agrees with type environment]
\label{def:ValueEnvKPsiTypeEnv}
We say that \emph{value environment $\sigma$ agrees with  semantic type
environment $\Sigma$, with respect to the state $(k,\Psi)$}, if
$\forall x \in \mdom{\Sigma}.\  \sigma(x) 
:_{k,\Psi} \Sigma(x)$.
We denote this by $\sigma :_{k,\Psi} \Sigma$.
\end{defi}

\begin{defi}[Semantic typing judgement]
\label{def:SemanticTypingJudgement}
We say that a term $a$ (possibly with free variables, but not containing
locations), \emph{has type $\tau$} with respect to a semantic type environment
$\Sigma$, written as $\ttype{\Sigma}{a}{\tau}$, if after substituting 
well-typed values for the free variables of $a$, we obtain a closed term that
has type $\tau$ for any number of computation steps. More precisely:
\begin{align*}
\ttype{\Sigma}{a}{\tau}\ \mdef\ \mfv{a} \subseteq \mdom{\Sigma} \ \mand 
\forall k\geq 0. \; \forall \Psi.\; \forall \sigma :_{k,\Psi} \Sigma.\; 
\msub{a}{\sigma} :_{k,\Psi} \tau
\end{align*}
\end{defi}

By construction, the semantic 
typing judgment enforces that all terms that are typable with 
respect to it do not produce type errors when evaluated.

\begin{defi}[Safe for $k$ steps]
\label{def:SafeK}
We call a configuration \xcfg{\h}{a} {\em safe for $k$ steps}, if the term $a$ 
does not get stuck in less than $k$ steps when evaluated in the heap $h$, \IE
we define the set of all such configurations by
\begin{align*}
\tsafe_k = \mset{\xcfg{\h}{a}}{\forall j
< k.\; \forall \h',b.\ \tjredc{\h}{a}{j}{\h'}{b}\,
\mand \mirred{\xcfg{\h'}{b}}\ \ \mimpl\ b \in\msn{Val}}
\end{align*}
\end{defi}


\begin{defi}[Safety]
\label{def:Safe}
We call a configuration {\em safe} if it does not get stuck in any
number of steps, and let $\tsafe = \textstyle{\bigcap_{k\in\mnat} \tsafe_k }$.
\end{defi}


\begin{thm}[Safety]
\label{theorem:Safety}
For all programs $a$ such that $\,\ttype{\emptyset}{a}{\tau}$ and for
all heaps $\h$ we have that $\xcfg{\h}{a} \in \tsafe$.
\end{thm}
\proof
One first easily shows that, if $a:_{k,\Psi} \tau$ and $h :_k \Psi$, then 
$\xcfg{h}{a} \in \tsafe_k$. The theorem then follows by observing that any $\h$ 
is well-typed with respect to the empty heap typing, to any
approximation $k$.
\qed

This is much more direct than a subject reduction proof
\cite{Wright:Felleisen:94}.
However, unlike with subject reduction, the validity of the typing rules still
needs to be proved with respect to the semantics.
We do this in two steps.
In the remainder of this section we introduce the specific semantic
interpretations of types, and prove that they satisfy certain semantic typing 
lemmas. These proofs are similar in spirit to proving the `fundamental theorem'
of Kripke logical relations \cite{Mitchell:Moggi:91}.
Then, in Section~\ref{sec:soundness} we prove the soundness of the rules of the 
initial type system with respect to these typing lemmas. 

Even though the semantic typing lemmas are constructed so that they directly 
correspond to the rules of the original type system, there is a big difference
between the two.
While the semantic typing lemmas allow us to logically derive valid
semantic judgments using other valid judgments as premises,
the typing rules are just syntax that is used in the inductive
definitions of the typing and subtyping relations.

\subsection{Subtyping}
\label{subsec:subtyping}

Since types in the step-indexed interpretation are sets (satisfying some additional constraints), the
natural subtyping 
relation is set inclusion. This subtyping relation forms a complete lattice on 
semantic types, where infima and suprema are given by set-theoretic
intersections and unions, respectively. 
The least element is $\tbott = \emptyset$, while the greatest is
\begin{align*}
\ttopp &= 
\mset{\mtuple{j,\Psi,v}}
{j \in \mathbb{N}, \Psi \in 
\msn{HeapTyping}_j , v \in \msn{CVal}}.
\end{align*}
Obviously $\tbott$ and $\ttopp$ satisfy both the stratification invariant
(\IE they are pre-types) and the closure under state extension condition, so they are
indeed semantic types.

We can easily show the standard subsumption property
%
%
\begin{lem}[Subsumption]
\label{lemma:SemSub}
If $\ttype{\Sigma}{a}{\alpha}$ and $\alpha \tsub \beta$ 
then $\ttype{\Sigma}{a}{\beta}$.\qed
\end{lem}

While it is very easy to define subtyping in this way, the interaction between
subtyping and the other features of the type system, in particular the object
types, is far from trivial.
This point will be discussed further in Section~\ref{subsec:object-types}.

\subsection{Procedure Types}
\label{subsec:procedure-types}

Intuitively, a procedure has type $\tarr{\alpha}{\beta}$ for $k$
computation steps if, when applied to any well-typed argument of type $\alpha$,
it produces a result that has type $\beta$ for another $k-1$ steps. This
is because the procedure application itself takes one computation step, and the
only way to use a procedure is by applying it to some argument. 

Additionally, we have to
take into account that the procedure can also be applied after some computation
steps that extend the  heap. So, for every $j < k$ and for every heap typing 
$\Psi'$
such that $\textend{k}{\Psi}{j}{\Psi'}$, when applying the procedure to a value
in type $\alpha$ for $j$ steps with respect to $\Psi'$, the result must have
type $\beta$ for $j$ steps with respect to $\Psi'$.  This computational
intuition nicely fits the
possible worlds reading of procedure types as intuitionistic implication.

\begin{defi}[Procedure types]
\label{def:ProcedureTypes}
If $\alpha$ and $\beta$ are semantic types, then 
$\tarr{\alpha}{\beta}$ consists of those triples $\mtuple{k, \Psi,
\mlam{x}{b}}$ such that for all $j < k$, heap typings $\Psi'$ and closed values
$v$:
\begin{align*}
 (\textend{k}{\Psi}{j}{\Psi'} 
 \mand \mtuple{j, \Psi', v} \in \alpha)\ \mimpl\ 
\msub{b}{\mmap{x}{v}} :_{j,\Psi'} \beta
\end{align*}
\end{defi}

\begin{figure}
\centering
\begin{align}
\tag{\textsc{SemLam}}
\qquad\ttype{\mext{\Sigma}{x}{\alpha}}{b}{\beta}\  &\Longrightarrow\  \ttype{\Sigma}{\mlam{x}{b}}{\tarr{\alpha}{\beta}}\\[1mm]
\tag{\textsc{SemApp}} 
(\ttype{\Sigma}{a}{\tarr{\beta}{\alpha}}
\mand\ttype{\Sigma}{b}{\beta}\  &\Longrightarrow\  \ttype{\Sigma}{\tapp{a}{b}}{\alpha}\\[1mm]
\tag{\textsc{SemSubProc}}
\alpha' \tsub \alpha\mand\beta \tsub \beta'\  &\Longrightarrow\ \tarr{\alpha}{\beta} 
\tsub \tarr{\alpha'}{\beta'}
\end{align}
\caption{\label{fig:typing-lemmas:procedure-types}Typing lemmas: procedure types}
\end{figure}

\begin{prop}
\label{prop:proc-extension-closed}
If $\alpha$ and $\beta$ are semantic types, then $\tarr{\alpha}{\beta}$ is also
a semantic type.\qed 
\end{prop}

Figure~\ref{fig:typing-lemmas:procedure-types} contains the semantic typing lemmas associated with procedure types. 
The procedure type constructor is of course contravariant in the argument type
and covariant in the result type.

\begin{lem}[Procedure types]
\label{lemma:all-proc-type-lemmas}
The three semantic typing lemmas shown in
Figure~\ref{fig:typing-lemmas:procedure-types} are valid implications.
\end{lem}
\proof[Proof sketch]
The validity of \textsc{(SemApp)} and \textsc{(SemLam)} is proved in  
\cite{Ahmed:04}. Verifying \textsc{(SemSubProc)} is simply a matter of 
unfolding the definitions.
\qed

\subsection{Revisiting Reference Types}
\label{subsec:reference-types}

While our calculus does not have references syntactically, we will use
the model of references from \cite{Ahmed:04, Ahmed:Appel:Virga:03}
in our construction underlying object types.
In order to interpret the variance annotations in object types,
we additionally introduce readable reference types and writable
reference types, with covariant and  contravariant subtyping, respectively 
\cite{Pierce:Sangiorgi:96, Reynolds:96}. 

A heap typing associates with each allocated location the precise type that can be
used when reading from it and writing to it. So all heap locations support both
reading and writing at a certain type, and we do not have read-only or write-only
locations.
Intuitively, for the readable reference types and the writable ones the precise
type of the locations is only partially known, so that without additional
information only one of the two operations is safe at a meaningful type.





We first recall the definition of reference types from
\cite{Ahmed:04, Ahmed:Appel:Virga:03}.
\begin{defi}[Reference types]
\label{def:ReferenceTypes}
If $\tau$ is a semantic type then
\[\trefv{\tinv}{\tau} =
\mset{\mtuple{k,\Psi,l}}{\tapprox{\Psi(l)}{k} = \tapprox{\tau}{k}}\]
\end{defi}

According to this definition,
a location $l$ has type $\trefv{\tinv}{\tau}$ if the type associated with $l$
by the heap typing $\Psi$ is approximately $\tau$. Semantic approximation is used
to satisfy the stratification invariant, and is operationally justified
by the fact that reading from a location or writing to it takes one
computation step. So, $l$ has type $\trefv{\tinv}{\tau}$ for $k$ steps if all
values that are read from $l$ or written to $l$ have type $\tau$ for $k-1$
steps.

The readable reference type \trefv{\tcov}{\tau} is similar to
\trefv{\tinv}{\tau}, but poses less constraints on the heap typing
$\Psi$: it only requires that $\Psi(l)$ is a subtype of $\tau$, as before up to
some approximation.
\begin{defi}[Readable reference types]
If $\tau$ is a semantic type then
\[\trefv{\tcov}{\tau} = \mset{\mtuple{k, \Psi, l}}{
\tapprox{\Psi}{k}(l) \tsub \tapprox{\tau}{k}}\]
\end{defi}

The value stored at location $l$ also has type $\tau$ by subsumption, and 
therefore  can be read and safely used as a value of type $\tau$.
However, the true type of location $l$ is in general unknown, so writing any
value to it could be unsafe (the true type of $l$ might be the empty type
$\tbott$).
Nevertheless, knowing that a location has type $\trefv{\tcov}{\tau}$  does
not mean that we cannot write to it: it simply means that we
do not know
the type of the values that can be written to it, so in the absence of further
information no writing can be guaranteed to be type safe\footnote{This
is conceptually different from the immutable reference types modeled in
\cite{Ahmed:04} using singleton types.}.

Dually, the type $\trefv{\tcon}{\tau}$ of writable references contains all
those locations $l$ whose type associated by $\Psi$ is a supertype of $\tau$.
\begin{defi}[Writable reference types]
If $\tau$ is a semantic type then
\[\trefv{\tcon}{\tau} = \mset{\mtuple{k, \Psi,
l}}{ \tapprox{\tau}{k}  \tsub \tapprox{\Psi}{k}(l)}\]
\end{defi}

We can safely write a value of type $\tau$ to a location of type
\trefv{\tcon}{\tau}, since this value also has the real type of
location $l$ by subsumption. However, the real type of such locations can be
arbitrarily general. In particular it can be $\ttopp$, the type of all values.
Thus a location about which we only know that it has type \trefv{\tcon}{\tau}
can only be read safely at type $\ttopp$.

\begin{figure}
\begin{align}
\tag{\textsc{SemSubCovRef}}
& \alpha \tsub \beta \quad\Longrightarrow\quad \trefv{\tcov}{\alpha} \tsub
\trefv{\tcov}{\beta}\\
\tag{\textsc{SemSubConRef}}
& \beta \tsub \alpha \quad\Longrightarrow\quad \trefv{\tcon}{\alpha} \tsub
\trefv{\tcon}{\beta}\\
\tag{\textsc{SemSubVarRef}}
& \trefv{\tinv}{\alpha} \tsub \trefv{\tvar}{\alpha}, \text{ where }
\tvar \in \{\tinv, \tcov, \tcon \}
\end{align}
\caption{\label{fig:subtyping-reference-types}Subtyping reference types}
\end{figure}

With these definitions in place, the usual reference type from
Definition~\ref{def:ReferenceTypes} can be recovered as the intersection of a
readable and a writable reference type:
\[\trefv{\tinv}{\tau} = \trefv{\tcov}{\tau} \cap \trefv{\tcon}{\tau}\]
Hence \trefv{\tcov}{\tau} and \trefv{\tcon}{\tau} are both supertypes
of \trefv{\tinv}{\tau}.
It can also be easily shown that the readable reference type constructor is
covariant, the writable reference type constructor is contravariant
(Figure~\ref{fig:subtyping-reference-types}),
while the usual reference types are obviously invariant.
For a variance annotation $\tvar \in \{ \tinv, \tcov, \tcon \}$ we use
$\trefv{\tvar}$ to stand for the reference type constructor with this variance. 

Note that, strictly speaking, the set \trefv{\tvar}{\tau} is not a semantic type
since for our calculus locations are not values (although locations appear in
object values $\svobj{d}$; see Section~\ref{subsec:operational-semantics}). In
fact, the definition of object types  (Definition~\ref{def:ObjectTypes} in the next
section) will not depend on \trefv{\tvar}{\tau} being a semantic type.
However, in order for the object type constructor to yield semantic types, it
is crucial that $\trefv{\tvar}\tau$ is closed under state extension.

\begin{prop}
\label{prop:ref-extension-closed}
If $\tau$ is a  semantic type, then $\trefv{\tvar}\tau$ is closed under
state extension.\qed 
\end{prop}

\subsection{Object Types}
\label{subsec:object-types}

Giving a semantics to object types is much more challenging than for the other
types. The typing rules from  Section~\ref{sec:calculus} indicate why this is
the case.
First, an adequate interpretation of object types must permit subtyping both
in width and in depth, taking the variance annotations into account.
Second, in contrast to all the other types we consider that have just a single
elimination rule, once constructed, objects support three different operations:
invocation, update, and cloning.
The definition of object types must ensure the consistent use of an object
through all possible future operations.
That is, all the requirements on which invocation, update or cloning
rely must already be established at object creation time.

Before defining the object types, it is instructive to consider some
simpler variants that do not fulfill all the requirements we have for object
types.


  Our decision to store methods in the heap as procedures, together
  with the `self-application' semantics of method invocation ({\sc
  Red-Inv} in Figure~\ref{fig:reduction}), suggest that object types are
  somewhat similar to recursive types of records of references holding procedures
  that take the enclosing record as argument:
  \[\tobjt{\tau}{d}\ \sortofequals\  \mu(\alpha).\{\m_d :
  \trefv{\tinv}{(\tarr{\alpha}{\tau_d})}\}_{d\in D}\]
  However, the invariance of the reference type constructor blocks any form of
  subtyping, even in width.
  A look at typing rules for subtyping recursive types, such as
  Cardelli's  Amber rule \cite{Cardelli:86} (which appears as rule
  \textsc{SubRec} in Figure~\ref{fig:SyntacticSubtyping}), suggests
  that the position of the recursion variable should be covariant. 
For instance, when attempting to establish the subtyping 
$[\m_1:\tau_1,\m_2:\tau_2]\subseteq
[\m_1:\tau_1]$ by
the Amber rule one needs to show that 
$\trefv{\tinv}{(\tarr{\alpha}{\tau_1})}\subseteq
\trefv{\tinv}{(\tarr{\beta}{\tau_1})}$, for any $\alpha$ and $\beta$ such that 
$\alpha\subseteq\beta$. Clearly this does not hold.
  Even in a simpler setting without the
  reference types (\EG for the functional object calculus)
  the contravariance of the procedure type constructor in its first argument
  would  cause subtyping to fail.

  A combination of type recursion and an existential quantifier that uses
  the recursion variable as bound would allow us to enforce covariance for the
  positions of the recursion
  variable, and thus have subtyping in width:
  \[\tobjt{\tau}{d}\ \sortofequals\  \mu(\alpha).\exists\alpha'{\tsub}\alpha.\{\m_d :
  \trefv{\tinv}{(\tarr{\alpha'}{\tau_d})\}_{d\in D}}\]
  Intuitively $\alpha'$
  can be viewed as the `true' (\IE most precise) type of the object, while
  $\alpha$ is a more general type that can be given to it by subtyping.
  This is essentially the idea of the encodings of object types explored by
  Abadi \ETAL \cite{Abadi:Cardelli:96,Abadi:Cardelli:Viswanathan:96}.
  
  For subtyping in depth with respect
  to the variance annotations we simply use the readable and writable reference
  types we defined in the previous section:
  \[\tobjtv{\tau}{d}\ \sortofequals\  \mu(\alpha).\exists\alpha'{\tsub}\alpha.\{\m_d :
  \trefv{\tvar_d}{(\tarr{\alpha'}{\tau_d})\}_{d\in D}}\]
  Still,
  by keeping $\alpha'$ abstract, neither the typing rule for method invocation
  ({\sc Inv} in Figure~\ref{fig:SyntacticTermTyping}), nor the one for object
  cloning ({\sc Clone}) is validated.
  
  By explicitly enforcing in the definition of object types that the
  object value itself in fact belongs to this existentially quantified
  $\alpha'$, the assumptions become sufficiently strong to repair the
  invocation case. This is consistent with seeing $\alpha'$ as the `true'
  type of the object. Semantically, we can express this using an intersection
  of types:
  \[\tobjtv{\tau}{d}\ \sortofequals\ \mu(\alpha).\exists\alpha'{\tsub}\alpha. (
  \{\m_d : \trefv{\tvar_d}{(\tarr{\alpha'}{\tau_d})\}_{d\in D}}\cap \alpha')\]
  Forcing not only the current object value to be in $\alpha'$, but also all
  the `sufficiently similar' values (maybe not even created yet), covers the
  case of cloning. 
  The following definition formalizes this construction.

\begin{defi}[Object types]
\label{def:ObjectTypes} Let $\alpha = \tobjtv{\tau}{d}$ be defined as the set
of all triples $\mtuple{k, \Psi, \tvobj{e}}$ such that $D \subseteq E$ and
\begin{align}
\tag{{\sc Obj-1}}
&  \exists \alpha'.\ {\alpha'} \in \msn{Type} \mand 
\tapprox{\alpha'}{k} \tsub \tapprox{\alpha}{k} \\
\tag{{\sc Obj-2}}
& \wedge (\forall d \in D.\; \mtuple{k,\Psi,l_d} \in \trefv{\tvar_d}{(\alpha' \to
\tau_d)}) \\
\tag{{\sc Obj-3}}
& \wedge (\forall j<k.\; \forall \Psi'.\; \forall \tvobjx{l_e'}{e}. \\[-.5mm] 
\notag
& \qquad \;
\textend{k}{\Psi}{j}{\Psi'} \mand
(\forall e \in E. \; \tapprox{\Psi'}{j}(l_e') =
\tapprox{\Psi}{j}(l_e)) \\[-.5mm] 
\notag
& \qquad 
 \mimpl \mtuple{j, \tapprox{\Psi'}{j}, \tvobjx{l_e'}{e}} \in \alpha')
\end{align}
\end{defi}
The condition stating that $D\subseteq E$ ensures that all 
values in an object type provide \emph{at least} the required methods
listed by this type, but can also provide more. Clearly this is necessary for
subtyping in width.
Condition \tagref{{\sc Obj-1}} postulates the existence of a 
more specific type $\alpha'$, 
the `true' type of the object $\tvobj{e}$ (up to approximation $k$),
and the subsequent conditions are all
stated in terms of $\alpha'$ rather than $\alpha$.
Condition \tagref{{\sc Obj-2}} states the requirements for the methods in terms
of the reference type constructors introduced in Section~\ref{subsec:reference-types}. 
Since the existentially quantified $\alpha'$ might equal $\alpha$, one must take
care  that 
\tagref{{\sc Obj-2}} does not introduce a circularity.
However, due to the use of approximation in the definition of the reference type
constructors, the condition only depends on $\tapprox{\alpha'}{k}$, rather than
$\alpha'$.
This will ensure the well-foundedness of the construction.

As explained above, in order to invoke methods
we must know that $\tvobj{e}$ belongs to
the more specific type $\alpha'$ for $j < k$ steps (which suffices since
application consumes a step). 
In the particular case where $\Psi'$ is $\Psi$ and $\tvobjx{l_e'}{e}$
is $\tvobj e$ condition \tagref{{\sc Obj-3}} states exactly this.
We need the more general formulation in order to ensure that the clones of the
considered object also belong to the same type $\alpha'$.
Therefore we enforce that no matter how an object value $\tvobjx{l_e'}{e}$ is
constructed, it belongs to type $\alpha'$ provided that it satisfies the same 
typing assumptions as $\tvobj{e}$, with respect to a possibly extended heap
typing $\Psi'$. Allowing for state extension is necessary
since cloning itself allocates new locations not present in the original $\Psi$,
and also because cloning can be performed after some
intermediate computation steps that result in further allocations.

We show that this definition of object types actually 
makes sense, in that it defines a semantic type.
This is not immediately obvious because of the recursion.

\begin{prop}
\label{prop:object-types}
If $\tau_d\in\msn{Type}$ for all $d\in D$, then we also have that
$\tobjtv{\tau}{d}\in\msn{Type}$.
\end{prop}
\proof[Proof sketch]
We must show (1) that $\tobjtv{\tau}{d} $ is well-defined, \IE that the
recursive definition is well-founded, and (2) that it is closed under state
extension. 

To prove the well-definedness one can use general results about recursive 
types in step-indexed semantics \cite{Appel:McAllester:01}, since the object
type constructor is `contractive'. Alternatively, from the observation that
$\tau = \bigcup_k\tapprox{\tau}{k}$ for all types $\tau$, it suffices to
directly argue that Definition~\ref{def:ObjectTypes} defines
$\tapprox{\tobjtv{\tau}{d}}{k}$ only in terms of
$\tapprox{\tobjtv{\tau}{d}}{j}$ for $j<k$. The closure under state extension 
follows from the corresponding property of the types $\alpha'\to\tau_d$ 
 (Proposition~\ref{prop:proc-extension-closed}) and of the sets
$\trefv{\tvar_d}{(\alpha'\to\tau_d)}$
(Proposition~\ref{prop:ref-extension-closed}), and from the transitivity of state extension. 
\qed

\begin{figure*}
\flushleft{Let $\alpha = \tobjtv{\tau}{d}$.}

\centering
\begin{small}
\begin{align*}
\tag{\textsc{SemObj}}
(\forall d\in D.\ \ttype{\mext{\Sigma}{x_d}{\alpha}}{b_d}{\tau_d})
\quad &\Longrightarrow\quad
\ttype{\Sigma}{\tobj{d}}{\alpha}\\[1mm]
\tag{\textsc{SemInv}}
(
\ttype{\Sigma}{a}{\alpha} \mand
e\in D\mand 
\tvar_e \in \{\tcov, \tinv\})
\quad &\Longrightarrow\quad
\ttype{\Sigma}{a.\m_e}{\tau_e}\\[1mm]
\tag{\textsc{SemUpd}}
(\ttype{\Sigma}{a}{\alpha}\mand
e\in D\mand
\tvar_e \in \{\tcon,\tinv\}\phantom{)}\quad \\ 
\mand \ttype{\mext{\Sigma}{x}{\alpha}}{b}{\tau_e})
\quad &\Longrightarrow\quad
\ttype{\Sigma}{\tupd{a}{\m_e}{\tself{x}{b}}}{\alpha}\\[1mm]
\tag{\textsc{SemClone}}
\ttype{\Sigma}{a}{\alpha}
\quad &\Longrightarrow\quad
\ttype{\Sigma}{\tclone{a}}{\alpha}\\[1mm]
\tag{\textsc{SemSubObj}}
(E \subseteq D\mand 
(\forall e \in E.\ \tvar_e \in \{\tcov,\tinv\}\mimpl\alpha_e \tsub \beta_e)
\phantom{)}\quad\\
\mand
(\forall e \in E.\ \tvar_d \in \{\tcon,\tinv\}\mimpl\beta_e \tsub \alpha_e))
\quad &\Longrightarrow\quad
\tobjtv{\alpha}{d} \tsub \tobjtv{\beta}{e}\\[1mm]
\tag{\textsc{SemSubObjVar}}
(\forall d \in D.\ \tvar_d = \tinv \mor \tvar_d = \tvar'_d)
\quad &\Longrightarrow\quad
\tobjtv{\alpha}{d} \tsub \tobjtvv{\tvar'_d}{\alpha}{d}
\end{align*}
\end{small}
\caption{\label{fig:typing-lemmas:object-types}
Typing lemmas: object types}
\end{figure*}

Figure~\ref{fig:typing-lemmas:object-types} presents the semantic typing
and subtyping lemmas for object types.

\begin{lem}[Object types]
\label{lemma:object-types}
All the semantic typing lemmas shown in
Figure~\ref{fig:typing-lemmas:object-types} are valid implications.
\end{lem}

\proof[Proof sketch]
The semantic typing lemmas are proved independently. 
We sketch this for \textsc{(SemObj)}.  
A detailed proof, as well as the proofs of the other typing lemmas
are given in the Appendix.

For $\alpha = \tobjtv{\tau}{d}$  
and assuming $\ttype{\mext{\Sigma}{x_d}{\alpha}}{b_d}{\tau_d}$ for all $d\in D$,
we must show that $\ttype{\Sigma}{\tobj{d}}{\alpha}$. 
So let $k \geq 0$, $\sigma$ and $\Psi$ be  such
that  $\sigma :_{k, \Psi} \Sigma$. 
By 
Definition~\ref{def:SemanticTypingJudgement} (Semantic typing judgement)
we must prove that
$\msub{\tobj{d}}{\sigma} :_{k,\Psi} \alpha$, or equivalently (after suitable 
$\alpha$-renaming), that $\tobjx{\msub{b_d}{\sigma}}{d} :_{k,\Psi} \alpha$ 
holds. 
Now let  $h, h'$ and $b'$ be such that $h :_k \Psi$ and 
\[\tjredc{h}{\tobjx{\msub{b_d}{\sigma}}{d}}{j}{h'}{b'}\] 
for some $j < k$, and assume that $\xcfg{h'}{b'}$ is irreducible. 
From the operational semantics it is clear that $j=1$, 
$b' \meqsyn \tvobj{d}$ and that, for some locations $l_d \notin \mdom{h}$, 
\[h' = \mextx{h}{l_d}{\tlam{x_d}{\msub{b_d}{\sigma}}}{d \in D}\]
Choosing $\Psi' = \tapprox{\mextx{\Psi}{l_d}{(\tarr{\alpha}{\tau_d})}{d \in 
D}}{k-1}$ it is easily seen
that $\textend{k}{\Psi}{k-1}{\Psi'}$. Furthermore, from the hypothesis by
\textsc{(SemLam)} we have that $\ttype{{\Sigma}}{\tlam{x_d}{b_d}}{\tarr\alpha{\tau_d}}$ for 
all $d\in D$. From this and the assumption that
$h:_k\Psi$ it follows that 
$h':_{k-1} \Psi'$. 

\refstepcounter{thm}
By Definition~\ref{def:ClosedTermKPsiType} it remains to establish that
$\mtuple{k-1, \Psi', \tvobj{d}} \in \alpha$. This is achieved by proving the
following more general claim by induction on $j_0$:

\noindent\label{the-important-claim}\textbf{\textit{Claim~\thethm.}} 
For all $j_0\geq 0$, 
$\Psi^*$ and $\tvobjx{\loc^*_d}{d}$ we have that
\begin{align*}
\textend{k-1}{\Psi'}{j_0}{\Psi^*}
\mand  (\forall d{\in} D.\tapprox{\Psi^*}{j_0}(\loc^*_d) = 
\tapprox{\Psi'}{j_0}({\loc_d}))
\mimpl \ \mtuple{j_0,\tapprox{\Psi^*}{j_0}, \tvobjx{\loc^*_d}{d}}\in\alpha
\end{align*}
The key step is in choosing $\alpha'$ equal to $\tapprox{\alpha}{j_0}$, 
then verifying the three conditions of Definition~\ref{def:ObjectTypes}
(Object types), where the inductive hypothesis is used for showing
\tagref{{\sc Obj-3}}.
\qed

\begin{rem}
In the above proof, establishing $\mtuple{k{-}1, \Psi', \tvobj{d}} \in \alpha$
directly does not seem possible, and the generalization to
Claim~\ref{the-important-claim} arises naturally from a failed proof attempt: in
order to prove the three conditions of Definition~\ref{def:ObjectTypes} a
sensible choice for $\alpha'$ is $\alpha$, and for $E$ is $D$, after which
\textsc{(Obj-1)} and \textsc{(Obj-2)} follow easily. But \textsc{(Obj-3)} 
requires us to show that $\mtuple{j, \tapprox{\Psi''}{j}, \tvobjx{l_d'}{d}} \in
\alpha$, for any $j<k$, any $l_d'$, and any extension $\Psi''$ of $\Psi'$ with
$\tapprox{\Psi''}{j}(l_d') = \tapprox{\Psi'}{j}({\loc_d})$. This is just what
Claim~\ref{the-important-claim} states.

The fact that there is an inductive argument hidden in this proof does not
come as a surprise:
the induction on the step index $j_0$ resolves the recursion that is inherent to
objects due to the self application semantics of method invocation.
\end{rem}

\subsection{Bounded Quantified Types}
\label{subsec:bounded-quantified-types}

Impredicative quantified types were previously studied in a step-indexed setting
by Ahmed \ETAL \cite{Ahmed:04, Ahmed:Appel:Virga:03} for a lambda-calculus with general references, and we
follow their presentation.
However, unlike in the work of Ahmed \ETAL
our quantifiers have bounds, and we are also studying subtyping.
It is important to note that the impredicative second-order types were
the reason why  a semantic stratification of types was needed in the presence of
general references \cite{Ahmed:04},
as opposed to a syntactic one based on the nesting of reference types 
\cite{Ahmed:Appel:Virga:02}. In the setting we consider in this paper we need
the semantic stratification not only to explicitly accommodate quantified types,
but also because our interpretation of object types uses existential types
implicitly.

As in Appel and McAllester's work \cite{Appel:McAllester:01}, 
a type constructor $F$ (\IE a function from semantic types to semantic types)
is non-expansive if in order to determine whether a term has type $F(\tau)$
with approximation $k$, it suffices to know the type $\tau$ only to
approximation $k$. As we will later show (Lemma~\ref{lemma:sem:NonExp}), all
the type constructors we define in this paper are non-expansive.
\begin{defi}[Non-expansiveness]
A type constructor $F: \msn{Type} \to \msn{Type}$ is {\em non-expansive} if for
all types $\tau$ and  for all $k \geq 0$ we have that 
$\tapprox{F(\tau)}{k}=\tapprox{F(\tapprox{\tau}{k})}{k}$.
\end{defi}

The definitions of second-order types require that $\forall$ and $\exists$ are
only applied to non-expansive type constructors. The non-expansiveness
condition ensures that in order to determine
level $k$ of a universal or existential type, quantification over the types
in $\msn{PreType}_k$ suffices. This helps avoid the circularity that is
otherwise introduced by the {\em impredicative} quantification.
\begin{defi}[Bounded universal types]
\label{def:BoundedUniversalTypes}
If $F: \msn{Type} \to \msn{Type}$ is non-expansive and $\alpha\in\msn{Type}$,
then we define $\tforall{\alpha}{F}$ by $\mtuple{k, \Psi,
\ttlam{a}}\in\tforall{\alpha}{F}$ if and only if 
\begin{align*}
   \forall
j,\Psi'.\;
\forall \tau.\; \textend{k}{\Psi}{j}{\Psi'}
 \mand {\tau} \in
\msn{Type}  \mand \tapprox{\tau}{j} \tsub \tapprox{\alpha}{j}
 \mimpl
\forall i<j.\; a :_{i,\tapprox{\Psi'}{i}} F(\tau)
\end{align*}
\end{defi}

\begin{defi}[Bounded existential types]
\label{def:BoundedExistentialTypes}
For all  non-expansive $F: \msn{Type} \to \msn{Type}$ and $\alpha\in\msn{Type}$,
the set $\texists{\alpha}{F}$ is defined by $\mtuple{k, \Psi,
\tpack{v}}\in\texists{\alpha}{F}$ if and only if 
\begin{align*}
 \exists \tau. {\tau} \in \msn{Type} \mand\tapprox{\tau}{k} \tsub
\tapprox{\alpha}{k} \mand \forall j < k.\; \mtuple{j, \tapprox{\Psi}{j}, v} \in F(\tau)
\end{align*}
\end{defi}

\begin{prop}
If $\alpha\in\msn{Type}$ and $F:\msn{Type} \to \msn{Type}$ is non-expansive,
then $\tforall{\alpha}{F}$ and $\texists{\alpha}{F}$ are also types.\qed
\end{prop}

\proof[Proof sketch]
The proofs are minor modifications of those given in \cite{Ahmed:04}, to
additionally take the bounds into account. 
\qed

\begin{lem}[Bounded quantified types]
\label{lemma:bounded-quantified-types}
All the semantic typing lemmas shown in
Figure~\ref{fig:typing-lemmas:bounded-quantified-types} are valid implications.
\end{lem}

\proof[Proof sketch]
The first four implications are proved as in \cite{Ahmed:04}; the additional
precondition $\tau\tsub\alpha$ in \textsc{(SemTApp)} and \textsc{(SemPack)}
serves to establish the requirements for the bounds. 
The two subtyping lemmas \textsc{(SemSubUniv)} and \textsc{(SemSubExist)} are
easily proved by just unfolding the definitions. 
\qed

\begin{figure*}
\flushleft{For all non-expansive $F,G : \msn{Type}\to\msn{Type}$,}
 
\begin{small}
\centering
\begin{align*}
\tag{\textsc{SemTAbs}}
(\forall \tau \in \msn{Type}.\; \tau \tsub \alpha \mimpl 
\ttype{\Sigma}{a}{F(\tau)})
\quad &\Longrightarrow\quad
\ttype{\Sigma}{\ttlam{a}}{\tforall{\alpha}{F}}\\[1mm]
\tag{\textsc{SemTApp}}
(\ttype{\Sigma}{a}{\tforall{\alpha}{F}} \mand \tau \in 
\msn{Type} \mand \tau \tsub \alpha)
\quad &\Longrightarrow\quad
\ttype{\Sigma}{\ttapp{a}}{F(\tau)}\\[1mm]
\tag{\textsc{SemPack}}
(\exists \tau \in \msn{Type}.\; \tau \tsub \alpha \mand 
\ttype{\Sigma}{a}{F(\tau)})
\quad &\Longrightarrow\quad
\ttype{\Sigma}{\tpack{a}}{\texists{\alpha}{F}}\\[1mm]
\tag{\textsc{SemOpen}}
(\ttype{\Sigma}{a}{\texists{\alpha}{F}} \mand \forall \tau 
{\in} \msn{Type}.\qquad\qquad\quad \\ 
\tau \tsub \alpha \mimpl 
\ttype{\mext{\Sigma}{x}{F(\tau)}}{b}{\beta})
\quad &\Longrightarrow\quad
\ttype{\Sigma}{\topen{a}{x}{b}}{\beta}\\[1mm]
\tag{\textsc{SemSubUniv}}
(\beta \tsub \alpha \mand \forall \tau \in \msn{Type}.\; 
\tau \tsub \beta \mimpl F(\tau) \tsub G(\tau))
\quad &\Longrightarrow\quad
\tforall{\alpha}{F} \tsub \tforall{\beta}{G}\\[1mm]
\tag{\textsc{SemSubExist}}
(\alpha \tsub \beta \mand \forall \tau \in \msn{Type}.\; \tau \tsub \alpha
\mimpl F(\tau) \tsub G(\tau))
\quad &\Longrightarrow\quad
\texists{\alpha}{F} \tsub \texists{\beta}{G}
\end{align*}
\end{small}
\caption{\label{fig:typing-lemmas:bounded-quantified-types}
Typing lemmas: bounded quantified types}
\end{figure*}

\subsection{Recursive Types}
\label{subsec:recursive-types}

In contrast to most previous work on
step-indexed models, we consider iso-recursive rather than equi-recursive
types, so
folds and unfolds are explicit in our syntax and consume computation steps.
Iso-recursive types have been previously considered by
Ahmed for a step-indexed relational model of the lambda calculus
\cite{Ahmed:06}. Iso-recursion is simpler, and sufficient for our purpose. 
As a consequence, we  require type constructors to be only non-expansive, as
opposed to the stronger `contractiveness' requirement 
\cite{Appel:McAllester:01}.

\begin{defi}[Recursive types]
\label{def:RecursiveTypes} 
Let $F:\msn{Type}\to\msn{Type}$ be a non-expansive function. We define the set
$\mu F$ by
\begin{gather*}
\mtuple{k, \Psi, \tfold{v}}\in\mu F\ \mequiv\ \forall j < k.\; 
\mtuple{j, \Psi', v} \in F(\mu F)
\end{gather*}
\end{defi}

\begin{prop}
For all non-expansive $F:\msn{Type}\to\msn{Type}$, $\mu F\in\msn{Type}$ is
well-defined.
\end{prop}

\proof[Proof sketch]
The well-definedness follows from the observation that $\tapprox{\mu F}{k}$ is defined
only in
terms of $\tapprox{F(\mu F)}{j}$ for $j<k$, which by the non-expansiveness of
$F$ means that $\tapprox{\mu F}{k}$ relies only on $\tapprox{\mu F}{j}$. 
The closure under state extension is established by an induction, proving that
for each $k\geq 0$, $\tapprox{\mu F}{k}\in\msn{Type}$.
\qed

Figure~\ref{fig:typing-lemmas:recursive-types} presents the semantic typing lemmas 
for recursive types. As a consequence, we have the expected fixed point property
$\ttype{}a{F(\mu F)}\ \mequiv\ \ttype{}{\tfold{a}}{\mu F}$.

\begin{lem}[Recursive types]
\label{lemma:recursive-types}
All the semantic typing lemmas shown in Figure~\ref{fig:typing-lemmas:recursive-types} are valid implications.
\end{lem}

\proof[Proof sketch]
The validity of \textsc{(SemFold)} and \textsc{(SemUnfold)} are easy consequences
 of Definition~\ref{def:RecursiveTypes}.  
For  \textsc{(SemSubRec)}, one shows by induction on $k$ that 
$\tapprox{\mu F}{k}\tsub \tapprox{\mu G}{k}$, using the precondition of the rule
and the non-expansiveness of $F$ and $G$.
\qed

\begin{figure}
\flushleft{For all non-expansive $F,G : \msn{Type}\to\msn{Type}$,}

\centering
\begin{align}
\tag{\textsc{SemUnfold}}
\ttype{\Sigma}{a}{\mu F} \  &\Longrightarrow\ 
\ttype{\Sigma}{\tunfold{a}}{F(\mu F)}\\
\tag{\textsc{SemFold}} 
\ttype{\Sigma}{a}{F(\mu F)} \  &\Longrightarrow\ 
\ttype{\Sigma}{\tfold{a}}{\mu F}\\
\tag{\textsc{SemSubRec}}
(\forall \alpha, \beta
.\; \alpha \tsub \beta \mimpl
F(\alpha) \tsub G(\beta)) \  &\Longrightarrow\ 
\mu F \tsub \mu G
\end{align}
\caption{\label{fig:typing-lemmas:recursive-types}Typing lemmas: recursive types}
\end{figure}

\begin{lem}[Non-expansiveness]
\label{lemma:sem:NonExp}
All the considered type constructors  are non-expansive.
\end{lem}

\proof[Proof sketch]
It is easily seen that the definition of $\tapprox{\alpha\to\beta}{k}$ uses
only $\tapprox{\alpha}{j}$ and $\tapprox{\beta}{j}$ for $j<k$, and therefore
that $\tapprox{\alpha\to\beta}{k}=\tapprox{\tapprox{\alpha}{k}\to\tapprox{\beta}{k}}{k}$. 
A similar statement holds for the $k$-th approximation of  quantified types
$\tforall{\alpha}{F}$
and $\texists{\alpha}{F}$, since their definition only depends on $\tapprox\alpha
j$ and $\tapprox F j$ for $j<k$. In the case of object and recursive
types, the properties 
$\tapprox{\tobjtv{\tau}{d}}{k} = \tapprox{\left[\m_d
:_{\tvar_d}{\tapprox{\tau_d}{k}}\right]_{d\in D}}{k}$ and 
$\tapprox{\mu F}{k} = \tapprox{\mu \tapprox{F}{k}}{k}$ can be established by 
induction on $k$, using the non-expansiveness of $F$ in the latter case. 
\qed

\section{Semantic Soundness}
\label{sec:soundness}

\noindent In order to prove that well-typed terms are safe to evaluate
we relate the syntactic types to their semantic counterparts, and then
use the fact that the semantic typing judgement enforces safety by
construction (Theorem~\ref{theorem:Safety}).  This approach is
standard in denotational semantics.
In fact, none of the main statements or proofs in this section mentions
step-indices explicitly.

\begin{defi}[Interpretation of types and typing contexts]\
\label{MeaningOfTypes}
Let $\eta$ be a total function from type variables to semantic types.
\begin{enumerate}[(1)]
  \item The interpretation $\mngt A\eta$ of a type $A$ is given by the structurally
recursive meaning function defined in Figure~\ref{fig:MeaningFunction}.   
  \item The interpretation of a well-formed typing context $\Gamma$ with 
respect to $\eta$ is given by the function that maps $x$ to $\mngt A \eta$,
for every $x{:}A\in\Gamma$.
\end{enumerate}

\begin{figure}
\begin{align*}
\mngt{X}{\eta} &= \eta(X) & \mngt{\sobjtv{A}{d}}{\eta} &= 
\tobjtvx{\mngt{A_d}{\eta}}{d}\\
\mngt{\sbott}{\eta} &= \tbott &
\mngt{\smu{X}{A}}{\eta} &= \mu (\mlam{\alpha {\in}
\text{\msn{Type}}}{\mngt{A}{\mext{\eta}{X}{\alpha}}})\\
\mngt{\stopp}{\eta} &= \ttopp & \mngt{\sforall{X}{A}{B}}{\eta} &= 
\tforall{\mngt{A}{\eta}}{(\mlam{\alpha{\in}\msn{Type}}{\mngt{B}{\mext{\eta}{X}{\alpha}}})}\\
\mngt{\sarr{A}{B}}{\eta} &= \tarr{\mngt{A}{\eta}}{\mngt{B}{\eta}} &
 \mngt{\sexists{X}{A}{B}}{\eta} &= \texists{\mngt{A}{\eta}}{(\mlam{\alpha{\in}\msn{Type}}{\mngt{B}{\mext{\eta}{X}{\alpha}}})} 
 \end{align*}
\caption{\label{fig:MeaningFunction}Interpretation of types}
\end{figure}
\end{defi}

Note that in Figure~\ref{fig:MeaningFunction} the type constructors used on
the left-hand sides of the equations are simply syntax, while those on the
right hand-sides refer to the corresponding semantic constructions, as defined
in the previous section.

Recall that non-expansiveness is a necessary precondition for some
of the semantic typing lemmas. In particular, the well-definedness of $\mngt
A\eta$ depends on non-expansiveness, due to the use of $\mu$,
$\forall_{(\cdot)}$ and $\exists_{(\cdot)}$ in Figure~\ref{fig:MeaningFunction}.
So we begin by showing that the interpretation of types is a non-expansive map.

\begin{lem}[Non-expansiveness]
\label{lemma:syn:NonExp} 
$\mngt{A}{\eta}$ 
is non-expansive in $\eta$.
\end{lem}

\proof[Proof sketch]
We show that $\tapprox{\mngt{A}{\eta}}{k} =
\tapprox{\mngt{A}{\tapprox{\eta}{k}}}{k}$ holds by induction on the structure of
$A$, relying on Lemma~\ref{lemma:sem:NonExp} for the non-expansiveness of the
semantic type constructions.
\qed

\begin{defi}[$\eta \models \Gamma$]\
\label{EtaSatisfiesGamma}
Let $\Gamma$ be a well-formed typing context. We say that \emph{$\eta$
satisfies $\Gamma$}, written as $\eta \models \Gamma$, 
if  $\eta(X) \tsub \mngt{A}{\eta}$ holds for all $X \ssub A$ appearing in
$\Gamma$. 
\end{defi}

We show the soundness of the subtyping relation.

\begin{lem}[Soundness of subtyping] 
\label{SoundnessOfSubtyping}
If $\ssubtype{\Gamma}{A}{B}$ and $\eta \models \Gamma$ then
$\mngt{A}{\eta} \tsub \mngt{B}{\eta}$.
\end{lem}

\proof[Proof sketch]
By induction on the derivation of $\ssubtype{\Gamma}{A}{B}$ and  
case analysis on the last applied rule. Each case is immediately reduced to one
of the subtyping lemmas from Section~\ref{sec:model}.
\qed

Finally, we prove the semantic soundness of the syntactic
type system with respect to the model. 

\begin{thm}[Semantic soundness]
\label{SemanticSoundness}
Whenever $\stype{\Gamma}{a}{A}$ and $\eta \models \Gamma$ it follows that 
$\ttype{\mnge{\Gamma}{\eta}}{\erase{a}}{\mngt{A}{\eta}}$.
\end{thm}

\proof[Proof sketch]
By induction on the derivation of $\stype{\Gamma}{a}{A}$ and case analysis on
the last rule applied. Each case is easily reduced to one
of the semantic typing lemmas from Section~\ref{sec:model},
using a standard type substitution lemma for derivations ending with an
application of \textsc{(Fold)}, \textsc{(Unfold)}, \textsc{(TApp)}, or
\textsc{(Pack)}. \qed

By Theorems \ref{SemanticSoundness} (Semantic soundness) and
\ref{theorem:Safety} (Safety), we have a proof of safety for the type system
from Section~\ref{subsec:type-system}.

\begin{cor}[Type safety]
\label{TypeSafety}
Well-typed terms are safe to evaluate.\qed
\end{cor}

\section{Self Types} 
\label{sec:self-types}


\noindent\emph{Self types} have been proposed by Abadi and Cardelli 
\cite{Abadi:Cardelli:96} as a means to reconcile recursive object types with 
`proper' subtyping. Self types are interesting because they allow us to type 
methods that return the possibly modified host object, or a clone of it. For
instance, a type of list nodes, with a filter method that produces the sublist
of all elements satisfying a given predicate, is
\begin{align*}
\textit{List}_A &= \textit{Obj}(X)[\textnormal{hd}:_\tinv A,\, \textnormal{tl} 
:_\tinv X{+}\textit{Unit},\, \textnormal{filter}:_\tcov (A\to\textit{Bool})\to 
X,\ldots]
\end{align*}
Note that the similar recursive  type 
\begin{align*}
\mu(X)[\textnormal{hd}:_\tinv A, 
\textnormal{tl}:_\tinv X{+}\textit{Unit}, \textnormal{filter}:_\tcov 
(A\to\textit{Bool})\to X,\ldots]
\end{align*} 
does not satisfy the usual subtyping for
object types because of the invariance of the hd and tl fields.
 
\subsection{Semantics of Self Types} 
 Abadi and Cardelli \cite[Ch.~15]{Abadi:Cardelli:96} show how self types can be 
 understood in terms of recursive and existentially quantified object types via 
 an encoding. More precisely, the type $\textit{Obj}(X)\!\tobjtv{B}{d}$  where 
 $X$ may occur positively in  $B_d$, stands for the recursive type  
 $\mu(Y)\exists(X\ssub Y)\!\tobjtv{B}{d}$. The bounded existential quantifier 
 introduced by this encoding gives rise to the desired subtyping in width 
 and depth, despite the type recursion.

Since our type system features recursive and bounded existential types, self 
types  could be accommodated via this  encoding. However  a
treatment of self types can be achieved even more directly, without relying on
the encoding. In fact, almost everything is 
in place already: recall that the semantics of object types 
(Definition~\ref{def:ObjectTypes}) employs recursion and an existential 
quantification to refer to the `true' type of an object. Condition {\sc Obj-2} 
in Definition~\ref{def:ObjectTypes} can be changed to take advantage of this 
type:

\begin{defi}[Self types]
\label{def:SelfTypes} Assume $F_d : \msn{Type}\to\msn{Type}$ are monotonic and
non-expansive type
constructors, for all $d\in D$. Then let $\alpha = \tobjtv{F}{d}$ be defined as
the set
of all triples $\mtuple{k, \Psi, \tvobj{e}}$ such that $D \subseteq E$ and
\begin{align}
\tag{{\sc Obj-1}}
&  \exists \alpha'.\ {\alpha'} \in \msn{Type} \mand 
\tapprox{\alpha'}{k} \tsub \tapprox{\alpha}{k} \\
\tag{{\sc Obj-2-self}}
& \wedge (\forall d \in D.\; \mtuple{k,\Psi,l_d} \in \trefv{\tvar_d}{(\alpha'\to F_d(\alpha'))}) \\
\tag{{\sc Obj-3}}
& \wedge (\forall j<k.\; \forall \Psi'.\; \forall \tvobjx{l_e'}{e}. \\[-.5mm] 
\notag
& \qquad \;
\textend{k}{\Psi}{j}{\Psi'} \mand
(\forall e \in E. \; \tapprox{\Psi'}{j}(l_e') =
\tapprox{\Psi}{j}(l_e)) \\[-.5mm] 
\notag
& \qquad 
 \mimpl \mtuple{j, \tapprox{\Psi'}{j}, \tvobjx{l_e'}{e}} \in \alpha')
\end{align}
\end{defi}

\begin{figure*}
\flushleft{Let $\alpha = \tobjtv{F}{d}$ and $\beta = \tobjtv{G}{e}$ with 
$E\subseteq D$.}

\centering
\begin{small}
\begin{align*}
\tag{\textsc{SemObj-Self}}
(\forall d\in D.\ \ttype{\mext{\Sigma}{x_d}{\alpha}}{b_d}{F_d(\alpha)})
\quad &\Longrightarrow\quad
\ttype{\Sigma}{\tobj{d}}{\alpha}\\[1mm]
\tag{\textsc{SemInv-Self}}
(
\ttype{\Sigma}{a}{\alpha} \mand
e\in D\mand 
\tvar_e \in \{\tcov, \tinv\})
\quad &\Longrightarrow\quad
\ttype{\Sigma}{a.\m_e}{F_e(\alpha)}\\[1mm]
\tag{\textsc{SemUpd-Self}}
(\ttype{\Sigma}{a}{\alpha}\mand
e\in D\mand
\tvar_e \in \{\tcon,\tinv\}\phantom{)}\quad \\ 
\mand \forall\xi\tsub\alpha.\ \ttype{\mext{\Sigma}{x}{\xi}}{b}{F_e(\xi)})
\quad &\Longrightarrow\quad
\ttype{\Sigma}{\tupd{a}{\m_e}{\tself{x}{b}}}{\alpha}\\[1mm]
\tag{\textsc{SemClone-Self}}
\ttype{\Sigma}{a}{\alpha}
\quad &\Longrightarrow\quad
\ttype{\Sigma}{\tclone{a}}{\alpha}\\[1mm]
\tag{\textsc{SemSubObj-Self}}
(\forall e \in E.\ (
\tvar_e \in \{\tcov,\tinv\}\mimpl \forall \xi\tsub\alpha.\ F_e(\xi) \tsub
G_e(\xi))\phantom{)}\quad\\ 
\mand (\tvar_e \in 
\{\tcon,\tinv\}\mimpl \forall \xi\tsub\alpha.\ G_e(\xi) \tsub
F_e(\xi))) \quad &\Longrightarrow\quad \alpha \tsub \beta\\[1mm]
\tag{\textsc{SemSubObjVar-Self}}
(\forall d \in D.\ \tvar_d = \tinv \mor \tvar_d = \tvar'_d)
\quad &\Longrightarrow\quad
\tobjtv{F}{d} \tsub \tobjtvv{\tvar'_d}{F}{d}
\end{align*}
\end{small}
\caption{\label{fig:typing-lemmas:self-types}
Typing lemmas: self types}
\end{figure*}

As in Section~\ref{subsec:object-types} one shows that
Definition~\ref{def:SelfTypes}  uniquely determines a type. In this proof, the
non-expansiveness of the type functions $F_d$ is necessary in order to ensure
that $\tapprox{\tobjtv{F}{d}}{k}$ is defined in terms of $\tapprox{\tobjtv{F}{d}}{j}$
for $j<k$ only. Moreover, the proofs of the typing lemmas  for object types
(see Section~\ref{app:typing-lemmas-objects} in the Appendix) carry over with
minor modifications, to show that the semantic typing lemmas in 
Figure~\ref{fig:typing-lemmas:self-types} hold. Most cases are obtained by
replacing the result type $\tau_d$ by $F_d(\alpha')$ throughout the proof, for
$\alpha'$ the existentially quantified type from condition
\tagref{\textsc{Obj-1}} of Definition~\ref{def:SelfTypes}. The proof of
\tagref{\textsc{SemInv-Self}} uses the monotonicity of $F_e$, to conclude that
the result of the invocation has type $F_e(\alpha)$ from the fact that it has
type $F_e(\alpha')$, as given by condition \tagref{\textsc{Obj-2-Self}}.
In the proofs of \tagref{\textsc{SemUpd-Self}} and
\tagref{\textsc{SemSubObj-Self}}, the universally quantified $\xi$ from the
respective assumptions is instantiated by $\alpha'$. Since
\tagref{\textsc{Obj-1}} only gives  that
$\tapprox{\alpha'}{k}\tsub\tapprox{\alpha}{k}$ but not necessarily
$\alpha'\tsub\alpha$, these three proofs also use the non-expansiveness of $F_d$
in an essential way. Finally, given the non-expansiveness of each $F_d$, an
induction shows that $\tapprox{\tobjtv{F}{d}}{k} = \tapprox{\left[\m_d
:_{\tvar_d}{\tapprox{F_d}{k}}\right]_{d\in D}}{k}$ for all $k$. In other words,
$\tobjtv{F}{d}$, viewed as a type constructor, is non-expansive and 
Lemma~\ref{lemma:sem:NonExp} still holds.


The interpretation of syntactic type expressions given in 
Figure~\ref{fig:MeaningFunction} extends straightforwardly to self types using 
the new type constructor:
\begin{align*}
\mngt{\textit{Obj}(X)\!\sobjtv{A}{d}}{\eta} &=
\tobjtvx{\tlam{\alpha{\in}\msn{Type}}{\mngt{A_d}{\mext{\eta}{X}{\alpha}}}}{d}
\end{align*}
With this interpretation and the semantic typing lemmas from 
Figure~\ref{fig:typing-lemmas:self-types}, the soundness theorem from 
Section~\ref{sec:soundness} should extend to a syntactic type system for
objects with self types similar to the one derived by Abadi and Cardelli 
\cite[Ch.~15]{Abadi:Cardelli:96} for their encoding (but also including 
variance annotations and a typing rule for cloning).


\subsection{Limitations}
Note that with the exception of {\sc SemUpd-Self}, all the  semantic
typing lemmas for self types are stronger than their counterparts from
Figure~\ref{fig:typing-lemmas:object-types}.
This is already enough to typecheck many examples involving self types
\cite[Ch.~15]{Abadi:Cardelli:96}.
  
However, as for the encoding of Abadi and Cardelli, when updating methods one
usually does not have full information about the precise self type $\alpha'$ of
the host object, which may be a proper subtype of $\alpha$. Therefore the
statement \tagref{\textsc{SemUpd-Self}} about method update in
Figure~\ref{fig:typing-lemmas:self-types}  includes a quantification over all
subtypes $\xi$ of the known type $\alpha$ of the object~$a$, to ensure that the
updated method also works correctly for the precise type. As a consequence the
new method body $b$ must be sufficiently parametric in the type of its self
parameter $x$, which can be overly restrictive. Abadi and Cardelli
\cite[Ch.~17]{Abadi:Cardelli:96} demonstrate this limitation with an example of
objects that provide a backup and a retrieve method:
\begin{align*}
\textit{Bk} &= \textit{Obj}(X)\!
 [\textnormal{retrieve} :_\tinv X,\, \textnormal{backup} :_\tinv X, \ldots]
\end{align*}
A sensible definition of  the backup method updates the
retrieve method so that a subsequent invocation of retrieve yields a clone of
the current  object $x$:
\begin{align*}
\textit{backup}(x)\ &= \ 
{\tlet{z}{\tclone{x}}{\tupd{x}{\textnormal{retrieve}}{\tself{y}{z}}}}
\end{align*}
Here, the `$\,\tlet{z}{a}{b}\,$'  stands for the usual syntactic sugar
$(\tlam{z}{b})\,a$. Let $\beta=\mngt{\textit{Bk}}{\eta}$ denote the
interpretation of the syntactic type \textit{Bk}. While the
$\textnormal{backup}$ method has the correct operational behaviour, 
to typecheck the method update to $x$ in its   body using
\tagref{\textsc{SemUpd-Self}} we would need the
statement  $\ttype{\Sigma\left[x{:=}\beta,z{:=}\beta,y{:=}\xi\right]}{z}{\xi}$. 
But this statement 
does not hold for an arbitrary subtype $\xi\tsub\beta$. Therefore the semantic typing lemmas
stated above  are not strong enough to prove that
$\ttype{\Sigma\left[x{:=}\beta,z{:=}\beta\right]}{\tupd{x}{\textnormal{retrieve}}{\tself{y}{z}}}{\beta}$,
and thus that
$\ttype{\Sigma\left[x{:=}\beta\right]}{\textit{backup}(x)}{\beta}$
holds for the method body. This prevents us from typing an object that contains
this method
(\EG $\left[\textnormal{backup}=\tself{x}{\textit{backup}(x)},\ldots\right]$)
to type $\beta$ using the semantic typing lemmas.

Abadi and Cardelli \cite[Ch.~17]{Abadi:Cardelli:96} address this lack of
expressiveness by modifying the calculus in two respects.
First, they introduce a new syntax for method update, 
$\tupd{a}{\m}{(y,z=c)\tself{x}{b}}$. Operationally this new construct behaves 
just like
\begin{align}
\label{eqn:syntax-extension-encoding}
\tlet{y}{a}{\tlet{z}{c}{\tupd{y}{\m}{\tself{x}{b}}}}
\end{align}
but its  typing rule  is more powerful than the one induced by this encoding. 
When typing $c$ and the method body $b$, $y$ can be assumed to have the precise 
type of the object:
\begin{align*}
\mrule{Upd-Self}{A\meqsyn \textit{Obj}(Y)\!\sobjtv{A}{d}\\
\stype{\Gamma}{a}{A} \\ e \in D \\ \tvar_e \in \{\tcon,\tinv\}\\\\
\stype{\Gamma,X\ssub A,y{:}X}{c}{C}\\
\stype{\Gamma,X\ssub A,y{:}X,z{:}C,x{:}X}{b}{A_e}}
{\stype{\Gamma}{\tupd{a}{\m_e}{(y,z=c)\tself{x}{b}}}{A}}
\end{align*}
Second, in order to propagate this information, typing rules with  `structural' 
assumptions are introduced. For instance, the inference rule for object cloning 
takes the form
\begin{align*}
\mrule{Clone-Str}{A\ssub \textit{Obj}(Y)\!\sobjtv{A}{d}\\
\stype{\Gamma}{a}{A}}{\stype{\Gamma}{\sclone{a}}{A}}
\end{align*}
thus applying also in the case where $A$ is a type variable. In this modified 
system, the body of the \textnormal{backup} method can be rewritten as
\begin{align}
\label{eqn:example-with-syntax-extension-encoding}
{\textit{backup}}_\textit{mod}(x)\ &= \ 
{\tupd{x}{\textnormal{retrieve}}{(y,z=\tclone{y})\tself{x}{z}}}
\end{align}
and the judgement 
$\stype{\Gamma,x{:}\textit{Bk}}{{\textit{backup}}_\textit{mod}(x)}{\textit{Bk}}$ 
is derivable.

Even in the purely syntactic setting, the \ADHOC character of the syntax 
extension is not entirely satisfactory, but for the step-indexed semantics of 
types both modifications are in fact problematic. First, although it seems 
reasonable to expect that all the semantic typing lemmas from
Section~\ref{sec:model} continue to hold, a change of the calculus and its
operational semantics would require us to recheck the proofs about object types
in detail.
Fortunately,  the syntax extension does not seem necessary
from the semantic typing point of view; we can already prove the semantic
soundness of rule \tagref{\textsc{Upd-Self}} with respect to the encoding of
the new method update construct from \eqref{eqn:syntax-extension-encoding}:

\noindent
If $\alpha=\tobjtv{F}{d}$, $e\in D$, and $\tvar_e \in \{\tcon,\tinv\}$ then 
\begin{multline*}
\ttype{\Sigma}{a}{\alpha}
\mand
\forall \xi\tsub\alpha.\ \ttype{\mext{\Sigma}{y}{\xi}}{c}{\gamma}
\mand 
\forall \xi\tsub\alpha.\ \ttype{\Sigma\left[ y \mathrel{\mathop:}= \xi, 
 z\mathrel{\mathop:}=\gamma, x\mathrel{\mathop:}=\xi\right]}{b}{F_e(\xi)}\\
 \quad \Longrightarrow\quad
 \ttype{\Sigma}{\tlet{y}{a}{\tlet{z}{c}{\tupd{y}{\m}{\tself{x}{b}}}}}{\alpha}
\end{multline*}
However, by itself this rule does not help in typing the
body of the \textnormal{backup} method, and the introduction of rules with structural
assumptions presents a more severe difficulty. Soundness of these rules relies
 on the fact that every subtype of an object type is another object type. In 
other words, in the \tagref{\textsc{Clone-Str}} rule the type $A$ is assumed to 
range only over object types. Such structural assumptions are usually not 
valid in semantic models, and they are certainly not justified with respect to
our semantically defined subtype relation, which is just set inclusion.

\subsection{Self Types with Structural Assumptions}
To sum up the previous subsection, 
the problem is that
the semantic typing lemmas from Figure~\ref{fig:typing-lemmas:self-types} are too
weak to type certain examples such as the body of the backup method, but
the usual way to strengthen these rules in a syntactic setting is not
semantically sound in our model.
Still, $\ttype{\mext{\Sigma}{x}{\beta}}{\textit{backup}(x)}{\beta}$ 
is a valid typing judgement about the method body. This can be seen by taking a
closer look at the semantic definition of the self type $\beta=
\mngt{\textit{Bk}}{\eta}$: essentially, if for a suitable substitution
$\sigma:_{k,\Psi}\mext{\Sigma}{x}{\beta}$ the substitution instance
\begin{align*}
\sigma(\textit{backup}(x)) =
{\tlet{z}{\tclone{\sigma(x)}}{\tupd{\sigma(x)}{\textnormal{retrieve}}{\tself{y}{z}}}}
\end{align*} 
becomes irreducible in less
than $k$ steps, then $\sigma (x)$ must be an object value
$v=\tvobjx{l_d}{d}$ such that $\mtuple{k,\Psi,v}\in\beta$. Property
\tagref{\textsc{Obj-1}} of $\beta$ asserts the existence of a type $\alpha'$ such that 
$\tapprox{\alpha'}{k}\tsub\tapprox{\beta}{k}$, and property 
\tagref{\textsc{Obj-3}} entails that $z$ becomes bound to a value $v'$ of this
type $\alpha'$. Thus by \tagref{\textsc{Obj-2-self}} the eventual update of the 
\textnormal{retrieve} field of
$v$ is valid, since the new method $\tlam{y}{v'}$ has the expected type
${\alpha'}\to{\alpha'}$ to sufficient approximation.

Similar `manual' reasoning seems possible in other cases, but a more
principled approach will let us use typing lemmas that are strong enough and
avoid explicit reasoning about the operational semantics and step indices.  To
facilitate this, we develop a semantic counterpart to the structural
assumptions  that appear in the syntactic type system of Abadi and Cardelli
\cite[Ch.~17]{Abadi:Cardelli:96}.
More precisely, we introduce a relation $\alpha'\tsubself\alpha$ between
semantic types that strengthens the subtype relation: 
intuitively $\alpha'$ is the precise, recursive type of some collection 
of object values from the object type $\alpha$. The type $\alpha$ acts as an
interface that lists the permitted operations on these object values. 

\begin{defi}[Self type exposure] 
\label{def:self-type-exposure}
For $\alpha=\tobjtv{F}{d}$ and
$\alpha'\in\msn{Type}$ the relation $\alpha'\tsubself\alpha$  holds
if and only if $\alpha'\tsub\alpha$ and 
for all $E\supseteq D$ and $\mtuple{k,\Psi,\tvobjx{l_e}{e}}\in\alpha'$,
\begin{align}
\tag{{\sc Obj-2-self}}
&  (\forall d \in D.\; \mtuple{k,\Psi,l_d} \in \trefv{\tvar_d}{(\alpha'\to F_d(\alpha'))}) \\
\tag{{\sc Obj-3}}
\wedge \quad&  (\forall j<k.\; \forall \Psi'.\; \forall \tvobjx{l_e'}{e}.
\\[-.5mm]
\notag
& \qquad \;
\textend{k}{\Psi}{j}{\Psi'} \mand
(\forall e \in E. \; \tapprox{\Psi'}{j}(l_e') =
\tapprox{\Psi}{j}(l_e))  \\[-.5mm] 
 \notag
 & \qquad 
 \mimpl \mtuple{j, \tapprox{\Psi'}{j}, \tvobjx{l_e'}{e}} \in \alpha')
\end{align}
\end{defi}

Notice that $\alpha'\tsubself\alpha$ essentially states that $\alpha'$ is a 
type that can take the place of the existentially quantified `self type' in an 
object type (see Definition~\ref{def:SelfTypes}). It is immediate from this 
definition that $\alpha'\tsubself\alpha$ implies $\alpha'\tsub\alpha$. Note 
however that $\tsubself$ is not reflexive: in general, $\alpha'$  is
not an object type (\EG $\alpha'$ could be empty). 
Intuitively, the object type $\alpha$ is obtained as a union of such $\alpha'$.

\begin{figure*}
\flushleft{Let $\alpha = \tobjtv{F}{d}$.}

\centering
\begin{small}
\begin{align*}
\tag{\textsc{SemObj-Str}}
(\forall d{\in} D.\ \forall\xi{\in}\msn{Type}.\ \xi\tsubself\alpha\mimpl 
\ttype{\mext{\Sigma}{x_d}{\xi}}{b_d}{F_d(\xi)})
\quad &\Longrightarrow\quad
\ttype{\Sigma}{\tobj{d}}{\alpha}\\[1mm]
%
%
\tag{\textsc{SemInv-Str}}
(
\alpha'\tsubself\alpha
\mand \ttype{\Sigma}{a}{\alpha'} 
\mand e\in D
\mand \tvar_e \in \{\tcov, \tinv\})
\quad &\Longrightarrow\quad
\ttype{\Sigma}{a.\m_e}{F_e(\alpha')}\\[1mm]
\tag{\textsc{SemUpd-Str}}
(\alpha'\tsubself\alpha
\mand \ttype{\Sigma}{a}{\alpha'}\mand
e\in D\mand
\tvar_e \in \{\tcon,\tinv\}\phantom{)}\quad \\ 
\mand \ttype{\mext{\Sigma}{x}{\alpha'}}{b}{F_e(\alpha')})
\quad &\Longrightarrow\quad
\ttype{\Sigma}{\tupd{a}{\m_e}{\tself{x}{b}}}{\alpha'}\\[1mm]
\tag{\textsc{SemClone-Str}}
\alpha'\tsubself\alpha 
\mand \ttype{\Sigma}{a}{\alpha'} 
\quad &\Longrightarrow\quad
\ttype{\Sigma}{\tclone{a}}{\alpha'}\\[1mm]
%
%
\tag{\textsc{SemLet-Str}}
   \ttype{\Sigma}{a}{\alpha} 
   \mand (\forall\xi{\in}\msn{Type}.\
   \xi\tsubself\alpha\mimpl\ttype{\mext{\Sigma}{x}{\xi}}{b}{\beta})
   \quad &\Longrightarrow\quad
   \ttype{\Sigma}{\tlet{x}{a}{b}}{\beta}   
\end{align*}
\end{small}
\caption{\label{fig:typing-lemmas:self-types-strengthened}
Typing lemmas with structural assumptions: self types}
\end{figure*}

Figure~\ref{fig:typing-lemmas:self-types-strengthened} lists new typing lemmas 
for self types that exploit the relation $\tsubself$.  Compared to 
\tagref{\textsc{SemInv-Self}} and \tagref{\textsc{SemClone-Self}} from 
Figure~\ref{fig:typing-lemmas:self-types}, the typing lemmas 
\tagref{\textsc{SemInv-Str}} and \tagref{\textsc{SemClone-Str}} use the 
additional assumptions $\alpha'\tsubself\alpha$ and 
$\ttype{\Sigma}{a}{\alpha'}$ to establish a more precise typing for the result. 
Similarly, while \tagref{\textsc{SemUpd-Self}} universally quantifies over all 
$\xi\tsub\alpha$ in its premise, \tagref{\textsc{SemUpd-Str}} limits this to 
those $\xi\in\msn{Type}$ for which $\xi\tsubself\alpha$ holds. Finally, 
\tagref{\textsc{SemLet-Str}} lets us use an object $a$ within $b$ with the more 
precise type $\xi$ where $\xi\tsubself\alpha$, and similarly 
\tagref{\textsc{SemObj-Str}} lets us type the method bodies under the more 
informative assumption that $\xi\tsubself\alpha$. The latter two are the key 
lemmas to introduce an assumption $\alpha'\tsubself\alpha$ in proofs using these
semantic typing lemmas.


As an illustration, consider the example of the backup method again. In order 
to construct objects with the backup method, we will establish that
\begin{align}
\label{eqn:typing-the-backup-example}
 \forall\xi{\in}\msn{Type}.\ \xi\tsubself\beta\mimpl 
\ttype{\mext{\Sigma}{x}{\xi}}{\textit{backup}(x)}{\xi}
\end{align}
holds, where $\beta=\mngt{\textit{Bk}}{\eta}$ and ${\textit{backup}}(x)$
abbreviates
 `${\tlet{z}{\tclone{x}}{\tupd{x}{\textnormal{retrieve}}{\tself{y}{z}}}}$' as 
before. From this, by \tagref{\textsc{SemObj-Str}} it will follow that 
\[\ttype{\Sigma}{[\textnormal{backup}=\tself{x}{{\textit{backup}(x)}},\textnormal{retrieve}=\ldots]}{\beta}\]
If we desugar the let construct 
in $\textit{backup}(x)$ and apply lemmas 
\tagref{\textsc{SemApp}} and \tagref{\textsc{SemLam}},
we notice that in order to show
\eqref{eqn:typing-the-backup-example} it suffices to prove that
$\ttype{\Sigma[x{:=}\xi]}{\tclone{x}}{\xi}$ and 
$\ttype{\Sigma[x{:=}\xi,z{:=}\xi]}{\tupd{x}{\textnormal{retrieve}}{\tself{y}{z}}}{\xi}$. 
Using $\xi\tsubself\beta$ and $\ttype{\Sigma[x{:=}\xi]}{x}{\xi}$, the  validity 
of the former judgement is immediate by \tagref{\textsc{SemClone-Str}}. 
Similarly, the latter follows by \tagref{\textsc{SemUpd-Str}} from the fact 
that  $\xi\tsubself\beta$ and since the \textnormal{retrieve} method is
listed with variance annotation `$\tinv$' in $\beta$.

\begin{lem}[Self types: lemmas with structural assumptions] 
All the semantic typing lemmas shown in
Figure~\ref{fig:typing-lemmas:self-types-strengthened} are valid implications.
\end{lem}

\proof[Proof sketch]
The proofs of \tagref{\textsc{SemInv-Str}}, \tagref{\textsc{SemClone-Str}}, and 
\tagref{\textsc{SemUpd-Str}} are straightforward adaptations of those for 
\tagref{\textsc{SemInv}}, \tagref{\textsc{SemClone}}, and 
\tagref{\textsc{SemUpd}}. As an example, we give the proof of
\tagref{\textsc{SemUpd-Str}} as Lemma~\ref{app:lemma:SemUpd-Str} in the Appendix. 
More interestingly, \tagref{\textsc{SemLet-Str}} relies on the following
property of object types $\alpha$:
\begin{align*}
\mtuple{k,\Psi,v}\in\alpha\quad&\Longrightarrow\quad 
\exists \alpha'\in\msn{Type}.\ \alpha'\tsubself\alpha \mand
\mtuple{k-1,\tapprox{\Psi}{k-1},v}\in\alpha'
\end{align*}
In the proof of \tagref{\textsc{SemLet-Str}}, this $\alpha'$ is used to
instantiate the universally quantified $\xi$ in the premise
$\xi\tsubself\alpha\mimpl\ttype{\mext{\Sigma}{x}{\xi}}{b}{\beta}$. The full
proof is given as Lemma~\ref{app:lemma:SemLet-Str}  in the Appendix.

The proof of \tagref{\textsc{SemObj-Str}} is similar to the one of
\tagref{\textsc{SemObj}} (Lemma~\ref{app:lemma:SemObj} in the Appendix),
except that we use the  heap typing extension
$\Psi' = \tapprox{\mextx{\Psi}{l_d}{(\tarr{\beta}{F_d(\beta)})}{d \in D}}{k-1}$, where 
$\beta$ is a recursive record type satisfying 
conditions \tagref{\textsc{Obj-2-self}} and \tagref{\textsc{Obj-3}}, but not
validating any subtyping property. In verifying that the extended heap is
well-typed with respect to this $\Psi'$, one uses that  
$\beta\tsubself\alpha$, in order to instantiate the assumptions on the
method bodies and obtain  
$\ttype{\mext{\Sigma}{x_d}{\beta}}{b_d}{F_d(\beta)}$. Finally, to show that the 
generated object value has type $\alpha$, Claim \ref{the-important-claim} is 
strengthened to show that the object value is in fact in $\beta$, which is a
subtype of $\alpha$ since $\beta\tsubself\alpha$ 
(see Proposition~\ref{prop:rec-object-type-is-a-self-type} and 
Lemma~\ref{app:lemma:SemObj-Str} in the Appendix for the full proof).   
\qed

\begin{rem} The implication $\ttype{\Sigma}{a}{\alpha} \ \Longrightarrow\ 
\exists\alpha'\in\msn{Type}.\ \alpha'\tsubself\alpha\mand 
\ttype{\Sigma}{a}{\alpha'}$ for $\alpha = \tobjtv{F}{d}$ may appear reasonable 
 (and would entail both \tagref{\textsc{SemLet-Str}} and 
\tagref{\textsc{SemObj-Str}}), but we do not believe that it holds.  
The problem is that, while the 
premise $\ttype{\Sigma}{a}{\alpha}$ guarantees for each $k\geq 0$ the existence
of a type $\alpha_k'$  that satisfies the requirement 
$\alpha_k'\tsubself\alpha$, it is in general not possible to
construct a type $\alpha'$ `in the limit' from this sequence. 
For the same reason, the implication 
$\ttype{\Sigma}{\tpack{a}}{\texists{\alpha}{F}}\ \Longrightarrow\ \exists
\alpha'\tsub\alpha.\,  \ttype{\Sigma}{a}{F(\alpha')}$ is not valid. 
 The typing lemma \tagref{\textsc{SemLet-Str}} avoids 
this problem since $\alpha'$ is only  needed up to a fixed approximation,
and  so the choice of $\alpha'_k$ for sufficiently large $k$ suffices (\CF proof of
Lemma~\ref{app:lemma:SemLet-Str} in the Appendix). 
On the other hand, the \tagref{\textsc{SemObj-Str}} lemma avoids the
problem by instantiating $\xi$ with a particular type $\beta$, for which
$\beta\tsubself\alpha$ is already known. 
\end{rem}

In this section we showed that our semantics of object types naturally extends 
to self types, while avoiding any change to the syntax and operational 
semantics of the calculus. We proved a first set of typing lemmas that are
natural and apply to many examples (Figure~\ref{fig:typing-lemmas:self-types}).
These lemmas are however not sufficient to typecheck self-returning methods.
To achieve this, we developed a second set of typing lemmas that involve the
object's precise type,  through the relation $\alpha'\tsubself\alpha$ 
(Figure~\ref{fig:typing-lemmas:self-types-strengthened}).
Note that these latter lemmas do not fully subsume the former ones, since the
$\tsubself$ relation is not reflexive. 
We leave open the problem of relating the lemmas in
Figure~\ref{fig:typing-lemmas:self-types-strengthened} to a syntactic type system.


\section{Generalizing Reference and Object Types}
\label{sec:gen-obj}

\noindent The semantics described in this paper generalizes the
reference types from \cite{Ahmed:04, Ahmed:Appel:Virga:03} to readable
and writable reference types.  This can be generalized even
further. We can have a reference type constructor that takes
\emph{two} types as arguments: one that represents the most general
type that can be used when writing to the reference, and another for
the most specific type that can be read from it \cite{Pottier::98}.
This can be easily expressed using our readable and writable reference
types together with intersection types:
\[\treft{\tau^w}{\tau^r} = \trefv{\tcon}{\tau^w} \cap
\trefv{\tcov}{\tau^r}\]
After unfolding the definitions, this yields 
\[\treft{\tau^w}{\tau^r} = \mset{\mtuple{k,\Psi,l}}{
\tapprox{\tau^w}{k} \tsub \tapprox{\Psi(l)}{k} \tsub \tapprox{\tau^r}{k}}.\]
As one would expect, this generalized reference type constructor is
contravariant in the first argument and covariant in the second one:
\begin{gather}
\tag{\textsc{SemSubRef-Gen}}
\beta^w \tsub \alpha^w \quad\mand\quad \alpha^r \tsub \beta^r
\qquad\Longrightarrow\qquad \treft{\alpha^w}{\alpha^r} \tsub
\treft{\beta^w}{\beta^r} 
\end{gather}
Note that, if one takes these generalized reference types as primitive, then the
three reference types from Section~\ref{subsec:reference-types} are
obtained as special cases:
\[
\trefv{\tinv}{\tau} = \treft{\tau}{\tau}, \qquad
\trefv{\tcov}{\tau} = \treft{\tbott}{\tau}, \qquad
\trefv{\tcon}{\tau} = \treft{\tau}{\ttopp},
\]
and the subtyping properties from Figure~\ref{fig:subtyping-reference-types}
are still valid.

\begin{figure}
\centering
\includegraphics[width=.85\textwidth]{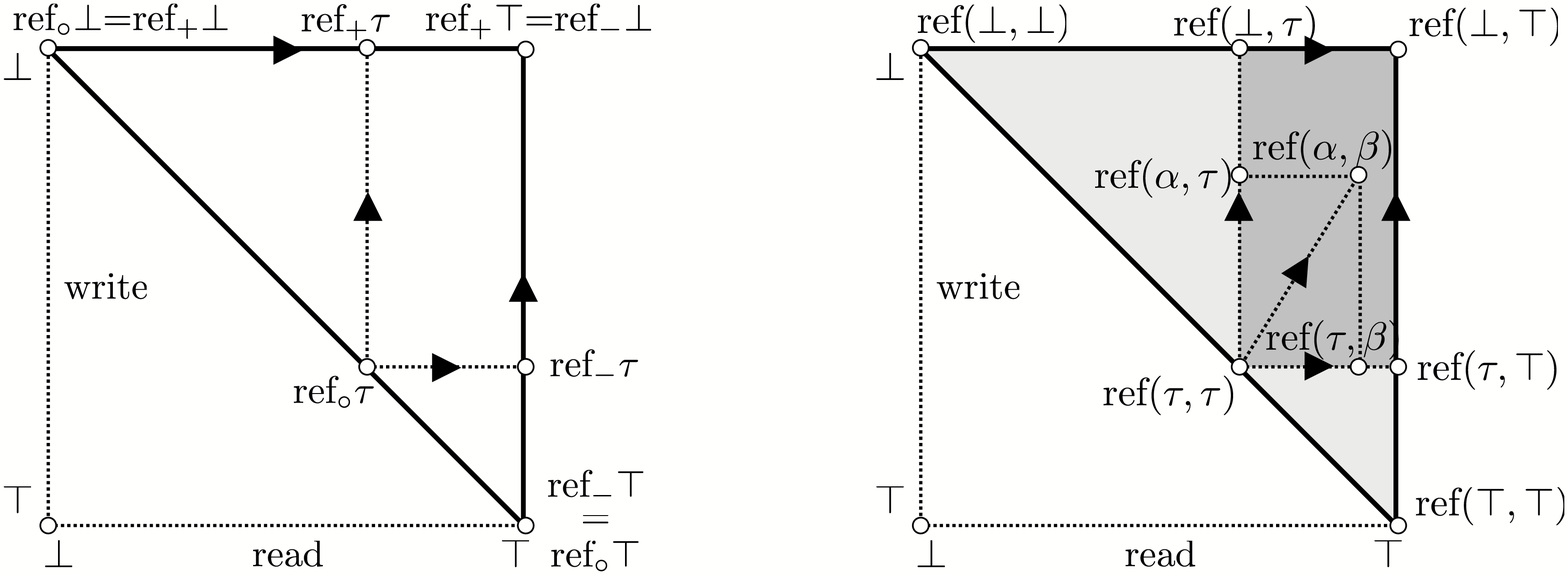}
\begin{minipage}[b]{.5\textwidth}
\caption{\label{fig:ext-ref-1} \small Readable/writable reference types}
\end{minipage}%
\begin{minipage}[b]{.5\textwidth}
\caption{\label{fig:ext-ref-2} \small Generalized reference types}
\end{minipage}
\end{figure}

Figures \ref{fig:ext-ref-1} and \ref{fig:ext-ref-2} give a graphical
representation of the different reference type constructors.
In both figures the horizontal axis represents the type at which a reference
can be read, while the vertical one gives the type at which it can be written.
Notice that because of the different variance the read axis goes from $\tbott$
to $\ttopp$ while the write axis from $\ttopp$ to $\tbott$.

Figure~\ref{fig:ext-ref-1} represents the usual, as well as the readable, and
the writable reference types as points on the three edges of a triangle.
Notice that the usual references can be read and written at the same type.
Without additional information, the readable references can  only be written
safely at type $\tbott$, and the writable ones can only be read at type
$\ttopp$. Subtyping is represented by arrows: covariant on the edge of the readable
reference types and contravariant on the writable reference types' edge.
An invariant reference type can only be subtyped either to a readable or to a
writable reference type.

Figure~\ref{fig:ext-ref-2} illustrates that our generalization of reference
types is indeed very natural. When generalizing, we take not only  
the points on the edges of the triangle, but also the points inside it to be
reference types. Furthermore, instead of having three different kinds of
reference types, we only have one. Subtyping is also more natural: the set of
all supertypes of a reference type cover the area of a rectangle which goes
from the point corresponding to this reference type to the `top' reference type
$\treft{\tbott}{\ttopp}$. For instance, the dark gray rectangle in
Figure~\ref{fig:ext-ref-2} contains all supertypes of $\treft{\tau}{\tau}$.

Applying this idea in the context of the imperative object calculus leads not
only to more expressive subtyping but also to simplifications, since the variance
annotations are no longer needed.
The extended object type $\tobjtxx{\tau^w_d}{\tau^r_d}{d}$ has two types
for each method $\m_d$:
$\tau^w_d$ is the most general type that can be used to
update the given method,
and $\tau^r_d$ is the most specific type that can be expected as a result
when invoking the method.
When defining the semantics of these generalized object types, the only
difference with respect to Definition~\ref{def:ObjectTypes}~(Object types) is
that condition ({\sc Obj-2}) is changed to use an extended reference type:
\begin{align}
\tag{{\sc Obj-2-Gen}}
\forall d \in D.\; \mtuple{k,\Psi,l_d} \in \treft{\alpha'
\to \tau^w_d}{\alpha' \to \tau^r_d}.
\end{align}

\begin{figure*}
\flushleft{Let $\alpha = \tobjtxx{\tau^w_d}{\tau^r_d}{d}$ and 
$\alpha' = \sobjtxx{\tau_d}{\tau_d}{d}$.
}

\centering
\begin{small}
\begin{align*}
\tag{\textsc{SemObj-Gen}}
(\forall d\in D.\ \ttype{\mext{\Sigma}{x_d}{\alpha'}}{b_d}{\tau_d})
\quad &\Longrightarrow\quad
\ttype{\Sigma}{\tobj{d}}{\alpha'}\\[1mm]
\tag{\textsc{SemInv-Gen}}
(
\ttype{\Sigma}{a}{\alpha} \mand
e\in D)
\quad &\Longrightarrow\quad
\ttype{\Sigma}{a.\m_e}{\tau^r_e}\\[1mm]
\tag{\textsc{SemUpd-Gen}}
(\ttype{\Sigma}{a}{\alpha}\mand
e\in D
\mand \ttype{\mext{\Sigma}{x}{\alpha}}{b}{\tau^w_e})
\quad &\Longrightarrow\quad
\ttype{\Sigma}{\tupd{a}{\m_e}{\tself{x}{b}}}{\alpha}\\[1mm]
\tag{\textsc{SemClone-Gen}}
\ttype{\Sigma}{a}{\alpha}
\quad &\Longrightarrow\quad
\ttype{\Sigma}{\tclone{a}}{\alpha}\\[1mm]
\tag{\textsc{SemSubObj-Gen}}
(E \subseteq D \mand 
(\forall e \in E.\ \beta^w_e \tsub \alpha^w_e \mand \alpha^r_e \tsub \beta^r_e))
\quad&\Longrightarrow\quad
\tobjtxx{\alpha^w_d}{\alpha^r_d}{d} \tsub 
\tobjtxx{\beta^w_e}{\beta^r_e}{e}
\end{align*}
\end{small}
\vspace{-0.5cm}
\caption{\label{fig:typing-lemmas:generalized-object-types}
Typing lemmas: generalized object types}
\end{figure*}

\begin{figure}
\flushleft{Let $A  = \sobjtxx{A^w_d}{A^r_d}{d}$ and 
$A' = \sobjtxx{A_d}{A_d}{d}$.
}

\centering
\begin{align*}
\mrule{Obj-Gen}{\forall d \in D.\; 
 \stype{\Gamma, x_d:A'}{b_d}{A_d}}{\stype{\Gamma}{\sobj{A'}{d}}{A'}} \qquad
\mrule{Clone-Gen}{\stype{\Gamma}{a}{A}}{\stype{\Gamma}{\tclone{a}}{A}}
\end{align*}\\[\inferenceruleskip]
\begin{align*}
\mrule{Inv-Gen}{\stype{\Gamma}{a}{A}
 \quad e \in D}{\stype{\Gamma}{a.\m_e}{A^r_e}} \qquad
\mrule{Upd-Gen}{\stype{\Gamma}{a}{A} \quad e \in D \quad
 \stype{\Gamma, x:A}{b}{A^w_e}}
 {\stype{\Gamma}{\supd{a}{\m_e}{\sself{x}{A}{b}}}{A}}
\end{align*}\\[\inferenceruleskip]
\begin{align*}
\mrule{SubObj-Gen}{E \subseteq D \quad \forall e \in E.\;
 \ssubtype{\Gamma}{B^w_e}{A^w_e} \quad \forall e \in E.\; 
 \ssubtype{\Gamma}{A^r_e}{B^r_e}}{
 \ssubtype{\Gamma}{\sobjtxx{A^w_d}{A^r_d}{d}}{\sobjtxx{B^w_e}{B^r_e}{e}}}
\end{align*}
\caption{\label{fig:TypingRulesGeneralizedObjectTypes}
The typing rules for generalized object types}
\end{figure}

Figure~\ref{fig:typing-lemmas:generalized-object-types} presents the
semantic typing lemmas that are validated by this definition of object types,
while Figure~\ref{fig:TypingRulesGeneralizedObjectTypes} gives the
corresponding syntactic typing rules.
Note that the complex and seemingly ad-hoc rules for subtyping object types given in 
Figure~\ref{fig:SyntacticSubtyping} or in \cite{Abadi:Cardelli:96} are
replaced by only one rule ({\sc SubObj-Gen}).

\begin{lem}[Generalized object types]
\label{lemma:generalized-object-types}
All the semantic typing lemmas shown in
Figure~\ref{fig:typing-lemmas:generalized-object-types} are valid implications.
\end{lem}
\proof[Proof sketch]
The proof of the subtyping lemma ({\sc SemSubObj-Gen}) follows
easily from the lemma for subtyping generalized reference types ({\sc SemSubRef-Gen}
above), and is therefore significantly simpler than when variance annotations are
involved (see Lemma~\ref{app:lemma:SemSubExtObj} in the Appendix).
For all the other semantic typing lemmas the proofs are basically
unchanged (see Section~\ref{app:typing-lemmas-objects} in the Appendix).
\qed


Note that the generalization of object types presented in this section is 
orthogonal to the extension to self types from  the previous section. The 
generalized object types lead to a type system that is both simpler and more 
expressive than the usual type systems for objects \cite{Abadi:Cardelli:96}.
%
Our generalized object types directly correspond to the
\emph{split types} of Bugliesi and Peric{\'a}s-Geertsen
\cite{Bugliesi:Pericas-Geertsen:02}, 
who have shown that these types are strictly more expressive than
object types with variance annotations
\cite[Example~4.3]{Bugliesi:Pericas-Geertsen:02}.



\section{Comparison to Related Work}
\label{sec:related}
    
\subsection{Domain-theoretic Models}
Abadi and Cardelli give a semantic model for the functional object calculus 
in \cite{Abadi:Cardelli:94, Abadi:Cardelli:96}.
Their type system is comparable to the one we consider here.
Types are interpreted as certain partial equivalence relations over an untyped 
domain-theoretic model of the calculus. No indication is given on how to adapt
this to the imperative execution model.

Based on earlier work by Kamin and Reddy \cite{Kamin:Reddy:94},
Reus \ETAL \cite{Reus:Schwinghammer:06, Reus:Streicher:04, 
Schwinghammer:Diss} construct domain-theoretic models for the 
imperative object calculus, with the goal of proving soundness for the logic of Abadi 
and Leino \cite{Abadi:Leino:04}. The  higher-order store exhibited by the 
object calculus requires defining the semantic domains by 
mixed-variant recursive equations. The dynamic allocation is then addressed by 
interpreting specifications of the logic as Kripke relations, indexed by store 
specifications, which are similar to the heap typings used here.

Building on work by Levy \cite{Levy:02}, an `intrinsically typed' model of the 
imperative object calculus is presented in the second author's PhD thesis 
\cite{Schwinghammer:Diss}, by solving the domain equations in a suitable 
category of functors. However, only first-order types are considered.

Compared to these domain-theoretic models, the step-indexed model we present 
not only soundly interprets a richer type language, but is also easier to work 
with. The way it is based on the operational semantics eliminates the need for 
explicit continuity conditions, and the admissibility conditions are replaced 
by the closure under state extension, which is usually very easy to check. All 
that is needed for the definition of iso-recursive and second-order types are 
non-expansiveness and the stratification invariant. What is missing from our
model is a semantic notion of equality that approximates program equivalence. 
For reasoning about program equivalence in an ML-like language, 
 Ahmed \ETAL \cite{Ahmed:Dreyer:Rossberg:09} have recently developed a
 \emph{relational}
 step-indexed model, and it could be interesting to adapt their work to an
 object-oriented setting.

Recently proposed 
models for polymorphism and general references 
\cite{Birkedal:Stovring:Thamsborg:09, Bohr:07,Bohr:Birkedal:06}
suggest that an adequate semantics for imperative objects with expressive
typing could in principle be developed also in a domain-theoretic setting.
A detailed comparison between step-indexed semantics and
domain-theoretic models would be useful, to make the similarities and
differences between the two approaches more precise. 

It is interesting to see how the object construction rule \textsc{(Obj)} is 
proved correct in each of the models described above. In the domain-theoretic
self-application models 
\cite{Reus:Schwinghammer:06, Reus:Streicher:04}, it directly corresponds to a 
recursive predicate whose well-definedness (\IE existence and uniqueness) must 
be established. This proof exploits properties of the underlying, recursively 
defined domain, and imposes some further restrictions on the semantic types: 
besides admissibility, types appearing in the defining equation of a recursive 
predicate need to satisfy an analogue of the contractiveness property 
\cite{Pitts:96}. In the typed functor category model \cite{Schwinghammer:Diss}, 
object construction is interpreted using a recursively defined function, and  
correspondingly \textsc{(Obj)} is proved by fixed point induction. In the 
step-indexed case, the essence of the proof is a more elementary induction on 
the step index,  with a suitably generalized induction hypothesis (see 
Claim~\ref{the-important-claim} in the proof sketch of 
Lemma~\ref{lemma:object-types} on page \pageref{lemma:object-types}, or the 
full proof in the Appendix).

\subsection{Interpretations of Object Types} 
Our main contribution in this paper is the novel interpretation of object types 
in the step-indexed model. The step-index-induced stratification permits the 
construction of mixed-variance recursive as well as impredicative, second-order 
types. Both are key ingredients in our interpretation of object types. The use 
of recursive and existentially quantified types is in line with the 
type-theoretic work on object encodings, which however has mainly focused on 
object calculi with a {functional} execution 
model~\cite{Bruce:Cardelli:Pierce:99}.

Closest to our work is the encoding of imperative objects into an imperative 
variant of system $F_{\ssub\mu}$ with updatable records, proposed by Abadi 
\ETAL \cite{Abadi:Cardelli:Viswanathan:96}. There, objects are interpreted as 
records containing references to the procedures that represent the methods. As 
in our case, these records have a recursive and existentially quantified record 
type. The difference is that two additional record fields are included in order 
to achieve invocation and cloning, and uninitialized fields are used to 
construct this recursive record. Subtyping in depth is considered in 
\cite{Abadi:Cardelli:Viswanathan:96} only for the encoding of the functional 
object calculus. However, if one added to the target language the readable and 
writable reference types we use in this paper, the encoding of the imperative 
object calculus would extend to subtyping in depth as well.

In the typing rules for self types, the structural assumptions about the subtype
relation play an important role \cite{Abadi:Cardelli:96}. In
Section~\ref{sec:self-types} we developed a semantic counterpart to such typing
rules with structural assumptions, in order to deal with the polymorphic update
of self-returning methods. This is, however, tailored specifically to object
types. Hofmann and Pierce \cite{Hofmann:Pierce:02} investigate the metatheory of
subtyping with structural assumptions in general, and 
give elementary presentations of two encodings of functional objects
in a variant of System $F_\leq$ with type destructors. 
It may be interesting to see if a step-indexed model of this variant of System
$F_\leq$ can be found.

\subsection{Step-indexed Models}
Step-indexed semantic models were introduced by Appel \ETAL in 
the context of foundational proof-carrying code. Their 
goal was to construct more elementary and modular proofs of type soundness 
that can be easily checked automatically. They were primarily interested in 
low-level languages, however they also applied their technique to a pure 
$\lambda$-calculus with recursive types \cite{Appel:McAllester:01}. Later Ahmed
\ETAL successfully extended it to general references and impredicative
polymorphism  \cite{Ahmed:04, Ahmed:Appel:Virga:03}. The
step-indexed semantic model we
present extends the one by Ahmed \ETAL with object types and subtyping. In order to
achieve this, we refine the reference types from \cite{Ahmed:04} to
readable and writable reference types.

Subtyping in a step-indexed semantic model was previously considered by Swadi 
who studied Typed Machine Language \cite{Swadi:03}. Our setup is 
however much different. In particular, the subtle issues
concerning the subtyping of object types are original to our work.

The previous work on step-indexing focuses on `semantic type
systems', \IE the semantic typing lemmas can directly be used for
type-checking programs \cite{Ahmed:Appel:Virga:03, 
Appel:McAllester:01, Appel:Richards:Swadi:02}. However, when one considers more complex type systems
with subtyping, recursive types or polymorphism, the semantic typing lemmas no
longer directly correspond to the usual syntactic rules.
These discrepancies can be fixed, but usually at the cost of more complex
models, like the one developed by Swadi to track type variables
\cite{Appel:Richards:Swadi:02, Swadi:03}.
In Swadi's model an additional `semantic kind system' is used to track the
contractiveness and non-expansiveness of types with free type variables.
We avoid having a more complex model (\EG one that tracks type variables) by 
considering iso-recursive rather than equi-recursive types.
An equi-recursive type is well-defined if its argument is contractive,
and some of the type constructors are not contractive in general (\EG the
identity as well as the equi-recursive type constructor itself).
On the other hand, an iso-recursive type is well-defined under the weaker
assumption that the argument is non-expansive, and all our type constructors
are indeed non-expansive (see Lemma~\ref{lemma:sem:NonExp}).
It is then relatively straightforward to use the semantic typing lemmas in
order to prove the soundness of the standard, syntactic type system we consider
(see Theorem~\ref{SemanticSoundness}).

\subsection{Type Safety Proofs}
Abadi and Cardelli use subject reduction to prove the safety of several type 
systems very similar to the one considered in this paper 
\cite{Abadi:Cardelli:96}. Those purely syntactic proofs are very different from
the `semantic' type safety proof we present (for detailed discussions
about the differences see \cite{Appel:McAllester:01, Wright:Felleisen:94}). 
Since type safety is built into the model, our safety proof neither relies
on a preservation property, nor can preservation be concluded from it.  

Constructing a step-indexed semantics is  more challenging than proving 
progress and preservation. However, for our particular semantics we could reuse 
the model by Ahmed \ETAL and extend it to suit our needs, even though the 
calculus we are considering is quite different. So one would expect that once 
enough general models are constructed  (\EG \cite{Ahmed:04, 
Appel:McAllester:01, Appel:Melli:Richards:Vouillon:07}), it will become easier 
to build new models just by mixing and matching. Assuming the existence of an 
adequate step-indexed model, the effort needed to prove the semantic typing 
lemmas using `pencil-and-paper' is somewhat comparable to the one required for 
a subject reduction proof. Since each of the semantic typing lemmas is proved 
in isolation, the resulting type soundness proof is more modular; the 
extensions we consider in Sections~\ref{sec:self-types} and 
\ref{sec:gen-obj} illustrate this aspect rather well. According to Appel's 
original motivation, the advantages of step-indexing should become even more 
apparent when formalizing the proofs in a proof assistant 
\cite{Appel:McAllester:01}.

\subsection{Generalized Reference and Object Types}
The readable  and the writable reference types we define in 
Section~\ref{subsec:reference-types} and use for modeling object types
in Section~\ref{subsec:object-types} are similar to the reference types
in the Forsythe programming language \cite{Reynolds:96} and to the channel
types of \cite{Castagna:Nicola:Varacca:08, Pierce:Sangiorgi:96}.
The generalization to a reference type constructor taking two arguments
described in Section~\ref{sec:gen-obj} is quite natural, and also appeared in
Pottier's thesis \cite{Pottier::98}, where it facilitated type inference by
allowing meets and joins to distribute over reference types.
This idea has recently been applied by Craciun \ETAL for inferring
variant parametric types in Java \cite{CraciunCHQ::09}.

The generalized object types we introduce in Section~\ref{sec:gen-obj} directly
correspond to the \emph{split types} of Bugliesi and
Peric{\'a}s-Geertsen \cite{Bugliesi:Pericas-Geertsen:02}.
Split types are also motivated by type inference,  
since they guarantee the existence of more precise upper and lower bounds.
In particular, Bugliesi and Peric{\'a}s-Geertsen show that split types are
strictly more expressive than first-order object types with variance
annotations \cite[Example~4.3]{Bugliesi:Pericas-Geertsen:02}.
They establish the soundness of a type system with split types by subject
reduction, with respect to a functional semantics of the object calculus.

\subsection{Functional Object Calculus}
Our initial experiments 
on the current topic
were done in the context of the
functional object calculus \cite{Hritcu:07}.
Even though in the functional setting the semantic model is much simpler,
both models satisfy the same semantic
typing lemmas. Even more, the syntactic type system we considered  for the
functional calculus is exactly the same as the one in this paper, so all the
results in Section~\ref{sec:soundness} directly apply to the functional object
calculus:
well-typed terms do not get stuck, no matter whether they are evaluated in a
functional or an imperative way.  
It would not be possible to directly 
prove such a result using subject reduction, since for subject reduction the
syntactic typing judgment for the imperative calculus would also depend on a heap typing,
and thus be different from the judgment for the functional calculus.
However, since we are not using subject reduction, we do not need to 
type-check partially evaluated terms that contain heap locations.

\section{Conclusion}
\label{sec:conclusion}

\noindent We have presented a step-indexed semantics for Abadi and
Cardelli's imperative object calculus, and used it to prove the safety
of a type system with object types, recursive and second-order types,
as well as subtyping.  We showed how this semantics can be extended to
self types and typing lemmas with structural assumptions; and
generalized in a way that eliminates the need for variance annotations
and at the same time simplifies the subtyping rules for objects.

The step-indexing technique is however not limited to type safety proofs, and
has already been employed for more general reasoning about programs.
Based on previous work by Appel and McAllester \cite{Appel:McAllester:01}, 
Ahmed built a step-indexed partial equivalence relation model for the lambda
calculus with recursive and impredicative quantified types, and showed that
her relational interpretation of types is sound for proving  contextual
equivalences \cite{Ahmed:06}. 
Recently, this was 
extended 
significantly to reason about program equivalence in the presence of
 general references  \cite{Ahmed:Dreyer:Rossberg:09}. 
Benton also used step-indexing as a technical device, together with a notion of 
orthogonality relating expressions to contexts, to show the soundness of a 
compositional program logic for a   simple stack-based abstract machine 
\cite{Benton:05}. He also employed step-indexing in a Floyd-Hoare-style 
framework based on relational parametricity for the specification and 
verification of machine code programs \cite{Benton:06}.

We hope that our work paves the way for similarly compelling, semantic
investigations of program logics for the imperative object calculus:
 using a step-indexed model it should be possible to prove the
soundness of more expressive program logics for this calculus.

\section*{Acknowledgements}
We express our gratitude to the anonymous reviewers  for their detailed and
constructive comments on the preliminary versions of this article.
We also thank Andreas Rossberg for pointing us to the work of John C.\ Reynolds
on Forsythe.
C\u{a}t\u{a}lin Hri\c{t}cu is supported by a fellowship from Microsoft Research
and the International Max Planck Research School for Computer Science.


\bibliographystyle{plain}
\bibliography{step-indexing}

\appendix
\section{}
\label{sec:appendix}

\subsection{Auxiliary Propositions}

\begin{prop}[Preorder]
\label{app:prop:P0}
The state extension relation, $\sqsubseteq$, is reflexive and transitive.
\qed
\end{prop}

\begin{prop}[Information-forgetting extension]
\label{app:prop:P1}
If $j \leq k$ then $\textend{k}{\Psi}{j}{\tapprox{\Psi}{j}}$.
\qed
\end{prop}

\begin{prop}[Relation between $\mtuple{k,\Psi, v} \in \tau$ 
and $v :_{k,\Psi}\tau$] Let $v$ be a closed value. 
\label{app:prop:P3}
\begin{enumerate}[\em(1)]
  \item If  $\mtuple{k,\Psi, v} \in \tau$ 
then $v :_{k,\Psi}\tau$.   
\item If  $v :_{k,\Psi}\tau$, $k>0$, and there
exists some $\h$ such that $\h:_k \Psi$, then  $\mtuple{k,\Psi, v} \in
\tau$.
\qed
\end{enumerate}
\end{prop}

\subsection{Typing Lemmas for Object Types}
\label{app:typing-lemmas-objects}

\begin{lem}[{\sc SemObj}: Object construction]
\label{app:lemma:SemObj}
For all object types  $\alpha = \tobjtv{\tau}{d}$, if
for all $d \in D$ we have $\ttype{\mext{\Sigma}{x_d}{\alpha}}{b_d}{\tau_d}$,
then $\ttype{\Sigma}{\tobj{d}}{\alpha}$. 
\end{lem}
\proof
Let $\alpha = \tobjtv{\tau}{d}$ and assume that 
$\forall d \in D.\;\ttype{\mext{\Sigma}{x_d}{\alpha}}{b_d}{\tau_d}$. 
We must show that 
$\ttype{\Sigma}{\tobj{d}}{\alpha}$.
Thus, let $k \geq 0$, $\sigma$ be a value environment and $\Psi$ be a heap typing
such that $\sigma :_{k, \Psi} \Sigma$.
By the definition of the semantic typing judgement
(Definition~\ref{def:SemanticTypingJudgement}) we
need to show that 
$\msub{\tobj{d}}{\sigma} :_{k,\Psi} \alpha$. Equivalently (after suitable
$\alpha$-renaming), we show that
\begin{equation*}
\tobjx{\msub{b_d}{\sigma}}{d} :_{k,\Psi} \alpha
\end{equation*}
   
Suppose $j < k, h, h'$ and $b'$ are such that the following three conditions are
fulfilled: 
\begin{gather}
\label{app:sem-obj:A1}
h :_k \Psi\quad \wedge\quad 
\tjredc{h}{\tobjx{\msub{b_d}{\sigma}}{d}}{j}{h'}{b'}\quad \wedge\quad 
\mirred{\xcfg{h'}{b'}} 
\end{gather} 
By the operational semantics  \textsc{Red-Obj} is the only rule that applies,
which means that necessarily $j=1$ and for some distinct $l_d \not \in \mdom{h}$
we have $b' = \tvobj{d}$ and 
\begin{align}
\label{app:sem-obj:A4}
h' & = \mextx{h}{l_d}{\tlam{x_d}{\msub{b_d}{\sigma}}}{d \in D}
\end{align}
We choose
\begin{equation}
\label{app:sem-obj:A6}
\Psi' = \tapprox{\mextx{\Psi}{l_d}{(\tarr{\alpha}{\tau_d})}{d \in D}}{k-1}
\end{equation}
and show that
\begin{gather}
\label{app:sem-obj:S1}
\textend{k}{\Psi}{k-1}{\Psi'}\quad\wedge\quad 
h' :_{k-1} \Psi'\quad\wedge\quad
\mtuple{k-1, \Psi', b'} \in \alpha
\end{gather}

That the first conjunct of \eqref{app:sem-obj:S1} holds is immediate from the
construction of $\Psi'$ \eqref{app:sem-obj:A6}. 

In order to show the second conjunct, by
Definition~\ref{def:WellTypedHeap} (Well-typed heap) we first need
to show that $\mdom{\Psi'} \subseteq \mdom{h'}$. From the first conjunct of 
\eqref{app:sem-obj:A1} and Definition~\ref{def:WellTypedHeap} it is clear that $\mdom{\Psi} \subseteq
\mdom{h}$. Thus from the shape of $h'$ \eqref{app:sem-obj:A4} and the definition
of $\Psi'$ \eqref{app:sem-obj:A6} we obtain the required inclusion.

Next, let $i < k-1$ and $l \in
\mdom{\Psi'}$. 
To establish $h' :_{k-1} \Psi'$ in \eqref{app:sem-obj:S1} we now need to show
that 
$\mtuple{i, \tapprox{\Psi'}{i}, h'(l)} \in \Psi'(l)$. 
We distinguish two cases:

\begin{enumerate}[$\bullet$]
\item Case $l = l_d$ for some $d \in D$. 
From \eqref{app:sem-obj:A4} and
\eqref{app:sem-obj:A6} respectively we get that
\begin{gather*}
h'(l) = \tlam{x_d}{\msub{b_d}{\sigma}}
\quad\wedge\quad
\Psi'(l) = \tapprox{\tarr{\alpha}{\tau_d}}{k-1}
\end{gather*}
Thus we need to show that
\begin{align}
\label{app:sem-obj:B}
\mtuple{i, \tapprox{\Psi'}{i},
\tlam{x_d}{\msub{b_d}{\sigma}}} \in \tapprox{\tarr{\alpha}{\tau_d}}{k-1}
\end{align}
By {\sc SemLam} in Figure~\ref{fig:typing-lemmas:procedure-types}
(Lemma~\ref{lemma:all-proc-type-lemmas}) and the assumption
$\ttype{\mext{\Sigma}{x_d}{\alpha}}{b_d}{\tau_d}$ we already know that
$\ttype{\Sigma}{\tlam{x_d}{b_d}}{\tarr{\alpha}{\tau_d}}$ for all $d\in D$. 
From this and $\sigma :_{k, \Psi} \Sigma$ by
Definition~\ref{def:SemanticTypingJudgement} (Semantic typing judgement) we
obtain
\begin{equation}
\label{app:sem-obj:CH}
\forall d\in D.\ \tlam{x_d}{\msub{b_d}{\sigma}}\; :_{k,\Psi}\;{\tarr{\alpha}{\tau_d}}
\end{equation}
Since $k>1$ and from \eqref{app:sem-obj:A1} $h :_k \Psi$,
Proposition~\ref{app:prop:P3} shows that
 \eqref{app:sem-obj:CH} implies 
\begin{align}
\label{app:sem-obj:B0'}
\forall d\in D.\ \mtuple{k, {\Psi},
\tlam{x_d}{\msub{b_d}{\sigma}}} \in \tarr{\alpha}{\tau_d} 
\end{align}
By Proposition \ref{app:prop:P1} we get that
$\textend{k-1}{\Psi'}{i}{\tapprox{\Psi'}{i}}$, which together with the first
conjunct of \eqref{app:sem-obj:S1} and the transitivity of $\sqsubseteq$ yields
$\textend{k}{\Psi}{i}{\tapprox{\Psi'}{i}}$. 
Since each $\tarr{\alpha}{\tau_d}$ is closed under state extension, the latter
property and 
 \eqref{app:sem-obj:B0'} imply the required \eqref{app:sem-obj:B}.

\item Case $l \in \mdom{\Psi}$. 
From \eqref{app:sem-obj:A4} and
\eqref{app:sem-obj:A6} respectively we get
that $h'(l) = h(l)$ and $\Psi'(l) = \tapprox{\Psi(l)}{k-1}$, so we actually need
to show
that $\mtuple{i, \tapprox{\Psi'}{i}, h(l)} \in \tapprox{\Psi(l)}{k-1}$. From
$h :_k \Psi$ \eqref{app:sem-obj:A1} by Definition~\ref{def:WellTypedHeap} we get
that $\mtuple{k-1, \tapprox{\Psi}{k-1}, h(l)} \in \Psi(l)$. Since $\Psi(l)$ is
closed under state extension and
$\textend{k-1}{\tapprox{\Psi}{k-1}}{i}{\tapprox{\Psi'}{i}}$,
we obtain $\mtuple{i, \tapprox{\Psi'}{i}, h(l)} \in \Psi(l)$. 
\end{enumerate}

Finally, we need to show the third conjunct of \eqref{app:sem-obj:S1}, \IE
$\mtuple{k-1, \Psi', \tvobj{d}}
\in \alpha$. To this end, we prove the following more general claim:

\noindent \emph{Claim:}
For all $j_0 \geq 0$, for all $\Psi_0$ and for all $\tvobjx{\loc'_d}{d}$
\begin{multline}
\label{app:sem-obj:U}
\textend{k-1}{\Psi'}{j_0}{\Psi_0} 
\mand  (\forall d\in D.\;
\tapprox{\Psi_0}{j_0}(\loc'_d) = \tapprox{\Psi'}{j_0}({\loc_d}))\\
\mimpl 
\ \mtuple{j_0,\tapprox{\Psi_0}{j_0}, \tvobjx{\loc'_d}{d}}\in\alpha
\end{multline}
From this and $\tapprox{\Psi'}{k-1} = \Psi'$, \eqref{app:sem-obj:S1} follows by taking $j_0 =
k-1$, $\Psi_0 = \Psi'$, and $\loc'_d=\loc_d$ for all
$d\in D$, and by observing that $\sqsubseteq$  is
reflexive (Proposition~\ref{app:prop:P0}).

The claim is proved by complete induction on $j_0$. So assume $j_0 \geq 0$ and
$\Psi_0$ are such that 
\begin{gather}
\label{app:sem-obj:U0H}
\textend{k-1}{\Psi'}{j_0}{\Psi_0}
\end{gather} 
Moreover, for
all $d\in D$  let $\loc_d'\in\mdom{\Psi_0}$ such that
\begin{align}
\label{app:sem-obj:U1H}
\forall d\in D.\ \tapprox{\Psi_0}{j_0}(\loc'_d) = \tapprox{\Psi'}{j_0}(\loc_d)
\end{align}
We show that $\mtuple{j_0,\tapprox{\Psi_0}{j_0},\tvobjx{\loc'_d}{d}}\in\alpha$,
by checking that all the conditions obtained by unfolding the definition of
$\alpha = \tobjtv{\tau}{d}$ hold.
Choosing $\alpha' = \tapprox{\alpha}{j_0}$ yields \textsc{(Obj-1)}:
\begin{gather}
\label{app:sem-obj:U2}
\exists \alpha'.\alpha'\in\msn{Type}\ \wedge\ \tapprox{\alpha'}{j_0}\subseteq\tapprox{\alpha}{j_0}
\end{gather}
 
Next, by the construction of $\Psi'$ in \eqref{app:sem-obj:A6}, together with
\eqref{app:sem-obj:U0H}, \eqref{app:sem-obj:U1H}, and the non-expansiveness of
procedure types, it follows that for all $d \in D$
\begin{gather}
\label{app:sem-obj:U4-5} 
 \tapprox{\Psi_0}{j_0}(l_d') = \tapprox{\Psi'}{j_0}(l_d) =
 \tapprox{\tarr{\alpha}{\tau_d}}{j_0} = 
 \tapprox{\tarr{\tapprox{\alpha}{j_0}}{\tau_d}}{j_0} = 
 \tapprox{\tarr{\alpha'}{\tau_d}}{j_0}
\end{gather}
By the definition of reference types (Definition~\ref{def:ReferenceTypes}) this
implies that
\begin{gather}
\label{app:sem-obj:Star}
\forall d \in D.\; \mtuple{j_0, \tapprox{\Psi_0}{j_0}, l'_d} \in
\trefv{\tinv}{(\tarr{\alpha'}{\tau_d})}
\end{gather}
By the lemma for subtyping variance annotations ({\sc SemSubVarRef} in
Figure~\ref{fig:subtyping-reference-types}) we then  obtain property
\textsc{(Obj-2)}:
\begin{gather}
\label{app:sem-obj:StarPrime}
\forall d \in D.\; \mtuple{j_0, \tapprox{\Psi_0}{j_0}, l'_d} \in
\trefv{\tvar_d}{(\tarr{\alpha'}{\tau_d})}
\end{gather}

Finally, we must prove \textsc{(Obj-3)}, \IE that for all $j<j_0$, $\Psi_1$ and
$\tvobjx{\loc''_d}{d}$
\begin{multline}
\label{app:sem-obj:U3}  
\textend{j_0}{\Psi_0}{j}{\Psi_1}\
 \wedge\; 
(\forall d\in D.\ \tapprox{\Psi_1}{j}(\loc''_d) = \tapprox{\Psi_0}{j}(\loc_d'))\\
 \Rightarrow\ 
\mtuple{j,\tapprox{\Psi_1}{j},\tvobjx{\loc''_d}{d}}\in\alpha
\end{multline}
Note that this last condition holds trivially in the base case
of the induction, when $j_0=0$. So assume $j<j_0$ and
$\Psi_1$ and $\loc''_d$ are such that $\textend{j_0}{\Psi_0}{j}{\Psi_1}$ and 
$\tapprox{\Psi_1}{j}(\loc''_d) = \tapprox{\Psi_0}{j}(\loc_d')$ for all $d\in 
D$. Now $j<j_0$ and assumption \eqref{app:sem-obj:U1H} yield that for all $d\in
D$
\begin{align*}
\tapprox{\Psi_1}{j}(\loc''_d) = 
\tapprox{\Psi_0}{j}(\loc'_d) 
= \tapprox{\tapprox{\Psi_0}{j_0}(\loc'_d)}{j} =
\tapprox{\tapprox{\Psi'}{j_0}(\loc_d)}{j} = \tapprox{\Psi'}{j}(\loc_d)
\end{align*}
Moreover, from $({k-1},{\Psi'})\sqsubseteq (j_0,\Psi_0)$
\eqref{app:sem-obj:U0H} and $(j_0,\Psi_0)\sqsubseteq({j},{\Psi_1})$, by the
transitivity of $\sqsubseteq$ we have that $({k-1},{\Psi'}) \sqsubseteq
({j},{\Psi_1})$.
Since $j<j_0$, the induction hypothesis of the claim gives
\begin{align*}
\mtuple{j,\tapprox{\Psi_1}{j},\tvobjx{\loc''_d}{d}}\in\alpha
\end{align*}
and we have established \eqref{app:sem-obj:U3}. 

By Definition~\ref{def:ObjectTypes} applied to the object type $\alpha =
\tobjtv{\tau}{d}$ the properties $D\subseteq D$, \eqref{app:sem-obj:U2},
\eqref{app:sem-obj:StarPrime}, and \eqref{app:sem-obj:U3} establish that indeed
$\mtuple{j_0,\tapprox{\Psi_0}{j_0},\tvobjx{\loc'_d}{d}}\in\alpha$. This finishes the inductive
proof of claim \eqref{app:sem-obj:U}, and the proof of the lemma.
\qed

\begin{lem}[{\sc SemInv}: Method invocation]
\label{app:lemma:SemInv}
For all object types  $\alpha = \tobjtv{\tau}{d}$ and for all $e \in D$, if 
$\ttype{\Sigma}{a}{\alpha}$ and $\tvar_e \in \{\tcov, \tinv\}$, then 
$\ttype{\Sigma}{a.\m_e}{\tau_e}$.
\end{lem}
\proof
Let $\alpha = \tobjtv\tau d$. We assume that $e\in D$, $\tvar_e \in \{\tcov,
\tinv\}$, and $\ttype{\Sigma}{a}{\alpha}$ and show that
$\ttype{\Sigma}{a.\m_e}{\tau_e}$. 
To this end, let $k\geq 0$, $\sigma$ and $\Psi$ such that $\sigma :_{k, \Psi}
\Sigma$. From $\ttype{\Sigma}{a}{\alpha}$ by
Definition~\ref{def:SemanticTypingJudgement} we get that
\begin{equation}
\label{app:sem-inv:H3'}
\msub{a}{\sigma} :_{k, \Psi} \alpha 
\end{equation}
We need to show that
${\msub{a}{\sigma}}.\m_e :_{k, \Psi} \tau_e$. 
Thus, let $j < k$, and consider heaps $\h$ and $\h'$ and a term $b'$ such that the following
three conditions are fulfilled:
\begin{gather}
\label{app:sem-inv:A1} 
\h :_k \Psi
\quad\wedge\quad
\tjredc{\h}{{\msub{a}{\sigma}}.\m_e}{j}{\h'}{b'}
\quad\wedge\quad
\mirred{\xcfg{\h'}{b'}}
\end{gather}
From the second and third conjunct of \eqref{app:sem-inv:A1} by the
operational semantics we have that for some $i\leq j$, $\h^*$ and $b^*$
\begin{gather}
\label{app:sem-inv:A4}
\tjred{\xcfg{\h}{\msub{a}{\sigma}}}{i}{\mirred{\xcfg{h^*}{b^*}}}
\quad\wedge\quad
\tjredc{\h^*}{b^*.\m}{j-i}{\h'}{b'}
\end{gather}
From the first conjunct together with \eqref{app:sem-inv:H3'} and the first
conjunct of \eqref{app:sem-inv:A1}, by Definition~\ref{def:ClosedTermKPsiType} it
follows that there exists a heap typing $\Psi^*$ such that
\begin{gather}
\label{app:sem-inv:B1}
\textend{k}{\Psi}{k-i}{\Psi^*}
\quad\wedge\quad
\h^* :_{k-i} \Psi^*
\quad\wedge\quad
\mtuple{k-i,\Psi^*,b^*}\in\alpha = \tobjtv\tau d
\end{gather}
By the definition of object types, the latter shows that there exists
$C$ and $\alpha'$ such that $b^*= \tvobjx{l_c}{c}$,
$D\subseteq C$ and \textsc{(Obj-1)} and \textsc{(Obj-2)} hold:
\begin{gather}
\label{app:sem-inv:B5}
\alpha'\in\msn{Type}\quad\wedge\quad
\tapprox{\alpha'}{k-i}\subseteq\tapprox{\alpha}{k-i}\\
\label{app:sem-inv:B7}
\forall d \in D.\; \mtuple{k-i, \Psi^*, l_d} \in
\trefv{\tvar_d}{(\tarr{\alpha'}{\tau_d})}
\\
\intertext{as well as \textsc{(Obj-3)}: for all $j_0<k-i$, all $\Psi'$ and all
$\tvobjx{l_c'}{c}$,}
\label{app:sem-inv:B6}
(\textend{k-i}{\Psi^*}{j_0}{\Psi'} 
  \mand \forall c \in C. \; \tapprox{\Psi'}{j_0}(l_c') =
\tapprox{\Psi^*}{j_0}(l_c)) \mimpl \mtuple{j_0, \tapprox{\Psi'}{j_0},
\tvobjx{l_c'}{c}} \in \alpha'
\end{gather}
From $e \in D$ and $\tvar_e \in \{\tcov, \tinv\}$ using \eqref{app:sem-inv:B7} we deduce
that
$\tapprox{\Psi^*}{k-i}(l_e) \subseteq \tapprox{\alpha' \to \tau_e}{k-i}$.
So by expanding the definition of $\h^* :_{k-i} \Psi^*$ from 
\eqref{app:sem-inv:B1} for $k-i-1<k-i$ we have
\begin{gather}
\label{app:sem-inv:B10}
\mtuple{k-i-1,\tapprox{\Psi^*}{k-i-1},\h^*(l_e)} \in \tapprox{\alpha' \to \tau_e}{k-i}
\end{gather}
By the definition of the procedure type $\alpha'\to\tau_e$ this means in
particular that $\h^*(l_e)$ must be an abstraction, \IE for some $x$ and $a'$,
$\h^*(l_e) = \tlam x {a'}$.
Thus, since $\tvobjx{l_c}{c} \in \msn{CVal}$ and $e \in D \subseteq C$, by
\eqref{app:sem-inv:A4}, {\sc Red-Ctx}, {\sc Red-Inv}, {\sc Red-Beta} and the operational
semantics, we obtain a reduction sequence of the form 
\begin{gather}
\label{app:sem-inv:A4-1}
\begin{aligned}
\xcfg{\h}{{\msub{a}{\sigma}}.\m_e} &\to^{i}\xcfg{\h^*}{\tvobjx{l_c}{c}.\m_e}\\
&\to \xcfg{h^*}{\tapp{(\tlam x{a'})}{\tvobjx{l_c}{c}}}\\
&\to \xcfg{h^*}{\msub{a'}{\mmap{x}{\tvobjx{l_c}{c}}}}\\
&\to^{j-i-2} \xcfg{h'}{b'}
\end{aligned}
\end{gather}
Since $k-i-2<k-i$ by Proposition~\ref{app:prop:P1} we have that
\begin{gather}
\label{app:sem-inv:B12-0}
\textend{k-i}{\Psi^*}{k-i-2}{\tapprox{\Psi^*}{k-i-2}}
\end{gather}
We can now use the property \textsc{(Obj-3)} of the object type $\alpha$:
we instantiate \eqref{app:sem-inv:B6} with $l'_c
= l_c$, $j_0=k-i-2$, and $\Psi' = \tapprox{\Psi^*}{k-i-2}$ to obtain
\begin{gather}
\label{app:sem-inv:B12}
\mtuple{k-i-2,\tapprox{\Psi^*}{k-i-2}, \tvobjx{l_c}{c}} \in\alpha'
\end{gather}
From this using \eqref{app:sem-inv:B10} and from the definition of procedure types,
it follows that
\begin{gather}
\label{app:sem-inv:B13'}
\msub{a'}{\mmap{x}{\tvobjx{l_c}{c}}} :_{k-i-2,\tapprox{\Psi^*}{k-i-2}}\tau_e
\end{gather}
On the other hand, the second conjunct of \eqref{app:sem-inv:B1} implies 
\begin{gather}
\label{app:sem-inv:B14}
h^* :_{k-i-2}\tapprox{\Psi^*}{k-i-2}
\end{gather}
by Definition \ref{def:WellTypedHeap}, Proposition~\ref{app:prop:P1} and the
closure of types under state extension. Moreover, by \eqref{app:sem-inv:A4-1},
$\tjredc{\h^*}{\msub{a'}{\mmap{x}{\tvobjx{l_c}{c}}}} {j-i-2}{\h'}{b'} $,
which by  \eqref{app:sem-inv:A1} is irreducible. This, combined with
\eqref{app:sem-inv:B14} and \eqref{app:sem-inv:B13'}, by
Definition~\ref{def:ClosedTermKPsiType}, means that there exists  $\Psi''$
such that
\begin{gather}
\label{app:sem-inv:C1}
\textend{k-i-2}{\tapprox{\Psi^*}{k-i-2}}{k-j}{\Psi''}
\quad\wedge\quad
h' :_{k-j}\Psi''
\quad\wedge\quad
\mtuple{k-j,\Psi'',b'} \in\tau_e
\end{gather}
From the first conjunct  above, the first conjunct in \eqref{app:sem-inv:B1},
and \eqref{app:sem-inv:B12-0}, using the transitivity of state extension we
obtain
\begin{gather}
\label{app:sem-inv:C4}
\textend{k}{\Psi}{k-j}{\Psi''}
\end{gather}
From \eqref{app:sem-inv:A1},  
\eqref{app:sem-inv:C4}, and the second and third conjuncts of 
\eqref{app:sem-inv:C1}, by Definition~\ref{def:ClosedTermKPsiType} we can conclude that 
${\msub{a}{\sigma}}.\m_e :_{k, \Psi} \tau_e$ holds. This is what we needed to
show.
\qed

\begin{lem}[{\sc SemUpd}: Method update]
\label{app:lemma:SemUpd}
For all object types  $\alpha = \tobjtv{\tau}{d}$ and
for all $e \in D$, if $\ttype{\Sigma}{a}{\alpha}$ and 
$\ttype{\mext{\Sigma}{x}{\alpha}}{b}{\tau_e}$ and $\tvar_e \in \{\tcon,
\tinv\}$, then $\ttype{\Sigma}{\tupd{a}{\m_e}{\tself{x}{b}}}{\alpha}$.
\end{lem}
\proof[Proof sketch]
The proof is similar to that of Lemma~\ref{app:lemma:SemInv} (Method
invocation). The existence of some $\Psi^*$ such that 
\begin{align*}
\tjredc{h}{\tupd{\msub{a}{\sigma}}{\m_e}{\tself{x}{\msub{b}{\sigma}}}&}{j}{h^*}
{\tupd{\tvobj{e}}{\m_e}{\tself{x}{\msub{b}{\sigma}}}}
\end{align*}
with $\textend{k}{\Psi}{k-j}{\Psi^*}$, $h^* :_{k-j} \Psi^*$
and 
$\mtuple{k-j, \Psi^*, \tvobj{e}} \in \alpha$ follows from the existence of a
corresponding reduction sequence 
$\tjredc{h}{\msub{a}{\sigma}}{j}{h^*}{{\tvobj{e}}}$. Since the only reduction
from  $\xcfg{h^*}{\tupd{\tvobj{e}}{\m_e}{\tself{x}{\msub{b}{\sigma}}}}$ is by
\textsc{(Red-Upd)} and results in the configuration  
$\xcfg{h'}{\tvobj{e}}$ where 
\begin{align}
\label{app:sem-upd:B3}
h'\ & =\ \mextx{h^*}{l_e}{\tlam{x}{\msub{b}{\sigma}}}{}
\end{align}
the proof of the lemma is essentially a matter of showing that 
$h' :_{k-j-1} \tapprox{\Psi^*}{k-j-1}$. 

First, note that 
$\mdom{\Psi^*} \subseteq \mdom{h'}
= \mdom{h^*}$ holds, by $h^* :_{k-j} \Psi^*$. 
Next, let $i<k-j-1$, and let $l \in \mdom{\Psi^*}$. It remains to  show that
\begin{align}
\label{app:sem-upd:to-show}
\mtuple{i, \tapprox{\Psi^*}{i}, h'(l)} \in \tapprox{\Psi^*(l)}{k-j-1}
\end{align}
Note that by the definition of $\mtuple{k-j, \Psi^*, \tvobj{e}} \in \alpha$
(Definition~\ref{def:ObjectTypes}) it follows that there exists $\alpha'\in \msn{Type}$ such
that $\tapprox{\alpha'}{k-j} \subseteq\tapprox{\alpha}{k-j}$, that $D \subseteq E$, and that 
\begin{gather}
\label{app:sem-upd:D5}
\forall d \in D.\; \mtuple{k-j, \Psi^*,l_d} \in
\trefv{\tvar_d}{(\tarr{\alpha'}{\tau_d})}
\end{gather}
We now prove \eqref{app:sem-upd:to-show} by a case distinction on the location
$l$: 
\begin{enumerate}[$\bullet$]
  \item Case $l=l_e$. From \eqref{app:sem-upd:B3} we have that $h'(l_e) =
\tlam{x}{\msub{b}{\sigma}}$. Since $e \in D\subseteq E$  and $\tvar_e \in
\{\tcon,\tinv\}$ by assumption, \eqref{app:sem-upd:D5} yields 
$\tapprox{\alpha' \to \tau_e}{k-j} \subseteq \tapprox{\Psi^*}{k-j}(l_e)$. 
Since $\tapprox{\alpha'}{k-j} \subseteq\tapprox{\alpha}{k-j}$, the subtyping lemma 
\textsc{(SemSubProc)} and the non-expansiveness of procedure types yield
$\tapprox{\alpha \to \tau_e}{k-j} \subseteq \tapprox{\Psi^*}{k-j}(l_e)$. The 
 monotonicity of semantic approximation therefore entails 
\begin{equation}
\label{app:sem-upd:D5''} 
\tapprox{\alpha \to \tau_e}{k-j-1} \subseteq \tapprox{\Psi^*}{k-j-1}(l_e) 
\end{equation} 
Additionally, the assumption $\ttype{\mext{\Sigma}{x}{\alpha}}{b}{\tau_e}$ gives
 $\tlam{x}{\msub{b}{\sigma}} :_{k-j,\Psi^*}{\tarr{\alpha}{\tau_e}}$.  Since
$0\leq i<k-j$ and $h^*:_{k-j}\Psi^*$, Proposition~\ref{app:prop:P3} yields
$\mtuple{k-j, {\Psi^*}, \tlam{x}{\msub{b}{\sigma}}} \in \tarr{\alpha}{\tau_e}$.
By the closure under state extension, this implies 
$\mtuple{i, \tapprox{\Psi^*}{i}, \tlam{x}{\msub{b}{\sigma}}} \in
\tarr{\alpha}{\tau_e}$, from which \eqref{app:sem-upd:to-show} follows by
\eqref{app:sem-upd:D5''}.

  \item Case $l\neq l_e$. This case is easier since the value in
the heap does not change for this location, \IE $h'(l)  = h^*(l)$, so the result
follows from the closure under state extension of $\Psi^*(l)$. \qed    
\end{enumerate}

\begin{lem}[{\sc SemClone}: Object cloning]
\label{app:lemma:SemClone}
For all object types  $\alpha = \tobjtv{\tau}{d}$, if
$\ttype{\Sigma}{a}{\alpha}$ then $\ttype{\Sigma}{\tclone{a}}{\alpha}$.
\end{lem}
\proof[Proof sketch]
The proof is  similar to that of Lemma~\ref{app:lemma:SemUpd} (Method update).
Assuming $\sigma :_{k, \Psi} \Sigma$ and $h:_k\Psi$ such that
$\xcfg{h}{\tclone{\msub{a}{\sigma}}}$ halts in fewer than $k$ steps, by appealing
to the operational semantics and the assumption that $\ttype{\Sigma}{a}{\alpha}$
one obtains the existence of some $\Psi^*$ such that
\begin{align*}
\tjredc{h}{\tclone{\msub{a}{\sigma}}&}{j}{h^*}{\tclone{\tvobj{e}}}
\to \xcfg{h'}{b'}
\end{align*}
with $\textend{k}{\Psi}{k-j}{\Psi^*}$, $h^* :_{k-j} \Psi^*$
and 
$\mtuple{k-j, \Psi^*, \tvobj{e}} \in \alpha$. 
 Since the only reduction
from  $\xcfg{h^*}{\tclone{\tvobj{e}}}$ is by
\textsc{(Red-Clone)} it is clear that for some (distinct) $l'_e\notin\mdom{h^*}$
we have 
\begin{align}
\label{app:sem-clone:B3}
h'\ =\ \mextx{h^*}{l_e'}{h^*(l_e)}{e \in E}
\quad\wedge\quad
b'\ =\ \tvobjx{l_e'}{e}
\end{align}
If we set $\Psi' = \tapprox{\mextx{\Psi^*}{l_e'}{\Psi^*(l_e)}{e \in E}}{k-j-1}$
then it follows that $\textend{k}{\Psi}{k-j-1}{\Psi'}$, and to establish the
lemma it suffices to prove 
\begin{gather}
\label{app:sem-clone:S1}
h' :_{k-j-1} \Psi'
\quad\wedge\quad
\mtuple{k-j-1, \Psi', \tvobjx{l_e'}{e}} \in \alpha
\end{gather}

Observing that $\mdom{\Psi'} \subseteq \mdom{h'}$ is satisfied, the first conjunct is
proved by showing that $\mtuple{i,
\tapprox{\Psi'}{i}, h'(l)} \in \tapprox{\Psi'(l)}{k-j-1}$ holds for all
$i<k-j-1$ and all $l\in\mdom{\Psi'}$. This is done by 
a case distinction on whether $l\in\mdom{\Psi^*}$ or $l=l_e'$ for some $e$.
In both cases,  the relation follows from $h^* :_{k-j} \Psi^*$ and the closure
under state extension of types. 

As for the second conjunct of \eqref{app:sem-clone:S1}, we note that  $\Psi'$ is
constructed from  $\Psi^*$ such that $\tapprox{\Psi'(l_e')}{k-j-1} =
\tapprox{\Psi^*(l_e)}{k-j-1}$ holds for all $e\in E$. 
Therefore, by unfolding the definition of the object types for 
$\mtuple{k-j, \Psi^*,\tvobj{e}} \in \alpha$, condition \textsc{(Obj-3)} allows
us to conclude that 
$\mtuple{k-j-1, \tapprox{\Psi'}{k-j-1}, \tvobjx{l_e'}{e}} \in \tapprox{\alpha}{k-j-1}$.
Then the required  $\mtuple{k-j-1, \Psi', \tvobjx{l_e'}{e}} \in
\alpha$
follows from the fact that $\tapprox{\Psi'}{k-j-1} = \Psi'$ holds by definition
of $\Psi'$, and that  $\tapprox{\alpha}{k-j-1}\subseteq\alpha$.
%
 \qed

\subsection{Subtyping Lemmas for Object Types}

\begin{lem}[{\sc SemSubObj}: Subtyping object types]
\label{app:lemma:SemSubObj}
$E \subseteq D$ and for all $e \in E$ if $\tvar_e \in \{\tcov,\tinv\}$ then
$\alpha_e \tsub \beta_e$ and if $\tvar_e \in \{\tcon,\tinv\}$ then $\beta_e
\tsub \alpha_e$ imply that $\tobjtv{\alpha}{d} \tsub \tobjtv{\beta}{e}$.
\end{lem}
\proof
We denote $\alpha \meqnot \tobjtv{\alpha}{d}$ and $\beta \meqnot 
\tobjtv{\beta}{e}$. We assume that $E \subseteq D$ and
\begin{gather}
\label{app:sem-sub-obj:G2}
\forall e \in E.\ (\tvar_e \in 
\{\tcov,\tinv\} \mimpl \alpha_e \tsub \beta_e)\, \mand\,  (\tvar_e \in
\{\tcon,\tinv\} \mimpl \beta_e \tsub \alpha_e), 
\end{gather}
and prove that for all heap typings 
$\Psi$, for all values $v$ and all $k \geq 0$, if $\mtuple{k, \Psi, v} \in \alpha$ then 
$\mtuple{k, \Psi, v} \in \beta$, by complete induction on $k$.
The 
induction hypothesis is that for all $j < k$ if $\mtuple{j, \Psi, v} \in 
\alpha$ then $\mtuple{j, \Psi, v} \in \beta$, or equivalently
$\tapprox{\alpha}{k} \tsub \tapprox{\beta}{k}$.

If we assume that $\mtuple{k, \Psi, v} \in \alpha$, then by the definition of
$\alpha$ (Definition \ref{def:ObjectTypes}) we have that $v = \tvobj{c}$, $D
\subseteq C$ and there exists $\alpha' \in \msn{Type}$ such that
$\tapprox{\alpha'}{k} \tsub \tapprox{\alpha}{k}$ and
\begin{gather}
\label{app:sem-sub-obj:H4}
\forall d \in D.\;\mtuple{k, \Psi, l_d} \in \trefv{\tvar_d}{(\alpha' \to
\alpha_d)}
\end{gather} 
Moreover, condition \textsc{(Obj-3)} holds with respect to $\alpha$, \IE 
for all $j<k$, all $\Psi'$ and all $\tvobjx{l_e'}{e}$ such that 
$\textend{k}{\Psi}{j}{\Psi'}$, 
\begin{gather}
\label{app:sem-sub-obj:H5}
(\forall e \in E. \; \tapprox{\Psi'}{j}(l_e') = \tapprox{\Psi}{j}(l_e))\ 
 \mimpl\  \mtuple{j, \tapprox{\Psi'}{j}, \tvobjx{l_e'}{e}} \in \alpha'
\end{gather}

From $E \subseteq D$ and $D \subseteq C$ by transitivity $E \subseteq C$. 
From $\tapprox{\alpha'}{k} \tsub \tapprox{\alpha}{k}$ and the induction
hypothesis $\tapprox{\alpha}{k} \tsub \tapprox{\beta}{k}$  we get that
$\tapprox{\alpha'}{k} \tsub \tapprox{\beta}{k}$, \IE \textsc{(Obj-1)} holds. 
Moreover, \eqref{app:sem-sub-obj:H5} entails that condition \textsc{(Obj-3)} also 
holds with respect to the object type $\beta$. 
So  in
order to conclude that $\mtuple{k, \Psi, v} \in \beta$, and therefore that $\alpha
\tsub \beta$, all that remains to be proven is condition \textsc{(Obj-2)}:
\begin{equation*}
\forall e \in E.\;\mtuple{k, \Psi, l_e} \in
\trefv{\tvar_e}{(\tarr{\alpha'}{\beta_e})}\\
\end{equation*}

\noindent For this, we choose some $e$ in $E$ and do a case analysis on 
the variance annotation $\tvar_e$:
\begin{enumerate}[$\bullet$]
\item Case $\tvar_e = \tcov$. 
By \eqref{app:sem-sub-obj:G2} we deduce that $\alpha_e \tsub
\beta_e$,
thus by the covariance of the procedure type constructor in its second argument
({\sc SemSubProc} in Figure~\ref{fig:typing-lemmas:procedure-types}) we
get that $\tarr{\alpha'}{\alpha_e} \tsub \tarr{\alpha'}{\beta_e}$.
But since $E \subseteq D$ from \eqref{app:sem-sub-obj:H4} we know
that $\mtuple{k, \Psi, l_e} \in \trefv{\tcov}{(\alpha' \to \alpha_e)}$.
Since the type constructor $\trefv{\tcov}{}$ is covariant ({\sc
SemSubCovRef} in Figure~\ref{fig:subtyping-reference-types}) this implies
$\mtuple{k, \Psi, l_e} \in \trefv{\tcov}{(\tarr{\alpha'}{\beta_e})}$.

\item Case $\tvar_e = \tcon$. 
Similarly to the previous case, \eqref{app:sem-sub-obj:G2} gives
us that $\beta_e \tsub \alpha_e$.
Again by the covariance of $\mlam{\xi}{\tarr{\alpha'}{\xi}}$ ({\sc
SemSubProc}) we infer that $\tarr{\alpha'}{\beta_e} \tsub
\tarr{\alpha'}{\alpha_e}$.
From \eqref{app:sem-sub-obj:H4} $\mtuple{k, \Psi, l_e} \in
\trefv{\tcon}{(\tarr{\alpha'}{\alpha_e})}$, so by the contravariance of
$\trefv{\tcon}$ ({\sc SemSubConRef} in
Figure~\ref{fig:subtyping-reference-types}) we get that
$\mtuple{k, \Psi, l_e} \in \trefv{\tcon}{(\tarr{\alpha'}{\beta_e})}$.

\item $\tvar_e = \tinv$. Now \eqref{app:sem-sub-obj:G2} entails that $\alpha_e
= \beta_e$. Since $\mtuple{k, \Psi, l_e} \in
\trefv{\tinv}{(\tarr{\alpha'}{\alpha_e})}$ by \eqref{app:sem-sub-obj:H4} we immediately
obtain that also $\mtuple{k, \Psi, l_e}
\in\trefv{\tinv}{(\tarr{\alpha'}{\beta_e})}$.
\qed
\end{enumerate}

\begin{lem}[{\sc SemSubObjVar}: Subtyping object variances]
\label{app:lemma:SemSubObjVar}
If for all $d \in D$ we have $\tvar_d = \tinv$ or $\tvar_d = \tvar'_d$ then
$\tobjtv{\tau}{d} \tsub \tobjtvv{\tvar'_d}{\tau}{d}$.
\end{lem}
\proof
The proof proceeds similarly to the proof of Lemma~\ref{app:lemma:SemSubObj}.
Let us denote $\alpha \meqnot \tobjtv{\tau}{d}$ and $\alpha' \meqnot 
\tobjtvv{\tvar'_d}{\tau}{d}$.
We assume that
\begin{equation}
\label{app:sem-sub-obj-var:G}
\forall d \in D.\ \tvar_d = \tinv \mor \tvar_d = \tvar'_d
\end{equation}

Let $\Psi$ and $v$ be arbitrary.
We prove that for all $k \geq 0$, if $\mtuple{k, \Psi, v} \in \alpha$ then
$\mtuple{k, \Psi, v} \in \alpha'$, by complete induction on $k$. The 
induction hypothesis is that for all $j < k$ if $\mtuple{j, \Psi, v} \in 
\alpha$ then $\mtuple{j, \Psi, v} \in \alpha'$, or equivalently
$\tapprox{\alpha}{k} \tsub \tapprox{\alpha'}{k}$.

Assume that $\mtuple{k, \Psi, v} \in \alpha$, then by the definition of $\alpha$
 we have that $v = \tvobj{e}$, $D \subseteq E$, and there exists a type
 ${\alpha''}$ such that $\tapprox{\alpha''}{k} \tsub \tapprox{\alpha}{k}$ and
 \begin{gather}
\label{app:sem-sub-obj-var:H4}
\forall d \in D.\;\mtuple{k, \Psi, l_d} \in \trefv{\tvar_d}{(\alpha'' \to
\tau_d)}
\end{gather}
Moreover, condition \textsc{(Obj-3)} holds.

From $\tapprox{\alpha''}{k} \tsub \tapprox{\alpha}{k}$ and the induction
hypothesis $\tapprox{\alpha}{k} \tsub \tapprox{\alpha'}{k}$ by transitivity we
get that $\tapprox{\alpha''}{k} \tsub \tapprox{\alpha'}{k}$.
This choice of $\alpha''$ also shows that
\textsc{(Obj-3)} holds for $\mtuple{k, \Psi, v}$ with respect to $\alpha'$. 
So in order to show that $\mtuple{k, \Psi, v} \in \alpha'$, and therefore that
$\alpha \tsub \alpha'$, all that remains to be proven is that:
\begin{equation*}
\forall d \in D.\;\mtuple{k, \Psi, l_d} \in
\trefv{\tvar_d'}{(\tarr{\alpha''}{\tau_d})}\\
\end{equation*}

\noindent We show this by case analysis on the disjunction
in \eqref{app:sem-sub-obj-var:G}. Both cases are trivial:
\begin{enumerate}[$\bullet$]
\item Case $\tvar_d = \tinv$. From \eqref{app:sem-sub-obj-var:H4} and $\trefv{\tinv}{(\alpha'' \to
\tau_d)} \tsub \trefv{\tvar_d'}{(\tarr{\alpha''}{\tau_d})}$ ({\sc SemSubVarRef}
in Figure~\ref{fig:subtyping-reference-types}) it is immediate that
$\mtuple{k, \Psi, l_d} \in \trefv{\tvar_d'}{(\tarr{\alpha''}{\tau_d})}$.
\item Case $\tvar_d = \tvar'_d$, then the required statement is the same
as \eqref{app:sem-sub-obj-var:H4}.
\qed
\end{enumerate}

\subsection{Typing Lemmas with Structural Assumptions for Self Types}

\begin{lem}[{\sc SemUpd-Str}: Method update with structural assumptions]
\label{app:lemma:SemUpd-Str}
For all object types  $\alpha = \tobjtv{F}{d}$ and all $\alpha'\in\msn{Type}$ 
such that $\alpha'\tsubself\alpha$, if $e \in D$, $\tvar_e \in \{\tcon, 
\tinv\}$ and $\ttype{\Sigma}{a}{\alpha'}$ and 
$\ttype{\mext{\Sigma}{x}{\alpha'}}{b}{F_e(\alpha')}$, then 
$\ttype{\Sigma}{\tupd{a}{\m_e}{\tself{x}{b}}}{\alpha'}$.
\end{lem}
\proof
The proof is an adaptation of the proof given for Lemma~\ref{app:lemma:SemUpd} 
(Method update)  above. Assume $\alpha = \tobjtv{F}{d}$, $e \in D$, and 
$\tvar_e \in \{\tcon, \tinv\}$, and let $\alpha'\in\msn{Type}$ such that 
$\alpha'\tsubself\alpha$. Moreover assume that $\ttype{\Sigma}{a}{\alpha'}$ and 
$\ttype{\mext{\Sigma}{x}{\alpha'}}{b}{F_e(\alpha')}$ hold. We show that 
$\ttype{\Sigma}{\tupd{a}{\m_e}{\tself{x}{b}}}{\alpha'}$.

Let $k \geq 0$, $\sigma$ be a value environment and $\Psi$ be a heap typing 
such that $\sigma :_{k, \Psi} \Sigma$. We must prove that 
$\msub{\tupd{a}{\m_e}{\tself{x}{b}}}{\sigma} :_{k,\Psi}\alpha'$, 
so let $h$ and $j<k$ be such that 
\begin{gather}
\label{app:sem-upd-str:A1} 
\h :_k \Psi
\quad\wedge\quad
\tjredc{\h}{\tupd{\msub{a}{\sigma}}{\m_e}{\tself{x}{\msub{b}{\sigma}}}}{j}{\h'}{a'}
\quad\wedge\quad
\mirred{\xcfg{\h'}{a'}}
\end{gather}
By the operational semantics, this sequence is induced by 
$\tjredc{\h}{{\msub{a}{\sigma}}}{i}{\h''}{a''}$ for $i\leq j$ and some $h''$ and
$a''$, and by the
assumption $\ttype{\Sigma}{a}{\alpha'}$ there exists some $\Psi''$ such that 
\begin{gather}
\label{app:sem-upd-str:B1}
\textend{k}{\Psi}{k-i}{\Psi''}
\quad\wedge\quad
\h'' :_{k-i} \Psi''
\quad\wedge\quad
\mtuple{k-i,\Psi'',a''}\in\alpha' \subseteq \tobjtv F d
\end{gather}
In particular, $a''$ is of the form $\tvobj{e}$ for some $E\supseteq D$, and by
the operational semantics 
$\tjredc{\h''}{\tupd{a''}{\m_e}{\tself{x}{\msub{b}{\sigma}}}}{}{\h'}{a''}$. In
particular, $a'$ is $a''$ and $h'$ is
$\mext{h''}{l_e}{\tlam{x}{\msub{b}{\sigma}}}$. 
By choosing $\Psi' = \tapprox{\Psi''}{k-j}$, the first and last conjuncts of 
\eqref{app:sem-upd-str:B1} yield
\begin{gather*}
\textend{k}{\Psi}{k-j}{\Psi'}
\quad\wedge\quad
\mtuple{k-j,\Psi',a''}\in\alpha'
\end{gather*}
by Proposition~\ref{app:prop:P1} and transitivity, and by closure under state
extension of $\alpha'$.  To establish the lemma, it remains to show that 
$h':_k\Psi'$. For $l\in\mdom{\Psi'}-\{l_e\}$ this follows from the second
conjunct of \eqref{app:sem-upd-str:B1} by the closure under state extension. The
interesting case is when $l=l_e$ and we must prove $h'(l) =
\tlam{x}{\msub{b}{\sigma}} :_{k-j}\Psi'(l)$. 
Since $\alpha'\tsubself\alpha$ and $\mtuple{k-j,\Psi',a''}\in\alpha'$,
condition \tagref{\textsc{Obj-2-self}} in
Definition~\ref{def:self-type-exposure} (Self type exposure) yields
$\mtuple{k-j,\Psi',l_e} \in \trefv{\tvar_e}{(\alpha'\to F_e(\alpha'))}$.
By assumption, $\tvar_e\in\{\tcon, \tinv\}$ so
$\Psi'(l)\supseteq\tapprox{\alpha'\to F_e(\alpha')}{k-j}$ holds by the
definition of $\trefv{\tvar_e}{}$. Hence it suffices to prove that 
\begin{align*}
\tlam{x}{\msub{b}{\sigma}} :_{k-j,\Psi'}\alpha'\to F_e(\alpha')
\end{align*}
which follows from the assumption 
$\ttype{\mext{\Sigma}{x}{\alpha'}}{b}{F_e(\alpha')}$. 
\qed

\begin{prop}[Self type exposure]
\label{prop:self-type-exposure}
Let $\alpha$ be a self type and suppose $\mtuple{k,\Psi,v}\in\alpha$. Then
there exists $\alpha''\in\msn{Type}$ such that $\alpha''\tsubself\alpha$ and
$\mtuple{k-1,\tapprox{\Psi}{k-1},v}\in\alpha''$. 
\end{prop}
\proof
Suppose $\mtuple{k,\Psi,v}\in\alpha = \tobjtv{F}{d}$. By
Definition~\ref{def:SelfTypes} (Self types), this
means that there exists $\alpha'\in\msn{Type}$ such that
$\tapprox{\alpha'}{k}\subseteq\tapprox{\alpha}{k}$ and conditions
\tagref{\textsc{Obj-2-Self}} and \tagref{\textsc{Obj-3}} are satisfied.
Choosing  $\alpha'' = \tapprox{\alpha'}{k}$, it is clear that
$\alpha''\tsubself\alpha$ since all the conditions only rely on $\alpha'$ to
approximation $k$.  Moreover, by instantiating $\Psi'=\Psi$ and 
$\tvobjx{l_e'}{e} = v$ in \tagref{\textsc{Obj-3}} we obtain that
$\mtuple{k-1,\tapprox{\Psi}{k-1},v}\in\alpha''$.
This proves the proposition.
\qed

\begin{lem}[{\textsc{SemLet-Str}}: Introducing structural assumptions]
\label{app:lemma:SemLet-Str}
Let $\alpha = \tobjtv{F}{d}$ and suppose that $\ttype{\Sigma}{a}{\alpha}$ and
that $\ttype{\mext{\Sigma}{x}{\xi}}{b}{\beta}$ for all $\xi\in\msn{Type}$ with
$\xi\tsubself\alpha$. Then $\ttype{\Sigma}{\tlet{x}{a}{b}}{\beta}$.
\end{lem}
\proof
Let $\alpha = \tobjtv{F}{d}$ and suppose that $\ttype{\Sigma}{a}{\alpha}$ and
that $\ttype{\mext{\Sigma}{x}{\xi}}{b}{\beta}$ for all $\xi\in\msn{Type}$ with
$\xi\tsubself\alpha$. We must show that $\ttype{\Sigma}{\tlet{x}{a}{b}}{\beta}$.
Thus, let $k\geq 0$, $\Psi$ and $\sigma$ be such that $\sigma :_{k, \Psi}
\Sigma$. By the definition of the semantic typing judgement
(Definition~\ref{def:SemanticTypingJudgement}) we
must show that
$\sigma(\tlet{x}{a}{b}):_{k,\Psi}\beta$, or equivalently (after suitable
$\alpha$-renaming and removing the syntactic sugar) that  
\begin{align*}
\tapp{(\tlam{x}{\sigma(b)})}{\sigma(a)}:_{k,\Psi}\beta
\end{align*}
Suppose $j < k, h, h'$ and $b'$ are such that 
\begin{gather}
\label{app:sem-let-str:A1}
h :_k \Psi\quad \wedge\quad 
\tjredc{h}{\tapp{(\tlam{x}{\sigma(b)})}{\sigma(a)}}{j}{h'}{b'}\quad \wedge\quad 
\mirred{\xcfg{h'}{b'}} 
\end{gather} 
From the second and third conjunct of \eqref{app:sem-let-str:A1} by the
operational semantics we have that for some $i\leq j<k$, some $h''$ and some
$\Psi''$, 
\begin{gather}
\label{app:sem-let-str:A4}
\tjred{\xcfg{\h}{\msub{a}{\sigma}}}{i}{\mirred{\xcfg{h''}{a''}}}
\quad\wedge\quad
\tjredc{\h''}{\tapp{(\tlam{x}{\sigma(b)})}{a''}}{j-i}{\h'}{b'}
\end{gather}
From the first conjunct together with the assumption $\ttype{\Sigma}{a}{\alpha}$ and the first
conjunct of \eqref{app:sem-let-str:A1}, by Definition~\ref{def:ClosedTermKPsiType} it
follows that there exists a heap typing $\Psi''$ such that
\begin{gather}
\label{app:sem-let-str:B1}
\textend{k}{\Psi}{k-i}{\Psi''}
\quad\wedge\quad
\h'' :_{k-i} \Psi''
\quad\wedge\quad
\mtuple{k-i,\Psi'',a''}\in\alpha = \tobjtv{F}{d}
\end{gather}
In particular, $a''\in\msn{Val}$ and the operational semantics gives 
\begin{gather}
\label{app:sem-let-str:A5}
\tjred{\xcfg{\h}{\tapp{(\tlam{x}{\sigma(b)})}{(\sigma(a))}}}{i}{\xcfg{h''}{\tapp{(\tlam{x}{\sigma(b)})}{a''}}}\to
\tjredc{\h''}{\msub{b}{\mext{\sigma}{x}{a''}}}{j-i-1}{\h'}{b'}
\end{gather}
From the third conjunct of \eqref{app:sem-let-str:B1} by 
Proposition~\ref{prop:self-type-exposure} there exists  $\alpha'\in\msn{Type}$
such that $\alpha'\tsubself\alpha$ and
$\mtuple{k-i-1,\tapprox{\Psi''}{k-i-1},a''}\in\alpha'$.
From $\sigma :_{k, \Psi} \Sigma$, the first conjunct
of \eqref{app:sem-let-str:B1}, Propositions~\ref{app:prop:P0} and \ref{app:prop:P1} and the closure 
under state extension, this yields 
\begin{align}
\label{app:sem-let-str:A6}
\mext{\sigma}{x}{a''}:_{k-i-1,\tapprox{\Psi''}{k-i-1}}\mext\Sigma{x}{\alpha'}
\end{align} 
Since $\alpha'\tsubself\alpha$,
by instantiating the universally quantified type $\xi$ in 
the hypothesis on $b$ we obtain that 
$\ttype{\mext{\Sigma}{x}{\alpha'}}{b}{\beta}$. Therefore,  \eqref{app:sem-let-str:A6} gives 
$\msub{b}{\mext{\sigma}{x}{a''}} :_{k-i-1,\tapprox{\Psi''}{k-i-1}}\beta$. 
Clearly $h'':_{k-i-1}\tapprox{\Psi''}{k-i-1}$ by the second conjunct of 
\eqref{app:sem-let-str:B1}, so that the second conjunct of
\eqref{app:sem-let-str:A4} shows that there is some $\Psi'$ such that 
$\textend{k}{\Psi}{k-i-1}{\tapprox{\Psi''}{k-i-1}}\sqsubseteq({k-j},{\Psi'})$, 
$\h' :_{k-j} \Psi'$ and $\mtuple{k-j,\Psi',b'}\in\beta$, by
Definition~\ref{def:ClosedTermKPsiType}. This establishes that 
$\sigma(\tlet{x}{a}{b}):_{k,\Psi}\beta$ holds as required.
\qed

We next define a recursive type of records $\tbeta{F}{d}$, which is the type 
arising from the recursive record interpretation of (imperative) objects
\cite{Bruce:Cardelli:Pierce:99}. While this type does not give rise to non-trivial 
subtyping, we will show that it satisfies $\tbeta{F}{d} \tsubself
\tobjtv{F}{d}$. 
\begin{defi}
\label{def:RecRecTypes}
Assume $F_d : \msn{Type}\to\msn{Type}$ are monotonic and non-expansive type
constructors, for all $d\in D$. Then let $\beta = \tbeta{F}{d}$ be defined as
the set of all triples $\mtuple{k,\Psi,\tvobj{d}}$ such that
\begin{align*}
\tag{{\sc Rec-1}}
&  (\forall d \in D.\; \mtuple{k,\Psi,l_d} \in \trefv{\tvar_d}{(\beta\to F_d(\beta))}) \\
\tag{{\sc Rec-2}}
\wedge \quad&  (\forall j<k.\; \forall \Psi'.\; \forall \tvobjx{l_e'}{e}.
\\[-.5mm]
\notag
& \quad\  
\textend{k}{\Psi}{j}{\Psi'} \mand
(\forall d \in D. \; \tapprox{\Psi'}{j}(l_d') =
\tapprox{\Psi}{j}(l_d))  
 \mimpl \mtuple{j, \tapprox{\Psi'}{j}, \tvobjx{l_d'}{d}} \in \beta)
\end{align*} 
\end{defi}

\noindent
Note that the recursive specification  of $\beta$ is well-founded, \IE $\beta$ is
well-defined. Moreover, $\beta$ is a type, \IE it is closed under state
extension.

\begin{prop}
\label{prop:rec-object-type-is-a-self-type}
For all self types $\tobjtv{F}{d}$ we have that \[\tbeta{F}{d} \tsubself
\tobjtv{F}{d}\]
\end{prop}
\proof
Let $\alpha = \tobjtv{F}{d}$ and $\beta = \tbeta{F}{d}$ for some arbitrary
monotonic and non-expansive type constructors $F_d$. It is clear that for all
$\mtuple{k,\Psi,\tvobj{d}}\in\beta$, conditions \tagref{\textsc{Obj-2-Self}}
and \tagref{\textsc{Obj-3}} from Definition~\ref{def:self-type-exposure} are
satisfied, by the definition of $\beta$ (Definition~\ref{def:RecRecTypes}). It
remains to prove that $\beta\subseteq\alpha$. We establish this by showing that
for all $k\geq 0$, $\tapprox{\beta}{k} \subseteq \tapprox{\alpha}{k}$, by
complete induction on $k$.  Let $\mtuple{k, \Psi, \tvobj{d}} \in \beta$; we need
to show that $\mtuple{k, \Psi, \tvobj{d}} \in \alpha$. We have that $D
\subseteq D$ and we choose $\alpha' = \beta$ which is a type and fulfills
$\tapprox{\beta}{k} \subseteq \tapprox{\alpha}{k}$ \tagref{\textsc{Obj-1}}
by the induction hypothesis.
The conditions \tagref{\textsc{Obj-2-Self}} and
\tagref{\textsc{Obj-3}} in Definition~\ref{def:SelfTypes} (Self types) are
exactly the same as conditions \tagref{\textsc{Rec-1}} and
\tagref{\textsc{Rec-2}} in the definition of $\beta$
(Definition~\ref{def:RecRecTypes}), which concludes the proof.
\qed

\begin{lem}[{\textsc{SemObj-Str}}: Object construction with structural
assumptions]
\label{app:lemma:SemObj-Str}
Let $\alpha = \tobjtv{F}{d}$ and suppose that for all $d\in D$ and all
$\xi\in\msn{Type}$ with $\xi\tsubself\alpha$, 
$\ttype{\mext{\Sigma}{x}{\xi}}{b_d}{F_d(\xi)}$. 
Then $\ttype{\Sigma}{\tobj{d}}{\alpha}$.
\end{lem}

\proof
Let $\alpha = \tobjtv{F}{d}$ and assume that 
\begin{gather}
\label{app:sem-obj-str:HH}
\forall d{\in} D.\; \forall\xi{\in}\msn{Type}.\ \xi\tsubself\alpha\mimpl 
\ttype{\mext{\Sigma}{x_d}{\xi}}{b_d}{F_d(\xi)}
\end{gather} 
We must show that 
$\ttype{\Sigma}{\tobj{d}}{\alpha}$.
Thus, let $k \geq 0$, $\sigma$ be a value environment and $\Psi$ be a heap typing
such that $\sigma :_{k, \Psi} \Sigma$.
By Definition~\ref{def:SemanticTypingJudgement} we need to show that 
$\msub{\tobj{d}}{\sigma} :_{k,\Psi} \alpha$. Equivalently (after suitable
$\alpha$-renaming), we show that
\begin{equation*}
\tobjx{\msub{b_d}{\sigma}}{d} :_{k,\Psi} \alpha
\end{equation*}

\noindent
Suppose $j < k, h, h'$ and $b'$ are such that the following three conditions are
fulfilled: 
\begin{gather}
\label{app:sem-obj-str:A1}
h :_k \Psi\quad \wedge\quad 
\tjredc{h}{\tobjx{\msub{b_d}{\sigma}}{d}}{j}{h'}{b'}\quad \wedge\quad 
\mirred{\xcfg{h'}{b'}} 
\end{gather} 
By the operational semantics  \textsc{Red-Obj} is the only rule that applies,
which means that necessarily $j=1$ and for some distinct $l_d \not \in \mdom{h}$
we have $b' = \tvobj{d}$ and 
\begin{align}
\label{app:sem-obj-str:A4}
h' & = \mextx{h}{l_d}{\tlam{x_d}{\msub{b_d}{\sigma}}}{d \in D}
\end{align}

\noindent
Let $\beta = \tbeta{F}{d}$, as in Definition~\ref{def:RecRecTypes}.
We choose
\begin{equation}
\label{app:sem-obj-str:A6}
\Psi' = \tapprox{\mextx{\Psi}{l_d}{(\tarr{\beta}{F_d(\beta)})}{d \in D}}{k-1}
\end{equation}
and show that
\begin{gather}
\label{app:sem-obj-str:S1}
\textend{k}{\Psi}{k-1}{\Psi'}\quad\wedge\quad 
h' :_{k-1} \Psi'\quad\wedge\quad
\mtuple{k-1, \Psi', b'} \in \alpha
\end{gather}

The first conjunct of \eqref{app:sem-obj-str:S1} holds by the
construction of $\Psi'$ \eqref{app:sem-obj-str:A6}. 
In order to show the second conjunct, let $i < k-1$ and $l \in
\mdom{\Psi'}$. 
We now need to show that 
$\mtuple{i, \tapprox{\Psi'}{i}, h'(l)} \in \Psi'(l)$. 
In case $l \in \mdom{\Psi}$ the proof proceeds exactly as for
Lemma~\ref{app:lemma:SemObj}, so we only consider the case when 
$l = l_d$ for some $d \in D$.
From \eqref{app:sem-obj-str:A4} and
\eqref{app:sem-obj-str:A6} we get that
\begin{gather*}
h'(l) = \tlam{x_d}{\msub{b_d}{\sigma}}
\quad\wedge\quad
\Psi'(l) = \tapprox{\tarr{\beta}{F_d(\beta)}}{k-1}
\end{gather*}
Thus we need to show that
\begin{align}
\label{app:sem-obj-str:B}
\mtuple{i, \tapprox{\Psi'}{i},
\tlam{x_d}{\msub{b_d}{\sigma}}} \in \tapprox{\tarr{\beta}{F_d(\beta)}}{k-1}
\end{align}
By Proposition~\ref{prop:rec-object-type-is-a-self-type} we obtain that $\beta
\tsubself \alpha$, so we can instantiate the universally quantified $\xi$ in
\eqref{app:sem-obj-str:HH} with $\beta$ and obtain that
\begin{gather*}
\ttype{\mext{\Sigma}{x_d}{\beta}}{b_d}{F_d(\beta)}
\end{gather*}
By {\sc SemLam} in Figure~\ref{fig:typing-lemmas:procedure-types}
(Lemma~\ref{lemma:all-proc-type-lemmas}) this gives us that
\begin{gather*}
\ttype{\Sigma}{\tlam{x_d}{b_d}}{\tarr{\beta}{F_d(\beta)}}
\end{gather*} 
From this and $\sigma :_{k, \Psi} \Sigma$ by
Definition~\ref{def:SemanticTypingJudgement} we obtain
\begin{equation}
\label{app:sem-obj-str:CH}
\tlam{x_d}{\msub{b_d}{\sigma}}\; :_{k,\Psi}\;{\tarr{\beta}{F_d(\beta)}}
\end{equation}
Since $k>1$ and from \eqref{app:sem-obj-str:A1} $h :_k \Psi$,
Proposition~\ref{app:prop:P3} shows that
 \eqref{app:sem-obj-str:CH} implies
\begin{align}
\label{app:sem-obj-str:B0'}
\mtuple{k, {\Psi},
\tlam{x_d}{\msub{b_d}{\sigma}}} \in \tarr{\beta}{F_d(\beta)} 
\end{align}
By Proposition \ref{app:prop:P1} we get that
$\textend{k-1}{\Psi'}{i}{\tapprox{\Psi'}{i}}$, which together with the first
conjunct of \eqref{app:sem-obj-str:S1} and the transitivity of $\sqsubseteq$ yields
$\textend{k}{\Psi}{i}{\tapprox{\Psi'}{i}}$. 
Since each $\tarr{\beta}{F_d(\beta)}$ is closed under state extension, the latter
property and 
 \eqref{app:sem-obj-str:B0'} imply the required \eqref{app:sem-obj-str:B}.

Finally, we need to show the third conjunct of \eqref{app:sem-obj-str:S1}, \IE
$\mtuple{k-1, \Psi', \tvobj{d}}
\in \alpha$. To this end, we prove the following more general claim:

\noindent \emph{Claim:}
For all $j_0 \geq 0$, for all $\Psi_0$ and for all $\tvobjx{\loc'_d}{d}$
\begin{multline}
\label{app:sem-obj-str:U}
\textend{k-1}{\Psi'}{j_0}{\Psi_0} 
\mand  (\forall d\in D.\;
\tapprox{\Psi_0}{j_0}(\loc'_d) = \tapprox{\Psi'}{j_0}({\loc_d}))\\
\mimpl 
\ \mtuple{j_0,\tapprox{\Psi_0}{j_0}, \tvobjx{\loc'_d}{d}}\in\beta
\end{multline}
From this and $\tapprox{\Psi'}{k-1} = \Psi'$, the last conjunct of 
\eqref{app:sem-obj-str:S1}
follows by taking $j_0 = k-1$, $\Psi_0 = \Psi'$, and $\loc'_d=\loc_d$ for all
$d\in D$, and by observing that $\sqsubseteq$  is
reflexive (Proposition~\ref{app:prop:P0}) and $\beta \tsub \alpha$
(since $\beta \tsubself \alpha$).

The claim above is proved by complete induction on $j_0$. So assume $j_0 \geq 0$
and $\Psi_0$ are such that 
\begin{gather}
\label{app:sem-obj-str:U0H}
\textend{k-1}{\Psi'}{j_0}{\Psi_0}
\end{gather} 
Moreover, for
all $d\in D$  let $\loc_d'\in\mdom{\Psi_0}$ such that
\begin{align}
\label{app:sem-obj-str:U1H}
\tapprox{\Psi_0}{j_0}(\loc'_d) = \tapprox{\Psi'}{j_0}(\loc_d)
\end{align}
We show that $\mtuple{j_0,\tapprox{\Psi_0}{j_0},\tvobjx{\loc'_d}{d}}\in\beta$,
by checking that the two conditions from the definition of $\beta$
(Definition~\ref{def:RecRecTypes}) are satisfied.
By the construction of $\Psi'$ in \eqref{app:sem-obj-str:A6}, together with
\eqref{app:sem-obj-str:U0H} and \eqref{app:sem-obj-str:U1H}, it follows that for
all $d \in D$
\begin{gather}
\label{app:sem-obj-str:U4-5} 
 \tapprox{\Psi_0}{j_0}(l_d') = \tapprox{\Psi'}{j_0}(l_d) =
 \tapprox{\tarr{\beta}{F_d(\beta)}}{j_0}
\end{gather}
By the definition of reference types (Definition~\ref{def:ReferenceTypes}) this
implies that
\begin{gather}
\label{app:sem-obj-str:Star}
\forall d \in D.\; \mtuple{j_0, \tapprox{\Psi_0}{j_0}, l'_d} \in
\trefv{\tinv}{(\tarr{\beta}{F_d(\beta)})}
\end{gather}
By the lemma for subtyping reference types ({\sc SemSubVarRef}
in Figure~\ref{fig:subtyping-reference-types}) we then obtain property
\textsc{(Rec-1)}:
\begin{gather}
\label{app:sem-obj-str:StarPrime}
\forall d \in D.\; \mtuple{j_0, \tapprox{\Psi_0}{j_0}, l'_d} \in
\trefv{\tvar_d}{(\tarr{\beta}{F_d(\beta)})}
\end{gather}

Second, we must prove \textsc{(Rec-2)}, \IE that for all $j<j_0$, $\Psi_1$ and
$\tvobjx{\loc''_d}{d}$
\begin{multline}
\label{app:sem-obj-str:U3}  
\textend{j_0}{\Psi_0}{j}{\Psi_1}\
 \wedge\; 
(\forall d\in D.\ \tapprox{\Psi_1}{j}(\loc''_d) = \tapprox{\Psi_0}{j}(\loc_d'))\\
 \Rightarrow\ 
\mtuple{j,\tapprox{\Psi_1}{j},\tvobjx{\loc''_d}{d}}\in\beta
\end{multline}
Note that this last condition holds vacuously in the base case
of the induction, when $j_0=0$. So assume $j<j_0$ and
$\Psi_1$ and $\loc''_d$ are such that $\textend{j_0}{\Psi_0}{j}{\Psi_1}$ and 
$\tapprox{\Psi_1}{j}(\loc''_d) = \tapprox{\Psi_0}{j}(\loc_d')$ for all $d\in 
D$. Now $j<j_0$ and assumption \eqref{app:sem-obj-str:U1H} yield that for all $d\in
D$
\begin{align*}
\tapprox{\Psi_1}{j}(\loc''_d) = 
\tapprox{\Psi_0}{j}(\loc'_d) 
= \tapprox{\tapprox{\Psi_0}{j_0}(\loc'_d)}{j} =
\tapprox{\tapprox{\Psi'}{j_0}(\loc_d)}{j} = \tapprox{\Psi'}{j}(\loc_d)
\end{align*}
Moreover, from $({k-1},{\Psi'})\sqsubseteq (j_0,\Psi_0)$
\eqref{app:sem-obj-str:U0H} and $(j_0,\Psi_0)\sqsubseteq({j},{\Psi_1})$, by the
transitivity of $\sqsubseteq$ we have that $({k-1},{\Psi'}) \sqsubseteq
({j},{\Psi_1})$.
Since $j<j_0$, the induction hypothesis of the claim gives
\begin{align*}
\mtuple{j,\tapprox{\Psi_1}{j},\tvobjx{\loc''_d}{d}}\in\beta
\end{align*}
and we have established \eqref{app:sem-obj-str:U3}. 

By Definition~\ref{def:RecRecTypes} applied to the type $\beta =
\tbeta{F}{d}$ the properties
\eqref{app:sem-obj-str:StarPrime}, and \eqref{app:sem-obj-str:U3} establish that indeed
$\mtuple{j_0,\tapprox{\Psi_0}{j_0},\tvobjx{\loc'_d}{d}}\in\beta$. This finishes
the inductive proof of claim \eqref{app:sem-obj-str:U}, and the proof of the lemma.
\qed

\subsection{Subtyping Lemma for Generalized Object Types}

\begin{lem}[{\sc SemSubGen-Obj}: Subtyping generalized object types]
\label{app:lemma:SemSubExtObj}
If $E \subseteq D$ and for all $e \in E$ we have that
$\beta^w_e \tsub \alpha^w_e$ and $\alpha^r_e \tsub \beta^r_e$ 
then $\tobjtxx{\alpha^w_d}{\alpha^r_d}{d} \tsub
\tobjtxx{\beta^w_e}{\beta^r_e}{e}$.
\end{lem}
\proof
Denote 
$\alpha \meqnot \tobjtxx{\alpha^w_d}{\alpha^r_d}{d}$,
$\beta \meqnot \tobjtxx{\beta^w_e}{\beta^r_e}{e}$,
and assume $E \subseteq D$ and
\begin{gather}
\label{app:sem-sub-ext-obj:G2}
\forall e \in E.~ (\beta^w_e \tsub \alpha^w_e \mand \alpha^r_e \tsub \beta^r_e).
\end{gather}
We prove that for all heap typings 
$\Psi$, for all values $v$ and all $k \geq 0$, if $\mtuple{k, \Psi, v} \in \alpha$ then 
$\mtuple{k, \Psi, v} \in \beta$, by complete induction on $k$.
The 
induction hypothesis is that 
$\tapprox{\alpha}{k} \tsub \tapprox{\beta}{k}$.

If we assume that $\mtuple{k, \Psi, v} \in \alpha$, then by the definition of
$\alpha$ (Definition \ref{def:ObjectTypes} with condition \textsc{(Obj-2-Gen)}
instead of \textsc{(Obj-2)}) we have that $v = \tvobj{c}$, $D \subseteq C$ and there
exists $\alpha' \in \msn{Type}$ such that $\tapprox{\alpha'}{k} \tsub
\tapprox{\alpha}{k}$ and
\begin{gather}
\label{app:sem-sub-ext-obj:H4}
\forall d \in D.\;\mtuple{k, \Psi, l_d} \in \treft{\alpha' \to
\alpha^w_d}{\alpha' \to \alpha^r_d}
\end{gather} 
Moreover, condition \textsc{(Obj-3)} holds with respect to $\alpha$, \IE 
for all $j<k$, all $\Psi'$ and all $\tvobjx{l_e'}{e}$ such that 
$\textend{k}{\Psi}{j}{\Psi'}$, 
\begin{gather}
\label{app:sem-sub-ext-obj:H5}
(\forall e \in E. \; \tapprox{\Psi'}{j}(l_e') = \tapprox{\Psi}{j}(l_e))\ 
 \mimpl\  \mtuple{j, \tapprox{\Psi'}{j}, \tvobjx{l_e'}{e}} \in \alpha'
\end{gather}
From $E \subseteq D$ and $D \subseteq C$ by transitivity $E \subseteq C$. 
From $\tapprox{\alpha'}{k} \tsub \tapprox{\alpha}{k}$ and the induction
hypothesis $\tapprox{\alpha}{k} \tsub \tapprox{\beta}{k}$  we get that
$\tapprox{\alpha'}{k} \tsub \tapprox{\beta}{k}$, \IE \textsc{(Obj-1)} holds. 
Moreover, \eqref{app:sem-sub-ext-obj:H5} entails that condition \textsc{(Obj-3)} also 
holds with respect to the object type $\beta$. 
So  in
order to conclude that $\mtuple{k, \Psi, v} \in \beta$, all that remains to be
proven is condition \textsc{(Obj-2-Gen)}:
\begin{equation}
\label{app:sem-sub-ext-obj:TS}
\forall e \in E.\; 
\mtuple{k, \Psi, l_e} \in \treft{\alpha' \to
\beta^w_e}{\alpha' \to \beta^r_e}
\end{equation}

Let $e \in E$. Since the procedure type constructor is covariant in the result
type ({\sc SemSubProc} in Figure~\ref{fig:typing-lemmas:procedure-types})
assumption \ref{app:sem-sub-ext-obj:G2} implies that
\begin{gather*}
\label{app:sem-sub-ext-obj:G22}
\alpha' \to \beta^w_e \tsub \alpha' \to \alpha^w_e \mand
\alpha' \to \alpha^r_e \tsub \alpha' \to \beta^r_e
\end{gather*}
From this by ({\sc SemSubRef-Gen}) we get that
\begin{equation*}
\treft{\alpha' \to \alpha^w_e}{\alpha' \to \alpha^r_e}
\tsub \treft{\alpha' \to \beta^w_e}{\alpha' \to \beta^r_e}
\end{equation*}
This together with \ref{app:sem-sub-ext-obj:H4} and $E \subseteq D$
directly implies \ref{app:sem-sub-ext-obj:TS}, which concludes the proof.
\qed

\end{document}